\documentclass[a4paper,11pt]{article}
\pdfoutput=1 

\usepackage{jheppub} 

\usepackage[T1]{fontenc} 

\usepackage{combelow}
\usepackage{bm}
\usepackage{braket}
\usepackage{slashed}
\usepackage{tikz}

\def\hatt{{\hat{t}}}
\def\hvarphi{{\hat{\varphi}}}
\def\hrho{{\hat{\rho}}}
\def\hvrho{{\hat{\varrho}}}
\def\hatz{{\hat{z}}}
\def\halpha{{\hat{\alpha}}}
\def\hbeta{{\hat{\beta}}}
\def\hgamma{{\hat{\gamma}}}
\def\hsigma{{\hat{\sigma}}}
\def\hlambda{{\hat{\lambda}}}

\title{\boldmath Helical massive fermions under rotation}

\author[a]{Victor E. Ambru\cb{s},}

\affiliation[a]{Department of Physics, West University of Timi\cb{s}oara, \\
Bd.~Vasile P\^arvan 4, Timi\cb{s}oara 300223, Romania}

\emailAdd{victor.ambrus@e-uvt.ro}

\abstract{
The properties of a massive fermion field undergoing rigid rotation at finite temperature and 
chemical potential are discussed. The polarisation imbalance is taken into account by 
considering a helicity chemical potential, which is dual to the helicity charge operator.
The advantage of the proposed approach is that, as opposed to the axial current, the helicity charge 
current remains conserved at finite mass. A computation of thermal expectation values of the 
vector, helicity and axial charge currents, as well as of the fermion condensate and 
stress-energy tensor is provided. In all cases, analytic constitutive equations are derived for the 
non-equilibrium transport terms, as well as for the quantum corrections to the equilibrium 
terms (which are derived using an effective relativistic kinetic theory model for fermions with 
helicity imbalance) in the limit of small masses. In the context of the parameters 
which are relevant to relativistic heavy ion collisions, the expressions derived in the 
massless limit are shown to remain valid for masses up to the thermal energy, except for 
the axial charge conductivity in the azimuthal direction, which presents strong variations 
with the particle mass.
}
\begin{document} 
\maketitle
\flushbottom

\section{Introduction}
\label{sec:intro}

Quantum systems undergoing rigid rotation have been the 
object of academic study since the late '70s, when Vilenkin 
showed that, due to the spin-orbit coupling, more anti-neutrinos 
would be emitted on the direction which is parallel to the 
angular momentum vector of a rotating black hole. Conversely,
an excess of neutrinos would be emitted on the anti-parallel 
direction \cite{vilenkin78}. Recently, evidence 
from relativistic heavy--ion collisions experiments
suggests that a global polarisation of the medium formed 
following the collision can be highlighted by looking at the 
decay products of the $\Lambda$ hyperons \cite{star17nat,star18prc}.
Such measurements can provide quantitative validation of models 
describing polarization of strongly interacting matter induced 
through various mechanisms \cite{kharzeev16}, such as
the chiral vortical effect \cite{rogachevsky10,baznat13,baznat18}
and the chiral magnetic effect \cite{fukushima08,braguta14}.

The past decade has seen many attempts to extend the standard
thermal field theory, in which the canonical thermodynamic variables 
are the temperature four-vector $\beta^\mu = u^\mu / T$ (given as the 
ratio between the local four-velocity of the fluid, $u^\mu$, and its temperature, $T$) and
chemical potentials related to the electric or baryon charges,
to also incorporate polarisation-related charges. Some recent examples 
include the recent results in Ref.~\cite{buzzegoli18}, where the 
excess is taken into account by means of an axial 
chemical potential, $\mu_A$, coupled with the axial charge current;
and Ref.~\cite{becattini19}, where it is suggested that an 
extra chemical potential, called the spin potential, could 
explain a polarisation excess through a coupling with the 
spin tensor.

There are two major drawbacks of the aforementioned approaches. First, 
the axial current $J^\mu_A = \overline{\psi} \gamma^\mu \gamma^5 \psi$ 
can be used to define a classically conserved axial charge, $Q_A$, 
only for massless fermions. In the case of massive 
particles, $J^\mu_A$ is no longer conserved. 
Second, the spin tensor (or its contraction with 
a constant spin potential) does not commute with the 
Dirac equation, which implies that its eigenfunctions 
are not solutions of the Dirac equation.

In this paper, an alternative to the axial charge current and spin tensor extensions of 
canonical thermal field theory is employed, which is based on the 
helicity charge current, introduced in Ref.~\cite{helican}. 
The advantage of the present approach is that the 
helicity charge current remains conserved for massive 
fermions and thus, one can consider 
the helicity operator $h$ (related to the temporal component of the 
Pauli-Lubanski vector operator) as a natural alternative
to the chirality operator for the regime of non-vanishing mass. 
It is defined through:
\begin{equation}
 h = \frac{\bm{S} \cdot \bm{P}}{p},
 \label{eq:h}
\end{equation}
where $\bm{S}$ and $\bm{P} = -i\nabla$ are the spin and momentum operators, while 
$p = |\bm{P}| = \sqrt{-\Delta}$ is the momentum magnitude (for a discussion 
on the properties of the fractional Laplacian, see Ref.~\cite{pozrikidis16}).
The eigenvalues of $h$ are $1/2$ and $-1/2$, distinguishing between right-handed 
and left-handed helicity particles, respectively.
At the relativistic quantum mechanics level, $h$ commutes 
with the Hamiltonian $H$ of the free Dirac field. 
With the help of $h$, it is possible to introduce the helicity 
charge current (HCC),
\begin{equation}
 J^\mu_H = \overline{\psi} \gamma^\mu h \psi + 
 \overline{h \psi} \gamma^\mu \psi,
\end{equation}
which is classically conserved when $\psi$ is a solution of the free Dirac equation. The associated time-independent charge, $Q_H$, can be associated to a helicity chemical potential $\mu_H$, which accounts for an overall helicity bias.

Despite the apparent non-locality of $h$ due to the factor $p^{-1}$ in 
its definition, the helicity can be regarded as a 
``good'' quantum number in quantum field theory. It has been routinely employed in canonical 
second quantisation to characterise particle states and for scattering processes in 
quantum electrodynamics (QED) \cite{itzykson80,weinberg95,peskin95}.
In quantum chromodynamics (QCD), the total quark helicity is conserved under the 
interactions due to the vector couplings with the gluon fields when the quarks are 
massless \cite{brodsky89,kapusta20arxiv}, which is a good approximation in the 
high-temperature quark-gluon plasma (QGP) formed in heavy-ion collisions (HIC).
Furthermore, the effects due to the helicity and spin imbalance in the QGP can 
be expected to be of similar importance, since their equilibration times are of 
a similar order of magnitude \cite{ruggieri16prd,kapusta20prc}.

This paper presents an analysis of the quantum thermal expectation values (t.e.v.s) of the vector and helicity charge currents (VCC and HCC), axial charge current (ACC), fermion condensate (FC) and stress-energy tensor (SET) corresponding to the free massive Dirac field in rigid rotation with angular velocity $\bm{\Omega}$, at finite vector ($\mu_V$) and helicity ($\mu_H$) chemical pontentials. 
It comes as an extension to the finite mass regime of Ref.~\cite{helican}, where the 
same quantities were investigated for massless fermions, in the presence of 
the vector ($\mu_V$), axial ($\mu_A$) and helical ($\mu_H$) chemical potentials. 
However, since the axial current $J^\mu_A = \overline{\psi} \gamma^\mu \gamma^5 \psi$
is not conserved for fermions of nonvanishing mass $M$ ($\partial_\mu J^\mu_A = 2 i M 
\overline{\psi} \gamma^5 \psi$), the axial chemical potential cannot be 
consistently introduced as a thermodynamic variable \cite{ruggieri20} and 
is thus assumed to vanish in this paper.

The t.e.v.s mentioned in the previous paragraph are computed as traces over 
Fock space, weighted by the density operator 
$\hvrho$ defining the thermal state. For global equilibrium states, such as the rigid 
rotation, $\hvrho$ can be written down in a straightforward manner with respect to 
conserved operators \cite{vilenkin80,itzykson80,kapusta89,laine16,mallik16}. 
Specifically, when $\bm{\Omega}$ is parallel to the $z$ axis, $\hvrho$ is given by
\cite{vilenkin80}
\begin{equation}
 \hvrho = \exp\left[-\beta_0 (\widehat{H} - 
 \Omega \widehat{M}^z - \mu_{V;0} \widehat{Q}_V - 
 \mu_{H;0} \widehat{Q}_H)\right],\label{eq:rho}
\end{equation}
where $\widehat{H}$, $\widehat{\bm{M}}$, $\widehat{Q}_V$ and $\widehat{Q}_H$ are 
the Hamiltonian, total angular momentum, vector charge and helicity charge operators,
respectively.
In non-equilibrium situations, the Zubarev formalism can be employed to
consistently define the density operator 
\cite{zubarev79,becattini15,mallik16,buzzegoli18,buzzegoli20phd}.
Several by-now traditional approaches can be employed for the analysis of 
rotating systems at finite temperature, including direct mode sums 
\cite{vilenkin80,casals12,ambrus14plb}, point splitting \cite{birrell82,duffy03,panerai16,ambrus16prd}
and the perturbative approach based on Kubo formulae in the
imaginary \cite{jeon95,kharzeev09prd,landsteiner11prl,laine16} or 
real \cite{schwinger61,keldysh65,danielewicz84,mallik16} time formalism. 
Full advantage may be taken of the standard diagramatics techniques 
in the perturbative approach by employing the Fourier transform to the momentum 
space. However, this comes at the cost that the rotation part of $\hvrho$,
$-\beta_0 \Omega \widehat{M}^z$, must be treated perturbatively, since 
$\widehat{M}^z$ does not commute with 
$\widehat{\bm{P}}$ and is therefore not diagonal with respect to the momentum 
eigenmodes \cite{landsteiner13lnp,becattini15,buzzegoli18}.

In this paper, the mode sum approach is employed to exactly compute the Fock 
space trace weighted by $\hvrho$ introduced in Eq.~\eqref{eq:rho}.
The key ingredient is the expansion of the field operator $\widehat{\Psi}$ 
with respect to the set of modes which are simultaneous eigenfunctions of the 
Hamiltonian $H$, linear ($P^z$) and angular momentum 
($M^z$) operators along $\bm{\Omega}$ and helicity operator $h$
\cite{ambrus14plb,jiang16,ambrus16prd,ebihara17plb,chernodub17njl,chernodub17edge,ambrus19lnp}. 
With respect to this basis, $\hvrho$ is diagonal and the thermal average 
is straightforward to compute.
Without resorting to perturbative expansions, the present approach yields 
exact integral expressions which are amenable to numerical integration for 
general values of the fermion mass $M$. In addition, analytic results 
are derived for all quantities, which are either exact at small mass 
at any distance from the rotation axis, or they are exact for any mass on the 
rotation axis. A similar method was used in Ref.~\cite{helican} for massless 
fermions.

A consequence of taking the non-perturbative approach is that the vacuum 
state must be carefully defined. As pointed out in the 
early '80s by Iyer \cite{iyer82}, the vacuum state of a fermion field 
corresponding to an observer undergoing rigid rotation differs in general 
from the stationary, Minkowski vacuum. The rotating vacuum proposed by Iyer 
treats as particle modes those modes for which the co-rotating energy
$\widetilde{E} = E - \Omega m$ (where $E$ is the Minkowksi energy and 
$m$ is the eigenvalue of $M^z$) is positive. 
In general, this choice includes modes with negative Minkowksi energy, $E < 0$.
This is contrary to the case of the 
Klein-Gordon field, where the stationary vacuum is the only possible choice 
for observers undergoing rigid rotation \cite{letaw80,letaw81,duffy03,ambrus14plb}. 
Moreover, the non-perturbative analysis of rigidly-rotating thermal states 
at finite temperature of the scalar field is not possible, since the t.e.v.s 
diverge due to the behaviour of the Bose-Einstein distribution 
$(e^{\widetilde{E} / T} -1)^{-1}$ when $\widetilde{E} \rightarrow 0$ 
\cite{kay91,ottewill00prd,ottewill00pla,ambrus14plb}.\footnote{It is noteworthy 
that this difficulty is not encountered in the perturbative 
approach \cite{becattini15}.}
When the rotating and Minkowski vacua do not coincide, the t.e.v.s computed 
with respect to the former exhibit temperature-independent contributions
\cite{vilenkin79,vilenkin80,ambrus14plb,ambrus19lnp}.
It is worth mentioning that the modes with $E \widetilde{E} < 0$ (giving rise to 
the difference between the stationary and rotating vacua) can be eliminated 
with appropriate boundary conditions, e.g. by enclosing the system inside a 
boundary \cite{duffy03,ambrus16prd} or when 
the space-time itself is bounded and the rotation parameter is sufficiently 
small \cite{nicolaevici01,ambrus16prd,ambrus16prd2,ambrus19apa}.

As a classical (i.e., non-quantum) reference theory, a simple kinetic model based on the Fermi-Dirac distribution is proposed, which takes into account the helicity bias by means of $\mu_H$. Unsurprisingly, the classical analysis predicts a perfect fluid behaviour, as expected since rigid-rotation is a global thermodynamic equilibrium solution in relativistic kinetic theory. This thermal equilibrium state is fully characterised by means of the vector $Q_V^{\rm RKT}$ and helicity $Q_H^{\rm RKT}$ charge densities, the energy density $E_{\rm RKT}$ and the pressure $P_{\rm RKT}$.

At the quantum level, the hydrodynamic content of the charge currents and the SET is extracted with the aid of the $\beta$ (thermometer) frame \cite{van12,van13,landsteiner13lnp,becattini15epjc}, in which the four velocity is that corresponding to rigid rotation.
The Landau frame can also be defined \cite{landsteiner11jhep,ambrus17plb,buzzegoli18}, 
but it is more cumbersome to use in the present context.
In the $\beta$ frame, the quantum corrections $\Delta Q_{V/H}(M)$, $\Delta E(M)$ and $\Delta P(M)$, taken with respect to the classical equilibrium quantities $Q^{\rm RKT}_{V/H}(M)$, $E_{\rm RKT}(M)$ and $P_{\rm RKT}(M)$, are investigated as functions of $M$. These quantum corrections are quadratic with respect to the vorticity parameter $\Omega$. The terms that deviate from the perfect fluid form describe anomalous transport phenomena in the form of rest frame charge and heat fluxes. These fluxes are characterised using the vortical charge and heat conductivities, $\sigma^\omega_\ell$ ($\ell \in \{V,A,H\}$) and $\sigma^\omega_\varepsilon$, as well as the circular charge and heat conductivities, $\sigma^\tau_\ell$ and $\sigma^\tau_\varepsilon$. These conductivities are computed analytically and numerically, reducing for $M = 0$ to the $\mu_A = 0$ limit of the results derived in Ref.~\cite{helican}. The robustness of the analytic expressions for the quantum corrections and conductivities is probed in the case of non-vanishing mass in the parameter regime which is prevalent in ultrarelativistic heavy ion collisions (labelled ``HIC'' throughout this paper), namely temperature $T = 150\ {\rm MeV}$ \cite{jacak12}, vector chemical potential $\mu_V = 30\ {\rm MeV}$ \cite{huang12} (considering that $\mu_V = \mu_B / 3$, where $\mu_B$ is the baryonic chemical potential) and $\Omega = 6.6\ {\rm MeV}$ (corresponding to $10^{22}\,{\rm s}^{-1}$ \cite{star17nat,wang17}). 

Before ending the introduction, it is worth pointing out that, since the helicity 
for a massive particle is not a Lorentz invariant property, the theory involving the 
helicity chemical potential becomes frame-dependent. The 
relevance of such a formulation can be seen in the case of systems that explicitly
break Lorentz invariance, such as the confined state encountered in a single nucleus 
or the quark-gluon plasma (QGP) undergoing rigid rotation. 

The outline of the paper is as follows. Preliminaries regarding the kinematics of 
rigidly-rotating states are introduced in Sec.~\ref{sec:RR}, where the 
kinematic tetrad consisting of the local velocity $u^\mu$, acceleration $a^\mu$, vorticity $\omega^\mu$ and a fourth vector $\tau^\mu$ is introduced. The helicity current is introduced in Sec.~\ref{sec:hel}. The kinetic theory model taking into account the helicity bias is formulated in Sec.~\ref{sec:RKT}. The finite temperature field theory formalism employed in this paper is summarised in Sec.~\ref{sec:therm} and the t.e.v.s of the  vector and helicity charge currents (VCC and HCC), axial charge current (ACC), fermion condensate (FC) and stress-energy tensor (SET) are discussed in Sections~\ref{sec:CC}, \ref{sec:AC}, \ref{sec:FC} and \ref{sec:SET}, respectively. 
Section~\ref{sec:conc} concludes this paper. Planck units are employed throughout this paper, such that $\hbar = k_B = c = 1$, while the Minkowski metric signature is taken as $(+,-,-,-)$. The Levi-Civita symbol 
is defined such that $\varepsilon^{0123} = (-g)^{-1/2}$.

\section{Kinematics of rigid rotation}\label{sec:RR}

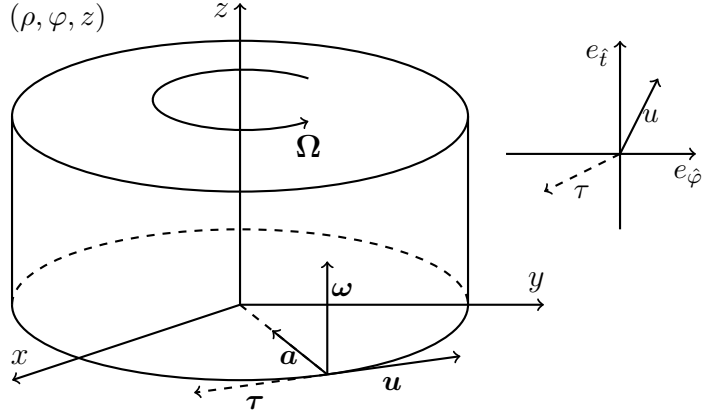
\begin{figure}
    \centering
\begin{tikzpicture}[x=1cm,y=1cm]
    \draw[thick] (0,0.5) ellipse (3 and 1);
    \draw[thick] (-3,-2) arc (180:360:3 and 1);
    \draw[dashed,thick] (-3,-2) arc (180:0:3 and 1);
    \draw[thick] (-3,0.5) -- (-3,-2);
    \draw[thick] (3,0.5) -- (3,-2);
    \draw[thick,->] (0.9,1) arc (45:315:1.2 and 0.4);
    \draw (0.9,0.1) node {\large $\bm{\Omega}$};
    \draw[thick,->] (0,-2)--(0,2);
    \draw (-0.25,1.9) node {\large $z$};
    \draw[thick,->] (0,-2)--(-3,-3);
    \draw (-2.9,-2.7) node {\large $x$};
    \draw[thick,->] (0,-2)--(4,-2);
    \draw (3.9, -1.7) node {\large $y$};
    \draw (-2.4,1.8) node {$(\rho, \varphi, z)$};
    \draw[thick,dashed] (0,-2)--(1.148,-2.924);
    \draw[thick,->] (1.148,-2.924) -- (0.422,-2.340);
    \draw[thick,->] (1.148,-2.924) -- (2.901,-2.682);
    \draw[thick,->] (1.148,-2.924) -- (1.148,-1.424);
    \draw[thick,dashed,->] (1.148,-2.924) -- (-0.605,-3.166);
    \draw (0.64,-2.7) node {${\bm{a}}$};
    \draw (2,-3.05) node {${\bm{u}}$};
    \draw (1.35,-1.8) node {${\bm{\omega}}$};
    \draw (0.2,-3.3) node {${\bm{\tau}}$};
    \draw[thick,->] (5,-1)--(5,1.5);
    \draw[thick,->] (3.5,0)--(6,0);
    \draw[thick,->] (5,0)--(5.5,1);
    \draw[thick,dashed,->] (5,0)--(4,-0.5);
    \draw (5.9,-0.3) node {$e_\hvarphi$};
    \draw (4.7,1.3) node {$e_\hatt$};
    \draw (5.4,0.5) node {$u$};
    \draw (4.5,-0.5) node {$\tau$};
\end{tikzpicture}
\caption{Schematic representation of the kinematic tetrad corresponding to a fluid undergoing rigid rotation with respect to the vertical axis, with angular velocity $\bm{\Omega} = \Omega \bm{e}_z$. The drawing on the left shows the spatial components of the vectors 
$\bm{u}$, $\bm{a}$, $\bm{\omega}$ and $\bm{\tau}$
forming the orthogonal tetrad, pointing along the azimuthal, radial, vertical and again azimuthal directions, respectively. The drawing on the right depicts the spacetime representation of the four-velocity ($u$) and the vector $\tau$.}
\label{fig:frame}
\end{figure}

The analysis in this paper is focussed on states undergoing rigid rotation about the $z$ axis. The properties of such states are most conveniently expressed with respect to cylindrical coordinates. For future convenience, the following tetrad is introduced:
\begin{align}
 e_\hatt =& \partial_t, &
 e_\hrho =& \partial_\rho, &
 e_\hvarphi =& \rho^{-1} \partial_\varphi, &
 e_\hatz = \partial_z,\nonumber\\
 \omega^\hatt =& dt, &
 \omega^\hrho =& d\rho, &
 \omega^\hvarphi =& \rho\, d\varphi, &
 \omega^\hatz = dz.
 \label{eq:RR_frame} 
\end{align}
In what follows, hatted indices are employed to refer to vector (or tensor) components expressed with respect to the above tetrad.

The four-velocity of a fluid undergoing rigid rotation can be written as:
\begin{equation}
 u = \Gamma(\partial_t + \Omega \partial_\varphi) = 
 \Gamma(e_\hatt + \rho \Omega e_\hvarphi),
 \label{eq:RR_u}
\end{equation}
such that its components with respect to the coordinate 
indices and with respect to the tetrad satisfy
\begin{equation}
 u^\hatt = u^t = \Gamma, \qquad 
 u^\hvarphi = \rho u^\varphi = \rho \Omega \Gamma.
 \label{eq:RR_u_comps}
\end{equation}
The Lorentz factor $\Gamma$,
\begin{equation}
 \Gamma = (1 - \rho^2 \Omega^2)^{-1/2},\label{eq:RR_Gamma}
\end{equation}
diverges as $\rho \rightarrow \rho_{\rm SLS}$, where
$\rho_{\rm SLS}$ is the distance from the rotation axis 
to the speed of light surface (SLS), on which the 
rigidly-rotating fluid rotates at the speed of light. 
It is given by:
\begin{equation}
 \rho_{\rm SLS} = \Omega^{-1}.\label{eq:RR_SLS}
\end{equation}

Aside from the tetrad introduced in Eq.~\eqref{eq:RR_frame}, it is convenient to define another tetrad, comprised of kinematic quantities derived from the four-velocity \cite{becattini15,becattini15epjc}.
Starting from the expression \eqref{eq:RR_u} for $u = u^\halpha e_\halpha$, the four-acceleration can be defined via:
\begin{equation}
 a = \nabla_u u = a^\hrho e_\hrho, \qquad 
 a^\hrho = -\rho \Omega^2 \Gamma^2.
 \label{eq:RR_a}
\end{equation}
It can be seen that $a \cdot u = 0$ by construction.
Taking into account the following expression for the gradient of the velocity vector,
\begin{equation}
 \nabla_\halpha u_\hbeta = \Omega \Gamma
 \begin{pmatrix}
  0 & 0 & 0 & 0 \\
  \rho \Omega \Gamma^2 & 0 & -\Gamma^2 & 0 \\
  0 & 1 & 0 & 0\\
  0 & 0 & 0 & 0
 \end{pmatrix},
\end{equation}
the kinematic vorticity vector, $\omega = \omega^\halpha e_\halpha$, can be defined through:
\begin{equation}
 \omega = \frac{1}{2} \varepsilon^{\halpha\hbeta\hgamma\hsigma}
 e_{\halpha} u_\hbeta \nabla_\hgamma u_\hsigma = 
 \omega^\hatz e_\hatz, \qquad 
 \omega^\hatz = \Gamma^2 \Omega,
 \label{eq:RR_omega}
\end{equation}
where the Levi-Civita tensor satisfies $\varepsilon^{\hat{0}\hat{1}\hat{2}\hat{3}} = +1$.
It can be seen that $\omega \cdot u = 0$ by construction. Moreover, 
in the case of rigid rotation, when the acceleration is given 
by Eq.~\eqref{eq:RR_a}, $\omega$ is orthogonal to both $u$ and $a$. 
A fourth vector which is orthogonal to $u$, $a$ and 
$\omega$ is (our definition differs from the one in 
Refs.~\cite{becattini15,becattini15epjc} by a minus sign)
\begin{equation}
 \tau^\halpha = 
 -\varepsilon^{\halpha\hbeta\hgamma\hsigma} 
 \omega_\hbeta a_\hgamma u_\hsigma, \qquad 
 \tau = -\rho \Omega^3 \Gamma^5(\rho \Omega e_\hatt + e_\hvarphi).
 \label{eq:RR_tau}
\end{equation}
It can be seen that the vectors $u$, $a$, $\omega$ 
and $\tau$ comprise an orthogonal tetrad. The squared norms of 
these vectors are given below:
\begin{align}
 u^2 =& 1, & 
 \tau^2 =& -\Omega^4 \Gamma^6(\Gamma^2 - 1),\nonumber\\
 \omega^2 =& -\bm{\omega}^2 = -\Omega^2 \Gamma^4, &
 a^2 =& -\bm{a}^2 = -\Omega^2 \Gamma^2(\Gamma^2 - 1).
 \label{eq:vecs_squares}
\end{align}
Figure~\ref{fig:frame} shows a schematic representation of these four vectors.

\section{Charge currents and conserved charges}\label{sec:hel}

\subsection{Classical theory}\label{sec:hel:class}

The free Dirac field is described by the following Lagrangian:
\begin{equation}
 L_{\rm free} = \frac{i}{2} (\overline{\psi} \gamma^\mu \partial_\mu \psi - 
 \overline{\partial_\mu \psi} \gamma^\mu \psi) - M \overline{\psi}
 \psi,\label{eq:L}
\end{equation}
where $M$ is the mass of the field quanta, 
$\psi$ is the Dirac 4-component spinor, 
$\overline{\psi} = \psi^\dagger \gamma^0$ is its Dirac adjoint
and $\gamma^\mu$ are the $4\times 4$ gamma matrices. 
For definiteness, these matrices together with the 
fifth gamma matrix, $\gamma^5 = i \gamma^0 \gamma^1 \gamma^2 \gamma^3$, 
are taken in the Dirac representation:
\begin{equation}
 \gamma^0 =
 \begin{pmatrix}
  1 & 0 \\ 0 & -1
 \end{pmatrix}, \qquad 
 \gamma^i = 
 \begin{pmatrix}
  0 & \sigma^i \\ -\sigma^i & 0,
 \end{pmatrix}, \qquad 
 \gamma^5 = 
 \begin{pmatrix}
  0 & 1 \\ 1 & 0 
 \end{pmatrix},
\end{equation}
where $\sigma^i$ are the Pauli matrices, given by:
\begin{equation}
 \sigma^1 = 
 \begin{pmatrix}
  0 & 1 \\ 1 & 0
 \end{pmatrix}, \qquad \sigma^2 = 
 \begin{pmatrix}
  0 & -i \\ i & 0
 \end{pmatrix}, \qquad \sigma^3 =
 \begin{pmatrix}
  1 & 0 \\ 0 & -1
 \end{pmatrix}.
 \label{eq:Dirac:representation}
\end{equation}
Noting that the spin matrix satsfies 
\begin{equation}
 S^k = \frac{i}{8} \varepsilon^{0ijk} [\gamma_i, \gamma_j]
 = \frac{1}{2} 
 \begin{pmatrix}
  \sigma^k & 0 \\ 0 & \sigma^k
 \end{pmatrix},\label{eq:Sdef}
\end{equation}
the helicity $h$, introduced in Eq.~\eqref{eq:h} 
can be shown to satisfy \cite{helican}
\begin{equation}
 h = \frac{1}{p} W^0 = \frac{\bm{S} \cdot \bm{P}}{p}
 = \gamma^5 \gamma^0 \frac{\bm{\gamma} \cdot \bm{P}}{2p},
 \label{eq:h_def}
\end{equation}
where $p = |\bm{p}|$ represents the momentum magnitude.
In Eq.~\eqref{eq:h_def}, $W^0$ is the zeroth component of the Pauli-Lubanski vector, 
defined as \cite{itzykson80}
\begin{equation}
 W^\mu = -\frac{1}{2} \varepsilon^{\mu \alpha\beta\lambda} M_{\alpha\beta} P_\lambda,
\end{equation}
which is expressed in terms of the total angular momentum operator 
$M_{\alpha\beta} = x_\alpha P_\beta - x_\beta P_\alpha + S_{\alpha\beta}$ 
and the momentum operator $P_\lambda = i \partial_\lambda$, where 
$S_{\alpha\beta} = \frac{i}{4}[\gamma_\alpha, \gamma_\beta]$. 
It can be shown that the eigenvalues of $h$ are $\lambda = \pm 1/2$ \cite{itzykson80}. 

The Dirac equation for the spinor $\psi$ and its adjoint 
$\overline{\psi}$ can be obtained using the Euler-Lagrange 
formalism, starting from the Lagrangian in Eq.~\eqref{eq:L}:
\begin{equation}
 i \slashed{\partial} \psi = M \psi, \qquad 
 i \overline{\slashed{\partial} \psi} = -M \overline{\psi},
 \label{eq:diraceq}
\end{equation}
where $\slashed{a} \equiv \gamma^\mu a_\mu$ is the Feynman slash notation.
The Dirac inner product of two 4-spinors $\psi$ and $\chi$ can be introduced as:
\begin{equation}
 \braket{\psi, \chi} = \int d^3x \, \overline{\psi}
 \gamma^0 \chi,\label{eq:scprod}
\end{equation}
being time independent when $\psi$ and $\chi$ satisfy 
the Dirac equation \eqref{eq:diraceq}.

The vector (VCC), axial (ACC) and helical (HCC) charge currents, 
denoted $J^\mu_V$, $J^\mu_A$ and $J^\mu_H$, respectively, 
can be introduced as follows \cite{helican}:
\begin{equation}
 J^\mu_V = \overline{\psi} \gamma^\mu \psi, \qquad 
 J^\mu_A = \overline{\psi} \gamma^\mu \gamma^5 \psi,\qquad 
 J^\mu_H = \overline{\psi} \gamma^\mu h \psi + \overline{h\psi} \gamma^\mu \psi.
 \label{eq:CC_class}
\end{equation}
To see the connection between the axial and helical charge currents, let  
$\psi = \psi_R + \psi_L = \psi_{\uparrow} + \psi_{\downarrow}$, 
where $\psi_{R/L} = \frac{1}{2}(1 \pm \gamma^5) \psi$ and 
$\psi_{\uparrow / \downarrow} = (\frac{1}{2} \pm h) \psi$. With this split, 
it can be seen that the axial and helical currents receive positive contributions from the 
right-chiral and right-helicity states, and negative contributions 
from the left-chiral and left-helicity states, respectively, as follows:
\begin{equation}
 J^\mu_A = \overline{\psi}_R \gamma^\mu \psi_R - \overline{\psi}_L \gamma^\mu \psi_L,\qquad 
 J^\mu_H = \overline{\psi}_{\uparrow} \gamma^\mu \psi_{\uparrow} - 
 \overline{\psi}_{\downarrow} \gamma^\mu \psi_{\downarrow}.
\end{equation}

Assuming that $\psi$ satisfies Eq.~\eqref{eq:diraceq}, it is straightforward 
to see that $\partial_\mu J^\mu_V = 0$ and thus, the vector current is 
classically conserved. For $J^\mu_A$, the mass term in Eq.~\eqref{eq:diraceq} 
breaks the conservation:
\begin{equation}
 \partial_\mu J^\mu_A = (i M \overline{\psi}) \gamma^5 \psi + 
 \overline{\psi} \gamma^5 (i M \psi) = 2i M \overline{\psi} \gamma^5 \psi.
 \label{eq:ACC_cons}
\end{equation}
Since $[P^\mu, W^\nu] = 0$, it can be seen that the 
divergence of $J^\mu_H$ is given by:
\begin{equation}
 \partial_\mu J^\mu_H = 
 \overline{\psi} [\gamma^\mu, h] \partial_\mu \psi + 
 \overline{[\gamma^\mu, h] \partial_\mu \psi} \psi= 0,
 \label{eq:HCC_cons}
\end{equation}
where the last equation follows after noting that $[\gamma^0, h] = 0$, while
\begin{equation}
 [\gamma^i, h] = -\frac{2i}{p} \gamma^5 \gamma^0 
 \varepsilon^{0ijk} P_j S_k = \frac{i}{p} \varepsilon^{0ijk} P_j \gamma_k.
\end{equation}
Thus, it can be seen that 
\begin{equation}
 [\gamma^\mu, h] \partial_\mu \psi = -i [\gamma^\mu, h] P_\mu \psi =
 \frac{1}{p} \varepsilon^{0ijk} P_j \gamma_k P_i \psi = 0.
\end{equation}

The associated charges can be computed by integrating the zeroth components of the 
currents with respect to $d^3x$. The following expressions are obtained:
\begin{equation}
 Q_V = \int d^3x\, \psi^\dagger \psi, \qquad 
 Q_A = \int d^3x\, \psi^\dagger \gamma^5 \psi, \qquad 
 Q_H = 2 \int d^3x\, \psi^\dagger h \psi.\label{eq:Q_def}
\end{equation}
To prove the last equality, it is convenient to consider the spatial 
Fourier decomposition of $\psi(t, \bm{x})$,
\begin{equation}
 \psi(t,\bm{x}) = \int \frac{d^3p}{(2\pi)^3} 
 e^{i \bm{p} \cdot \bm{x}} \psi_{\bm{p}}(t), \qquad 
 h \psi(t, \bm{x}) = \int \frac{d^3p}{(2\pi)^3} 
 e^{i \bm{p} \cdot \bm{x}} 
 \frac{\bm{S} \cdot \bm{p}}{p}
 \psi_{\bm{p}}(t).
\end{equation}
This allows $Q_H$ to be written as
\begin{equation}
 Q_H = \int \frac{d^3p}{(2\pi)^3} \left(\overline{\psi}_{\bm{p}} \gamma^0 
 \frac{\bm{S} \cdot \bm{p}}{p} \psi_{\bm{p}} + 
 \overline{\frac{\bm{S} \cdot \bm{p}}{p} \psi_{\bm{p}}} \gamma^0 \psi_{\bm{p}}
 \right). 
\end{equation}
Noting that $\overline{\bm{S}} = \bm{S}$ and $[\bm{S}, \gamma^0] = 0$, the second 
term in the above integrand can be seen to be equal to the first. Applying the inverse 
Fourier transform recovers Eq.~\eqref{eq:Q_def}.

The charges $Q_\ell$ ($\ell \in \{V, A, H\}$) can be shown to satisfy:
\begin{equation}
 \partial_t Q_V = 0, \qquad 
 \partial_t Q_A = 2i M \int d^3x\, \overline{\psi} \gamma^5 \psi, \qquad 
 \partial_t Q_H = 0.
 \label{eq:Q_cons}
\end{equation}
Thus, at the classical level and at arbitrary mass $M$, only the vector 
and helical charges, $Q_V$ and $Q_H$, are conserved.

\subsection{Charge operators}\label{sec:hel:Q}

After second quantization, $\psi$ is promoted to a quantum operator $\widehat{\Psi}$. 
The operator versions of the vector (VCC), axial (ACC) and helical (HCC) 
charge currents introduced in Eq.~\eqref{eq:CC_class} reads
\begin{equation}
 \widehat{J}^\mu_V = \frac{1}{2}[\widehat{\overline{\Psi}}, \gamma^\mu \widehat{\Psi}], \qquad 
 \widehat{J}^\mu_A = \frac{1}{2}[\widehat{\overline{\Psi}}, \gamma^\mu \gamma^5 \widehat{\Psi}], \qquad 
 \widehat{J}^\mu_H = \frac{1}{2}[\widehat{\overline{\Psi}}, \gamma^\mu h \widehat{\Psi}] + 
 \frac{1}{2}[\widehat{\overline{h \Psi}}, \gamma^\mu \widehat{\Psi}],
 \label{eq:CC_hat}
\end{equation}
where the commutators were introduced to avoid operator ordering ambiguities.
In order to compute the charge operators $\widehat{Q}_\ell$ ($\ell \in \{V,A,H\}$),
it is convenient to introduce a set of modes $U_j$ which are simultaneous eigenvectors 
of the Hamiltonian $H = i \partial_t$ and helicity $h$ operators:
\begin{equation}
 H U_j = E_j U_j, \qquad 
 h U_j = \lambda_j U_j,\label{eq:Uj_eigen}
\end{equation}
where $\lambda_j = \pm \frac{1}{2}$ is the helicity 
eigenvalue, while the energy $E_j$ is allowed to be negative. 
Since the chirality operator $\gamma^5$ does not commute with the 
Dirac Hamiltonian $H = i \partial_t = \gamma^0(-i \gamma^j \partial_j + M)$ 
for fermions of non-vanishing mass, its eigenvectors are not solutions 
of the Dirac equation. Starting from the Dirac equation, Eq.~\eqref{eq:diraceq},
it can be shown that
\begin{equation}
 h \psi = \frac{1}{2p} \gamma^5 H \psi - 
 \frac{M}{2p} \gamma^5 \gamma^t \psi.
\end{equation}
Manipulating the above expression for the case when $\psi \rightarrow U_j$, 
it can be seen that 
\begin{equation}
 \gamma^5 U_j = \frac{2\lambda_j}{p_j}(E_j - M \gamma^t) U_j,
 \label{eq:Uj_eigen_g5}
\end{equation}
explicitly showing that $U_j$ becomes an eigenfunction of $\gamma^5$ 
at non-vanishing mass.

Introducing now the anti-particle modes through the charge conjugation operation, 
$V_j = i \gamma^2 U_j^*$, it can be seen that Eqs.~\eqref{eq:Uj_eigen} and 
\eqref{eq:Uj_eigen_g5} entail:
\begin{equation}
 H V_j = -E_j V_j, \qquad 
 h V_j = \lambda_j V_j, \qquad 
 \gamma^5 V_j = -\frac{2\lambda_j}{p_j}(E_j + M \gamma^t) V_j.
 \label{eq:Vj_eigen}
\end{equation}
The modes $U_j$ and $V_j$ are assumed to be normalised with respect to 
the Dirac inner product \eqref{eq:scprod},
\begin{equation}
 \braket{U_j, U_{j'}} = \braket{V_j, V_{j'}} = \delta_{j,j'}, \qquad 
 \braket{U_j, V_{j'}} = 0.\label{eq:UV_norm}
\end{equation}

The field operator $\widehat{\Psi}$ can be expressed with respect to a complete set of 
particle ($U_j$) and anti-particle ($V_j$) modes, as follows:
\begin{equation}
 \widehat{\Psi} = \sum_j (U_j \hat{b}_j + V_j \hat{d}^\dagger_j),
 \label{eq:psi_hat}
\end{equation}
where $\hat{b}_j$ and $\hat{d}_j^\dagger$ are the 
particle annihilation and antiparticle creation operators,
respectively. These operators satisfy canonical anticommutation 
relations, i.e.
\begin{equation}
 \{\hat{b}_j, \hat{b}_{j'}^\dagger\} = \delta_{j,j'}, \qquad 
 \{\hat{d}_j, \hat{d}_{j'}^\dagger\} = \delta_{j,j'},
\end{equation}
while all other anticommutators vanish. Taking into account the normalisation 
relations in Eq.~\eqref{eq:UV_norm}, the charge operators can be expressed as:
\begin{align}
 :\widehat{Q}_V: =& \sum_j (\hat{b}_j^\dagger \hat{b}_j - 
 \hat{d}_j^\dagger \hat{d}_j),&
 :\widehat{Q}_H: =& 2\sum_j \lambda_j (\hat{b}_j^\dagger \hat{b}_j - 
 \hat{d}_j^\dagger \hat{d}_j),\label{eq:QVH_hat}
\end{align}
where the colons $::$ indicate normal (Wick) ordering, 
which, for operators that are quadratic in the one-particle 
operators $\hat{b}_j$ and $\hat{d}_j$, amounts to 
subtracting the vacuum expectation value:
\begin{equation}
 :\widehat{Q}_\ell: = \widehat{Q}_\ell - \braket{0 | \widehat{Q}_\ell | 0}.
 \label{eq:wick}
\end{equation}

In the case of massless fermions ($M = 0$), the eigenvectors $U_j$ and $V_j$ of the 
Hamiltonian and helicity operators are automatically eigenvectors of the 
chirality operator, $\gamma^5$, satisfying:
\begin{equation}
 \gamma^5 U_j = \chi_j U_j,\qquad 
 \gamma^5 V_j = -\chi_j V_j,
\end{equation}
where $\chi_j = \pm 1$ is the chirality eigenvalue. From 
Eqs.~\eqref{eq:Uj_eigen_g5} and \eqref{eq:Vj_eigen}, it can be seen 
that the helicity and chirality are linked through
\begin{equation}
 \chi_j = 2\lambda_j\, {\rm sgn}(E_j).
 \label{eq:lambdachi}
\end{equation}
The explicit action of the chirality operator on the helicity eigenvectors 
in the limit of vanishing mass is shown using explicit mode solutions in 
Eq.~\eqref{eq:therm:M0}. The helicity charge operator becomes diagonal, 
being given by
\begin{equation}
 :\widehat{Q}_A: = \sum_j \chi_j (\hat{b}_j^\dagger \hat{b}_j + 
 \hat{d}_j^\dagger \hat{d}_j).\label{eq:QA_hat}
\end{equation}

\subsection{Quantum anomalies} \label{sec:hel:anomal}

While conserved at the level of the classical theory, the charge currents 
$J_\ell^\mu$ ($\ell \in \{V,A,H\}$) are not guaranteed to be conserved at 
the quantum level, once interactions are taken into account. 
Typically, the conservation equations
are violated due to quantum anomalies, which are visible at the level of triangle 
Feynman diagrams. In the case of a theory which is symmetric under the 
$SU(3)_L \times SU(3)_R$ group, Bardeen \cite{bardeen69} showed that the 
conservation equation for the VCC can be achieved by adding counterterms to the 
action \cite{itzykson80,bertlmann96}, which have the unavoidable effect 
of breaking the conservation of the ACC.
Besides the effect of the terms uncovered by Bardeen, which originate from 
triangle, box and pentagon diagrams involving the vector ($V$) and
axial ($A$) vertices, the conservation of the ACC is violated also 
due to space-time metric fluctuations. This latter contribution 
can be revealed via the $ATT$ triangle diagram, involving two graviton 
($T$) vertices. For the special case of the $U(1)_V \times U(1)_A$ 
symmetry, the anomalous violation of the ACC conservation law can 
be put in the form \cite{itzykson80,bertlmann96,buzzegoli20phd}
\begin{align}
 \partial_\mu \widehat{J}^\mu_A =& 
 2i M \widehat{\overline{\Psi}}\gamma^5 \widehat{\Psi} + 
 \mathcal{A}_{AVV} + \mathcal{A}_{AAA} + \mathcal{A}_{ATT}, \nonumber\\
 \mathcal{A}_{AVV} =& -\frac{e_V^2}{16\pi^2} \varepsilon^{\mu\nu\alpha\beta}
 F^V_{\mu\nu} F^V_{\alpha\beta},\nonumber\\
 \mathcal{A}_{AAA} =& -\frac{e_A^2}{48\pi^2} \varepsilon^{\mu\nu\alpha\beta}
 F^A_{\mu\nu} F^A_{\alpha\beta},\nonumber\\
 \mathcal{A}_{ATT} =& \frac{1}{384\pi^2} \varepsilon^{\mu\nu\alpha\beta} R^\lambda{}_{\sigma\mu\nu} 
 \widetilde{R}^\sigma{}_{\lambda\alpha\beta},
 \label{eq:anomaly_A}
\end{align}
where $F^V_{\mu\nu}$ and $F^A_{\mu\nu}$ are the vector and axial field strengths, 
respectively, $e_V$ and $e_A$ are the corresponding charges, 
while $R^\lambda{}_{\sigma\mu\nu} = 
\Gamma^\lambda{}_{\alpha\nu} \Gamma^{\alpha}{}_{\sigma \mu} - 
\Gamma^\lambda{}_{\alpha\mu} \Gamma^{\alpha}{}_{\sigma \nu} +
\partial_\nu \Gamma^\lambda{}_{\sigma\mu} - 
\partial_\mu \Gamma^\lambda{}_{\sigma\nu}$ is the Riemann tensor,
$\Gamma^\lambda{}_{\alpha\beta} = \frac{1}{2} g^{\lambda\sigma}(\partial_\beta g_{\sigma\alpha} 
+ \partial_\alpha g_{\sigma\beta} - \partial_\sigma g_{\alpha\beta})$ is the Christoffel 
symbol and $g_{\mu\nu}$ is the space-time metric.
In Eq.~\eqref{eq:anomaly_A}, the anomalous contributions are written in terms of 
pieces coming from various triangle diagrams.

Aside from the non-conservation of the axial current,
the anomalous triangle diagrams can be related to anomalous transport laws, which 
can be revealed for fermions in an external electromagnetic field (e.g., the chiral 
magnetic effect \cite{fukushima08,braguta14}), or at finite vorticity (e.g., 
the chiral vortical effects \cite{rogachevsky10,baznat13,baznat18} or the 
helical vortical effects \cite{helican}). For concreteness, the 
results obtained in Ref.~\cite{helican} for the vortical charge and heat 
conductivities, $\sigma_\ell^\omega$ ($\ell \in \{V, A, H\}$) and $\sigma_\epsilon^\omega$,
are reproduced below:
\begin{gather}
 \sigma^\omega_V = \underbrace{\frac{2\mu_H T}{\pi^2} \ln 2}_{HTV?} + 
 \underbrace{\frac{\mu_V \mu_A}{\pi^2}}_{AVV}, \qquad 
 \sigma^\omega_A = \underbrace{\frac{T^2}{6}}_{ATT} + 
 \underbrace{\frac{\mu_V^2 + \mu_A^2 + \mu_H^2}{2\pi^2}}_{AVV, AAA, AHH?}, \qquad
 \sigma^\omega_H = \underbrace{\frac{2\mu_V T}{\pi^2} \ln 2}_{HTV?} + 
 \underbrace{\frac{\mu_H \mu_A}{\pi^2}}_{AHH?}, 
 \nonumber\\
 \sigma_\epsilon^\omega = \underbrace{\frac{\mu_A T^2}{3}}_{ATT} + 
 \underbrace{\frac{4\mu_V\mu_H T}{\pi^2} \ln 2}_{HTV?}+ 
 \underbrace{\frac{\mu_A^3 + 3 \mu_V^2 \mu_A + 3\mu_A \mu_H^2}{3\pi^2}}_{AAA, AVV, AHH?},
 \label{eq:anomalous_helican}
\end{gather}
where the terms of order $O(T^{-1})$ were omitted.
As pointed out by Landsteiner \cite{landsteiner11prl,landsteiner13lnp},
there is an intimate connection between each contribution appearing above 
and the triangle anomalies in the underlying quantum theory. The terms due 
to the $AVV$, $AAA$ and $ATT$ diagrams highlighted above are already known
\cite{landsteiner11prl,landsteiner13lnp}. The remaining terms involve the 
helical chemical potential $\mu_H$ or the helical vortical conductivity $\sigma^\omega_H$ 
and may originate from new anomalies related 
to triangle diagrams involving the $H$ vertex. In particular, the leading 
order terms in $\sigma^\omega_V$ and $\sigma^\omega_H$ together with the 
second term in $\sigma_\epsilon^\omega$ may originate from the $HTV$ diagram,
involving the $H$, $T$ and $V$ vertices. Similarly, the 
$\mu_H^2$ contribution to $\sigma^\omega_A$ and the $\mu_A \mu_H^2$ 
term appearing in $\sigma_\epsilon^\omega$ may trace their origin to the $AHH$ 
diagram \cite{helican}. A quantitative assessment of the 
structure of the anomalies involving the $H$ vertex is difficult 
to make in lack of an explicit computation, due to at least two 
factors. Firstly, the helical current is not a manifestly 
Lorentz-covariant quantity and therefore the anomaly may display 
non-covariant terms. Secondly, the helicity operator, and thus the 
HCC, is non-local in position space, due to the term $p^{-1} = (-\Delta)^{-1/2}$ 
in Eq.~\eqref{eq:h_def}, which may lead to unexpected divergences 
at the level of the triangle diagrams. A more thorough analysis 
of these new anomalies is beyond the scope of this work and is 
left as a subject for future work.


\section{Relativistic kinetic theory analysis}\label{sec:RKT}

Rigid body motion is fundamentally regarded as a
solution of the fluid equations for which dissipative 
processes are absent. From a kinetic theory perspective,
this corresponds to a state which in canonical 
thermodynamics is characterised via the temperature
four-vector, $\beta^\halpha = T^{-1} u^\halpha$. The 
state corresponds to global thermodynamic equilibrium 
when $\beta^\halpha$ satisfies the Killing equation. 
For rigid rotation characterised by the velocity in Eq.~\eqref{eq:RR_u}, this can be achieved when $T$ satisfies \cite{cercignani02}:
\begin{equation}
 T = \Gamma T_0,\label{eq:RKT_T}
\end{equation}
where $T_0$ is the temperature on the rotation axis 
and $\Gamma$ is the Lorentz factor given in Eq.~\eqref{eq:RR_Gamma}.

Recent works proposed the extension of the canonical 
formulation to account at the kinetic level for the degree 
of polarisation of the underlying quantum fluid. 
Starting from the Wigner function formalism,
discussed in, e.g., Ref.~\cite{degroot80},
Becattini {\it et al} proposed in Ref.~\cite{becattini13}
expressions which couple the thermal vorticity,
$\overline{\omega}_{\mu\nu} = -\frac{1}{2} (\nabla_\mu \beta_\nu - 
\nabla_\nu \beta_\mu)$,
to the spin operator. Recently, Florkowski {\it et al}
applied this formalism to obtain dynamic equations for 
the macroscopic polarisation in the frame of relativistic 
fluid dynamics with spin \cite{florkowski18a,florkowski18b},
however their analysis is performed in the frame of 
Boltzmann (classical) statistics.
Starting from the Wigner equation, Weickgenannt {\it et al} 
proposed an extension of the local equilibrium distribution 
as a function the collision invariants which takes into account 
the total angular momentum, expressed as the sum between the 
orbital and spin angular momenta \cite{weickgenannt19,weickgenannt20},
however, the equilibrium properties of the resulting system 
were not systematically explored.

In this section, a simple kinetic model is proposed to account 
for the equilibrium distribution of Fermi-Dirac particles with 
two possible helicities ($\lambda = 1/2$ and $-1/2$).
In addition to the vector chemical potential, $\mu_V$, which distinguishes between particles and anti-particles, a straightforward extension of the Fermi-Dirac distribution is considered 
that includes the helicity chemical potential, $\mu_H$. Requiring that $\mu_H$ distinguishes between polarisations as indicated by the corresponding quantum helicity charge operator $\widehat{Q}_H$, defined in Eq.~\eqref{eq:QVH_hat}, the following expression is obtained:
\begin{equation}
 f_{q/\overline{q}; \lambda}^{\rm (eq)} = \frac{1}{(2 \pi)^3} 
 \left\{\exp\left[\frac{1}{T}(p_\halpha u^\halpha \mp \mu_\lambda)\right] + 1\right\}^{-1}, \qquad 
 \mu_\lambda = \mu_V  + 2\lambda \mu_H,
 \label{eq:RKT_feq_gen}
\end{equation}
where $\mu_\lambda = \mu_V + 2 \lambda \mu_H$ is the total chemical
potential corresponding to right-handed ($\mu_+ = \mu_V + \mu_H$) and left-handed ($\mu_- = \mu_V -\mu_H$) particles. In the above, $p^\halpha$ and $\lambda$ represent the four-momentum
and helicity of the particle, which is assumed 
to have mass $M$ ($p^2 = M^2$).
Only one fermion species is considered, although multiple species or internal degrees of freedom, such as colour, can be accounted for at the kinetic level through an overall degeneracy factor, $g_s$ (the spin is already 
taken into account by considering $\lambda = \pm 1/2$).
It is worth pointing out that Eq.~\eqref{eq:RKT_feq_gen} loses 
Lorentz covariance since the helicity $\lambda$ is not 
a Lorentz scalar. It can be expected that a more fundamental 
formulation, starting from the theory of Wigner functions,
may lead to a manifestly covariant equilibrium distribution,
however such an analysis is beyond the scope of the 
present work. The model proposed herein serves just as 
a baseline to highlight the effects of taking into account 
a helicity bias using the standard chemical potential approach 
at the level of a classical theory.

In the absence of an established theory for the dynamics of the 
distribution of particles with spin, one can assume that  
$f_{q/\overline{q};\lambda} \equiv f_{q/\overline{q};\lambda}(\bm{x},\bm{p},t)$
obeys the relativistic Boltzmann equation \cite{cercignani02}:
\begin{equation}
 p^\mu \partial_\mu f_{q/\overline{q}; \lambda} = J[f_{q/\overline{q}; \lambda}].\label{eq:boltz}
\end{equation}
The collision operator $J[f]$ leading to the gas thermalisation should be implemented such that maximum entropy is ensured when $f_{q/\overline{q}; \lambda} = f^{\rm (eq)}_{q/\overline{q};\lambda}$, where the equilibrium distribution is given in Eq.~\eqref{eq:RKT_feq_gen}. In global thermodynamic equilibrium, $f_{q/\overline{q}; \lambda} = f^{\rm (eq)}_{q/\overline{q};\lambda}$ everywhere and the right hand side of Eq.~\eqref{eq:boltz} vanishes identically. Assuming that $\mu_H$ does not depend on the particle momentum $\bm{p}$, Eq.~\eqref{eq:boltz} is satisfied when
\begin{equation}
 \nabla_\hsigma (u_\halpha / T) + 
 \nabla_\halpha (u_\hsigma / T) = 0,\qquad
 \nabla_\halpha (\mu_V /T) = \nabla_\halpha (\mu_H /T) = 0.
 \label{eq:RKT_Killing}
\end{equation}
As mentioned at the beginning of this section, the above equations are satisfied when 
$u^\halpha /T$ is proportional to a Killing vector and the ratios $\mu_V / T$ and $\mu_H / T$ are constant.
Taking the solution corresponding to rigid rotation, given in Eq.~\eqref{eq:RR_u}, 
with the temperature given through Eq.~\eqref{eq:RKT_T}, it can be seen that 
the chemical potentials satisfy:
\begin{equation}
 \mu_V = \mu_{V;0} \Gamma, \qquad
 \mu_H = \mu_{H;0} \Gamma,
 \label{eq:RKT_mu}
\end{equation}
where $\mu_{V;0}$ and $\mu_{H;0}$ represent the values of the vector and helicity chemical potentials on the rotation axis.

The distributions \eqref{eq:RKT_feq_gen} can be specialised 
to the case summarised in Eq.~\eqref{eq:RR_u}:
\begin{align}
 f_{q/\overline{q}; \lambda}^{\rm (eq)} =& \frac{1}{(2 \pi)^3} 
 \left\{\exp\left[\frac{1}{T_0}(\widetilde{p^t} \mp \mu_{\lambda;0})\right] + 1\right\}^{-1},\nonumber\\
 \widetilde{p}^t =& p^t - \Omega M^z = p^t(1 - v^\hvarphi \rho \Omega),
 \label{eq:RKT_feq_rot}
\end{align}
where $\mu_{\lambda;0} = \mu_{V;0} + 2\lambda \mu_{H;0} 
= \mu_\lambda / \Gamma$ is the 
total chemical potential on the rotation axis.
In the above, $\widetilde{p}^t$ denotes the co-rotating energy of the 
particle, while the azimuthal velocity 
$v^\hvarphi = p^\hvarphi / p^\hatt$
is written in terms to the azimuthal component of the momentum,
$p^\hvarphi = -\sin\varphi p^x + \cos\varphi p^y$.
Since $-1 < v^\hvarphi < 1$, it can be seen that 
$\widetilde{p}^t$ satisfies:
\begin{equation}
 \widetilde{p}^t > 0,\label{eq:RKT_pt_positive}
\end{equation}
valid for any particle motion, as long as $\rho \Omega < 1$.
The above inequality will be essential in Sec.~\ref{sec:therm:vac}, 
where the second quantisation of the Dirac field is discussed.

In the zero temperature limit, Eq.~\eqref{eq:RKT_feq_rot} reduces to:
\begin{equation}
 \lim_{T \rightarrow 0} 
 f_{q/\overline{q};\lambda}^{\rm(eq)} =
 \frac{1}{(2\pi)^3} 
 \theta(\pm E^F_\lambda - \widetilde{p}^t), \qquad 
 E^F_{\lambda} = \mu_{\lambda;0}
 = \mu_{V;0} + 2 \lambda \mu_{H;0}, \label{eq:RKT_EF}
\end{equation}
where the Fermi level $E^F_{\lambda}$ is helicity-dependent.

Starting from the distributions \eqref{eq:RKT_feq_rot}, the 
vector charge current $J_V^{\halpha}$ (VCC),
helicity charge current $J_H^{\halpha}$ (HCC)
and stress-energy tensor $T^{\halpha\hbeta}$ (SET) 
can be computed as follows:
\begin{align}
 \begin{pmatrix}
  J^{{\rm RKT};\halpha}_V \\
  J^{{\rm RKT};\halpha}_H
 \end{pmatrix}
 =& \sum_\lambda
 \begin{pmatrix}
  1 \\ 2\lambda
 \end{pmatrix}
 \int \frac{d^3p}{p^\hatt} p^\halpha
 [f_{q;\lambda}^{\rm (eq)} - f_{\overline{q};\lambda}^{\rm (eq)}] = 
 \begin{pmatrix}
  Q^{\rm RKT}_V \\
  Q^{\rm RKT}_H 
 \end{pmatrix} u^\halpha, \nonumber\\
 T^{\halpha\hsigma}_{\rm RKT} =& \sum_\lambda 
 \int \frac{d^3p}{p^\hatt} p^\halpha p^\hsigma
 [f_{q;\lambda}^{\rm (eq)} + f_{\overline{q};\lambda}^{\rm (eq)}] = 
 (E_{\rm RKT} + P_{\rm RKT}) u^\halpha u^\hsigma - 
 P_{\rm RKT} \eta^{\halpha\hsigma},
 \label{eq:RKT_CC_SET}
\end{align}
where the charge densities $Q^{\rm RKT}_{V/H}$, 
energy density $E_{\rm RKT}$ and pressure $P_{\rm RKT}$ are given by:
\begin{align}
 \begin{pmatrix}
  Q^{\rm RKT}_V \\ Q^{\rm RKT}_H
 \end{pmatrix} =& \frac{1}{2\pi^2} \sum_{\lambda} 
 \begin{pmatrix}
  1 \\ 2\lambda
 \end{pmatrix}
 \int_0^\infty dp\, p^2 \left[
 \frac{1}{e^{(p^\hatt - \mu_\lambda)/T} + 1} - 
 \frac{1}{e^{(p^\hatt + \mu_\lambda)/T} + 1} \right],\nonumber\\
 \begin{pmatrix}
  E_{\rm RKT} \\ E_{\rm RKT} - 3P_{\rm RKT}
 \end{pmatrix}=& \frac{1}{2\pi^2} \sum_\lambda 
 \int_0^\infty \frac{p^2 dp}{p^\hatt} 
 \begin{pmatrix}
  (p^\hatt)^2 \\ M^2
 \end{pmatrix}
 \left[
 \frac{1}{e^{(p^\hatt - \mu_\lambda)/T} + 1} + 
 \frac{1}{e^{(p^\hatt + \mu_\lambda)/T} + 1}\right].
 \label{eq:RKT_gen}
\end{align}
The above results are obtained from Eq.~\eqref{eq:RKT_CC_SET} in two steps. The first step consists of contracting $J^{{\rm RKT};\halpha}_{p/h}$ with $u_\halpha$, as well as $T^{\halpha\hsigma}_{\rm RKT}$ with $u_\halpha u_\hsigma$ (for $E_{\rm RKT}$) and with $\eta_{\halpha\hsigma}$ (for $E_{\rm RKT} - 3P_{\rm RKT}$), respectively. Afterwards, a Lorentz transformation is performed, such that $u \cdot p \rightarrow p^\hatt$. This transformation is permitted by the Lorentz invariant integration measure ($d^3p / p^\hatt$), however, one must assume at this step that $\mu_{\lambda}/T = (\mu_V + 2\lambda \mu_H)/T$ is a Lorentz scalar. The vector part, $ \mu_V /T = \mu_{V;0}/T_0$, clearly satisfies this assumption. However, the second part, $2\lambda \mu_H/T = 2\lambda \mu_{H;0} / T_0$, is not a Lorentz scalar due to the polarisation prefactor, $2\lambda$. It is known that for massive fermions, the polarisation is frame dependent. This is generally true for the particles with azimuthal velocity $p^\hvarphi / m$ smaller than $u^\hvarphi = \rho \Omega$. In the vicinity of the rotation axis, these particles populate the infrared sector of the integral in Eq.~\eqref{eq:RKT_gen}, which makes small contributions due to the factor $p^2$ in the integrand. One can conclude that treating $2\lambda \mu_H / T$ as a Lorentz scalar can give a correct order of magnitude assessment of Eq.~\eqref{eq:RKT_CC_SET}, at least in the vicinity of the rotation axis. 

In the massless limit, $p^\hatt = p$,
the helicity becomes frame-independent, such that the approximation in Eq.~\eqref{eq:RKT_gen} becomes exact.
Using the relations in Eq.~\eqref{eq:FD}, the integrals in Eq.~\eqref{eq:RKT_gen} can be performed analytically:
\begin{align}
 \left.Q^{\rm RKT}_{V/H}\right\rfloor_{M\rightarrow 0} =& \frac{\mu_{V/H}}{3} 
 \left(T^2 + \frac{\mu_{V/H}^2 + 3\mu_{H/v}^2}{\pi^2} \right),\nonumber\\
 \left.E_{\rm RKT}\right\rfloor_{M\rightarrow 0} =&
 \frac{7\pi^2 T^4}{60}
 + \frac{T^2}{2}(\mu_V^2 + \mu_H^2) 
 + \frac{1}{4\pi^2}(\mu_{V}^4 + 
 6 \mu_{V}^2 \mu_{H}^2 + \mu_{H}^4), \nonumber\\
 \left.\frac{E_{\rm RKT} - 3P_{\rm RKT}}{M^2}\right\rfloor_{M\rightarrow 0} =& \frac{T^2}{6} + 
 \frac{1}{2\pi^2}(\mu_V^2 + \mu_H^2),
 \label{eq:RKT_M0}
\end{align}
where $T = T_0 \Gamma$, $\mu_V = \mu_{V;0} \Gamma$ 
and $\mu_H = \mu_{H;0} \Gamma$, 
while $E_{\rm RKT} = 3P_{\rm RKT}$ in general for massless constituents. 
For future reference, the massless 
limit of the ratio $(E -3P) / M^2$ between the trace 
of the SET and $M^2$ was also included. This quantity will be compared with 
the QFT equivalent (also equal to the ratio 
$\overline{\psi}\psi / M$ between the fermion condensate 
and $M$). It can 
be seen that in $Q_V^{\rm RKT}$ and $Q_H^{\rm RKT}$, the roles of 
the chemical potentials $\mu_V$ and $\mu_{H}$ 
are reversed, while the expressions for $E_{\rm RKT}$ and 
$(E_{\rm RKT} - 3P_{\rm RKT}) / M^2$ are symmetric under $\mu_V \leftrightarrow \mu_H$.
Furthermore, in the limit $\mu_H \rightarrow 0$ of 
vanishing helicity chemical potential, the results for 
$Q^{\rm RKT}_V$, $E_{\rm RKT}$ and $(E_{\rm RKT} - 3P_{\rm RKT}) / M^2$ coincide with those presented 
in Ref.~\cite{ambrus19lnp}, while $Q^{\rm RKT}_H$ vanishes.
For future reference, the results for the charge densities $Q_+$ and $Q_-$ corresponding 
to the right-handed and left-handed helicities, respectively, are presented below:
\begin{equation}
 Q^{\rm RKT}_\pm \equiv Q^{\rm RKT}_V \pm Q^{\rm RKT}_H,\qquad 
 \left.Q^{\rm RKT}_\pm \right\rfloor_{M\rightarrow 0} = \frac{\mu_\pm}{3} \left(T^2 + 
 \frac{\mu_\pm^2}{\pi^2}\right),\qquad 
 \mu_\pm = \mu_V \pm \mu_H.
 \label{eq:RKT_M0_Qpm}
\end{equation}

For small masses, a perturbative approach can be employed to derive the behaviour 
of the corrections to Eq.~\eqref{eq:RKT_M0} due to finite particle mass. 
First, the integration variable is switched from $p$ to $x = p^\hatt / M$ 
in Eq.~\eqref{eq:RKT_gen}:
\begin{align}
 Q^{\rm RKT}_\pm =& \frac{M^3}{\pi^2} \int_1^\infty dx\, x\sqrt{x^2 - 1} 
 \left[
 \frac{1}{e^{(xM - \mu_\pm)/T} + 1} - 
 \frac{1}{e^{(xM + \mu_\pm)/T} + 1} \right], \nonumber\\
 \begin{pmatrix}
  3P_{\rm RKT} \\ E_{\rm RKT} - 3 P_{\rm RKT}
 \end{pmatrix}
 =& \frac{M^4}{2\pi ^{2}} \sum_{\lambda} \int _{1}^{\infty } dx
 \begin{pmatrix}
 (x^2 - 1)^{3/2} \\ (x^2 - 1)^{1/2}
 \end{pmatrix}
 \left[\frac{1}{e^{(xM - \mu_\lambda)/T} + 1} + 
 \frac{1}{e^{(xM + \mu_\lambda)/T} + 1}\right].
 \label{eq:RKT_x}
\end{align}
When the chemical potentials are non-vanishing, 
the following expansions can be performed inside the integrands in 
Eq.~\eqref{eq:RKT_x}:
\begin{align}
 \frac{1}{2} \left[\frac{1}{e^{(xM - \mu_\lambda)/T} + 1} - \frac{1}{e^{(xM + \mu_\lambda)/T} + 1}\right]
 =& -\sum_{j = 0}^\infty \frac{(\mu_\lambda / M)^{2j+1}}{(2j+1)!}
 \frac{d^{2j+1}}{dx^{2j+1}} \left(\frac{1}{e^{x M/T} + 1}\right),\nonumber\\
 \frac{1}{2} \left[\frac{1}{e^{(xM - \mu_\lambda)/T} + 1} + \frac{1}{e^{(xM + \mu_\lambda)/T} + 1}\right]
 =& \sum_{j = 0}^\infty \frac{(\mu_\lambda / M)^{2j}}{(2j)!}
 \frac{d^{2j}}{dx^{2j}} \left(\frac{1}{e^{x M/T} + 1}\right). \label{eq:RKT_exp_aux1}
\end{align}
The Fermi-Dirac factor is now replaced using the following series representation:
\begin{equation}
 \frac{1}{e^{x M/T} + 1} = 
 \sum_{\ell = 1}^\infty (-1)^{\ell +1} e^{-\ell x M/T}. \label{eq:RKT_exp_aux2}
\end{equation}
The derivatives of orders $k = 2j+1$ and $2j$ with respect to $x$ appearing in Eq.~\eqref{eq:RKT_exp_aux1} can be computed automatically, giving factors of $(-\ell M/T)^k$.
Assuming that the sums over $j$ and over $\ell$ commute, the sum over $j$ can be performed by noting that:
\begin{align}
 \sum_{j = 0}^\infty \frac{z^{2j+1}}{(2j+1)!} =& \sinh z, &
 \sum_{j = 0}^\infty \frac{z^{2j}}{(2j)!} =& \cosh z.
\end{align}
This allows Eq.~\eqref{eq:RKT_exp_aux1} to be written as:
\begin{align}
 \frac{1}{2} \left[\frac{1}{e^{(xM - \mu_\lambda)/T} + 1} - \frac{1}{e^{(xM + \mu_\lambda)/T} + 1}\right]
 =& \sum_{\ell = 1}^\infty (-1)^{\ell+1}
 e^{-\ell x M/T} \sinh(\ell \mu_\lambda /T),\nonumber\\
 \frac{1}{2} \left[\frac{1}{e^{(xM - \mu_\lambda)/T} + 1} + \frac{1}{e^{(xM + \mu_\lambda)/T} + 1}\right]
 =& \sum_{\ell = 1}^\infty (-1)^{\ell+1}
 e^{-\ell x M /T} \cosh(\ell \mu_\lambda/T). \label{eq:RKT_exp_aux3}
\end{align}
It is clear that the above series converge only when $x M > \mu_\lambda$, however, 
the method can be further employed to recover the first correction due to small 
but non-vanishing $M$. After noting that
\begin{align}
 \sum_{\lambda = \pm \frac{1}{2}} 
 \begin{pmatrix}
  1 \\ 2\lambda 
 \end{pmatrix}
 \sinh(\ell \mu_\lambda/T) =& 
 2 \begin{pmatrix}
  \sinh(\ell \mu_V/T) \cosh(\ell \mu_H/T) \\
  \cosh(\ell \mu_V/T) \sinh(\ell \mu_H/T)
 \end{pmatrix},\nonumber\\
 \sum_{\lambda = \pm \frac{1}{2}} 
 \begin{pmatrix}
  1 \\ 2\lambda
 \end{pmatrix}
 \cosh(\ell \mu_\lambda/T) =& 2 
 \begin{pmatrix}
  \cosh(\ell \mu_V/T) \cosh(\ell \mu_H/T) \\
  \sinh(\ell \mu_V/T) \sinh(\ell \mu_H/T)
 \end{pmatrix},\label{eq:RKT_exp_aux4}
\end{align}
the following expressions are obtained:
\begin{align}
 Q^{\rm RKT}_\pm =& \frac{2M^2T}{\pi^2} 
 \sum_{\ell = 1}^\infty \frac{(-1)^{\ell + 1} }{\ell}
 \sinh(\ell \mu_\pm/T)
 \int_1^{\infty} dx \frac{2x^2 - 1}{\sqrt{x^2 - 1}} e^{-\ell x M/T},\nonumber\\
\begin{pmatrix}
 3P_{\rm RKT}\\ 
 E_{\rm RKT} - 3P_{\rm RKT}
\end{pmatrix} =& \frac{2M^4}{\pi^2}
\sum_{\ell = 1}^\infty (-1)^{\ell + 1} 
\frac{\cosh(\ell \mu_V/T)}{{\rm sech}(\ell \mu_H/T)}
\int_1^\infty dx 
\begin{pmatrix}
(x^2 - 1)^{3/2} \\ (x^2 - 1)^{1/2}
\end{pmatrix} e^{-\ell xM/T}.
\label{eq:RKT_xjlF}
\end{align}

The integrals with respect to $x$ in Eq.~\eqref{eq:RKT_xjlF} can be performed by employing \cite{olver10}:
\begin{equation}
 K_\nu(z) = \frac{\sqrt{\pi} (z/2)^{\nu}}{\Gamma(\nu + \frac{1}{2})} 
 \int_1^\infty dt\, e^{-zt} (t^2 - 1)^{\nu - \frac{1}{2}}.
\end{equation}
where $K_\nu(z)$ is a modified Bessel function of the third kind and 
$\nu$ is assumed to satisfy ${\rm Re}(\nu) > -1/2$.
Setting $\nu = n \in \mathbb{N}$ yields:
\begin{equation}
 \int_1^\infty dx\, (x^2 - 1)^{n - \frac{1}{2}} e^{-\ell x M/T} 
 = (2n -1)!! \frac{K_n(\ell M/T)}{(\ell M/T)^n},
 \label{eq:K_int}
\end{equation}
where $(2n -1)!! = 1 \cdot 3 \cdot 5 \cdots (2n-1)$ is the double factorial
and the convention that $(-1)!! = 1$ is used when $n = 0$. After employing the above
steps, Eq.~\eqref{eq:RKT_xjlF} can be put in 
the following form \cite{florkowski15jpg}:
\begin{align}
 Q^{\rm RKT}_\pm =& \frac{2 M^2 T}{\pi^2} 
 \sum_{\ell = 1}^\infty \frac{(-1)^{\ell + 1}}{\ell} 
 \sinh(\ell \mu_\pm/T)
 \left[\frac{2 K_1(\ell M/T)}{\ell M/T} + 
 K_0(\ell M/T)\right],\nonumber\\
 \begin{pmatrix}
  P_{\rm RKT} \\ E_{\rm RKT} - 3P_{\rm RKT}
 \end{pmatrix} =& \frac{2 M^4}{\pi^2} 
 \sum_{\ell = 1}^\infty (-1)^{\ell + 1} 
 \frac{\cosh(\ell \mu_V/T)}{{\rm sech}(\ell \mu_H/T)}
 \begin{pmatrix}
 K_2(\ell M/T) / (\ell M/T)^2\\
 K_1(\ell M/T) / (\ell M/T)
 \end{pmatrix}.
 \label{eq:RKT_expr}
\end{align}
The massless limit result given in Eq.~\eqref{eq:RKT_M0} can 
be recovered from Eq.~\eqref{eq:RKT_expr} using the limiting 
behaviour \cite{olver10}
\begin{align}
 K_0(z) =& -\gamma - \ln (z/2) + O(z^2),\nonumber\\
 z^{-1} K_1(z) =& \frac{1}{z^2} + \frac{1}{2} \left[-\frac{1}{2} + 
 \gamma + \ln (z/2)\right] + O(z^2),\nonumber\\
 z^{-2} K_2(z) =& \frac{2}{z^4} - \frac{1}{2z^2} + 
 \frac{1}{8} \left[\frac{3}{4} - \gamma - \ln(z/2)\right] 
 + O(z^2),\label{eq:K_smallz}
\end{align}
where $\gamma = 0.577216\dots$ is the Euler-Mascheroni constant.
The above small $z$ expansion can be used to compute the contributions
to Eq.~\eqref{eq:RKT_expr} corresponding to various orders of the 
mass, $M$. This procedure can be used only for the first few terms 
in this series, since the expansion in Eq.~\eqref{eq:K_smallz} is 
not uniformly convergent. Indeed, since $z = \ell M/T$, it is 
clear that the infinite sums over $\ell$ of the terms corresponding 
to positive powers of $\ell$ are divergent. Furthermore, 
the terms with $z^0$ in Eq.~\eqref{eq:K_smallz} are 
independent of $T$, $\mu_V$ and $\mu_H$. 
With the above discussion in mind, the following
$O(M^2)$ corrections to the massless results in Eq.~\eqref{eq:RKT_M0} can be obtained:
\begin{align}
  Q^{\rm RKT}_\pm
 =& \mu_\pm
 \left( \frac{T^2}{3} + \frac{\mu_\pm^2}{3\pi^2} 
 - \frac{M^2}{2\pi^2} \right),\nonumber\\
 \begin{pmatrix}
 P_{\rm RKT} \\ E_{\rm RKT}
 \end{pmatrix} =& 
 \begin{pmatrix}
  1/3 \\ 1
 \end{pmatrix}
 \left[\frac{7\pi^2 T^4}{60}
 + \frac{T^2}{4}(\mu_+^2 + \mu_-^2)
 + \frac{\mu_+^4 + \mu_-^4}{8\pi^2}\right]- \frac{M^2}{12} \left[T^2 + 
 \frac{3(\mu_+^2 + \mu_-^2)}{2\pi^2} \right],
 \label{eq:RKT_M_corr}
\end{align}
where $Q^{\rm RKT}_{\pm}$ and $\mu_\pm$ were introduced in Eq.~\eqref{eq:RKT_M0_Qpm}.
The correction to $(E_{\rm RKT} - 3P_{\rm RKT}) / M^2$ cannot be obtained
using this method. Since $M$ is a constant while $T = T_0 \Gamma$ and 
$\mu_\pm =  \mu_{\pm;0} \Gamma$ increase with the distance $\rho$ from the 
rotation axis, it can be seen that the mass corrections make subleading contributions 
in the vicinity of the SLS, where $\Gamma \rightarrow \infty$. 

\section{Quantum rigidly-rotating thermal states}
\label{sec:therm}

This section begins with a review of the mode solutions of the 
Dirac equation which are helicity eigenvectors, discussed in 
Subsec.~\ref{sec:therm:modes}. The massless limit and the rest 
frame limit are also discussed therein. The second quantisation 
procedure and a review of the stationary (Minkowski) and 
rotating vacua is provided in Subsec.~\ref{sec:therm:vac}. 
The formalism employed for the computation of thermal expectation 
values (t.e.v.s) is briefly reviewed in Subsec.~\ref{sec:therm:tevs}.
Some analytical techniques which will be employed in the analysis 
of the t.e.v.s are summarised in Subsec.~\ref{sec:therm:tricks}.

\subsection{Mode solutions}\label{sec:therm:modes}

At the relativistic quantum mechanics level, the Dirac equation can be solved in terms of a 
complete set of particle and anti-particle mode solutions, denoted using $U_j$ and $V_j$.
The particle mode solutions $U_j$ can be taken to simultaneously satisfy the following 
eigenvalue equations:
\begin{equation}
 H U_j = E_j U_j, \qquad
 P^z U_j = k_j U_j, \qquad
 M^z U_j = m_j U_j, \qquad
 h U_j = \lambda_j U_j,
\end{equation}
where $j = (E_j, k_j, m_j, \lambda_j)$ denotes collectively the eigenvalues of the Hamiltonian $H = i \partial_t$, $z$ components of the momentum $P^z = -i \partial_z$ and total angular momentum $M^z = -i\partial_\varphi + S^z$, and helicity operator $h = \bm{S} \cdot \bm{P} / p$. The explicit expressions for $U_j$ were obtained in Refs.~\cite{ambrus14plb,ambrus16prd} and are reproduced below without derivation:
\begin{gather}
 U_j = \frac{e^{-i E_j t + i k_j z}}{2\pi} u_j, \qquad 
 u_j = \frac{1}{\sqrt{2}} 
 \begin{pmatrix}
  \mathfrak{E}^+_j \phi_j \smallskip \\
  \frac{2\lambda_j E_j}{|E_j|} 
  \mathfrak{E}^-_j \phi_j
 \end{pmatrix},\nonumber\\
 \phi_j = \frac{1}{\sqrt{2}} 
 \begin{pmatrix}
  \mathfrak{p}^+_j e^{i (m_j - \frac{1}{2}) \varphi} 
  J_{m_j - \frac{1}{2}} (q_j \rho) \smallskip\\
  2i\lambda_j \mathfrak{p}^-_j e^{i (m_j + \frac{1}{2}) \varphi} 
  J_{m_j + \frac{1}{2}} (q_j \rho) \\
 \end{pmatrix},\nonumber\\
 \mathfrak{E}_j^\pm = \left(1 \pm \frac{M}{E_j}\right)^{1/2},\qquad
 \mathfrak{p}_j^\pm = \left(1 \pm \frac{2\lambda_j k_j}{p_j}\right)^{1/2},
 \label{eq:U}
\end{gather}
where $q_j = \sqrt{p_j^2 - k_j^2}$ and $p_j = \sqrt{E_j^2 - M^2}$.
In the above, the sign of the energy $E_j$ is left arbitrary.
The anti-particle modes $V_j$ are obtained via the 
charge conjugation operation:
\begin{equation}
 V_j = i \gamma^2 U_j^* = (-1)^{m_j - \frac{1}{2}}
 \frac{i E_j}{|E_j|} U_{\overline{\jmath}},
 \label{eq:Vj}
\end{equation}
where $\overline{\jmath} = (-E_j, -k_j, -m_j, \lambda_j)$ when 
$j = (E_j, k_j, m_j, \lambda_j)$.
The modes $U_j$ and $V_j$ are normalised with respect to
the Dirac inner product \eqref{eq:scprod} according to:
\begin{gather}
 \braket{U_j, U_{j'}} = \braket{V_j, V_{j'}} = \delta(j, j'), \qquad
 \braket{U_j, V_{j'}} = 0, \nonumber\\
 \delta(j,j') = \frac{\delta(E_j - E_{j'})}{|E_j|} 
 \delta(k_j - k_{j'}) \delta_{m_j, m_{j'}}
 \delta_{\lambda_j, \lambda_{j'}}.
 \label{eq:ortho}
\end{gather}

In the massless case, $\mathfrak{E}^\pm_j = 1$ and it is easy to see that the particle modes $U_j$ and their respective anti-particle modes become eigenstates of the chirality operator $\gamma^5$:
\begin{equation}
 \gamma^5 U_j\rfloor_{M=0} = \frac{2\lambda_j E_j}{|E_j|} U_j\rfloor_{M=0}, \qquad 
 \gamma^5 V_j\rfloor_{M=0} = -\frac{2\lambda_j E_j}{|E_j|} V_j\rfloor_{M=0},
 \label{eq:therm:M0}
\end{equation}
thereby confirming explicitly that the chirality and helicity eigenmodes coincide. It can be seen that the chirality eigenvalues $\chi_j$ satisfy $\chi_j = 2\lambda_j$ for the positive energy modes and $\chi_j =-2\lambda_j$ for the negative energy modes, confirming the result in Eq.~\eqref{eq:lambdachi}.

Before concluding this subsection, it is worth discussing the properties of $U_j$ and $V_j$ in the rest frame, where $k_j = q_j = 0$. Since the Bessel functions $J_n(z)$ vanish for $n =\pm 1, \pm 2, \dots$, only the modes with $m = \pm 1/2$ are non-trivial. In this case, the particle modes $U_j \equiv U^{\lambda_j}_{E_j,k_j,m_j}$ and their corresponding anti-particle modes are given by:
\begin{align}
 U^\lambda_{M,0, 1/2}(x)\rfloor_{\rm rest} =& \frac{e^{-iM t}}{4\pi} 
 \begin{pmatrix}
  \mathfrak{E}^+ \\ 0 \\ 2\lambda \mathfrak{E}^- \\ 0
 \end{pmatrix}, & 
 U^\lambda_{M,0,- 1/2}(x)\rfloor_{\rm rest} =& \frac{i e^{-iM t}}{4\pi} 
 \begin{pmatrix}
  0 \\ 2\lambda \mathfrak{E}^+ \\ 0 \\ \mathfrak{E}^-
 \end{pmatrix}, \nonumber\\
 V^\lambda_{M,0, 1/2}(x)\rfloor_{\rm rest} =& \frac{e^{iM t}}{4\pi} 
 \begin{pmatrix}
  0 \\ -2\lambda \mathfrak{E}^- \\ 0 \\ \mathfrak{E}^+
 \end{pmatrix}, & 
 V^\lambda_{M,0,- 1/2}(x)\rfloor_{\rm rest} =& -\frac{i e^{iM t}}{4\pi} 
 \begin{pmatrix}
  \mathfrak{E}^- \\ 0 \\ -2\lambda \mathfrak{E}^+ \\ 0
 \end{pmatrix}.
 \label{eq:therm:rest}
\end{align}
It can be seen that the above spinors become eigenmodes of the spin operator, $S^z$:
\begin{align}
 S^z U^\lambda_{M,0, \pm 1/2}(x)\rfloor_{\rm rest} =& 
 \pm\frac{1}{2} U^\lambda_{M,0, \pm 1/2}(x)\rfloor_{\rm rest}, \nonumber\\
 S^z V^\lambda_{M,0, \pm 1/2}(x)\rfloor_{\rm rest} =& 
 \mp\frac{1}{2} V^\lambda_{M,0, \pm 1/2}(x)\rfloor_{\rm rest}.
\end{align}
Thus, $m= \pm 1/2$ reduces to the spin quantum number. Even though the momentum vanishes in the rest frame, the helicity $\lambda = \pm1/2$ remains a well-defined quantum number, distinguishing between two linearly independent solutions, as predicted by the orthogonality relation in Eq.~\eqref{eq:ortho}.

\subsection{Second quantisation}\label{sec:therm:vac}

In order to promote the wave function $\psi(x)$ to the quantum operator $\widehat{\Psi}(x)$, it is necessary to define the vacuum state. Two vacuum states shall be considered: the Minkowski (non-rotating) vacuum state, denoted using $M$, and the co-rotating vacuum state, denoted using $\Omega$. The difference between these vacuum states is explained by Iyer in Ref.~\cite{iyer82}. For the stationary state, the natural definition of the vacuum state is to interpret the states with positive Hamiltonian eigenvalues, $E_j > 0$, as particle states. The states with $E_j < 0$ are then anti-particle states. This is achieved by expanding the field operator $\widehat{\Psi}$ with respect to the modes $(U_j, V_j)$ while considering only positive values of $E_j$:
\begin{align}
 \widehat{\Psi}(x) =& \sum_{\lambda = \pm \frac{1}{2}}
 \sum_{m = -\infty}^\infty
 \int_{|E| > M} dE \, |E|\,
 \int_{-p}^p dk \,\theta(E)
 \left[U_{E,k,m}^\lambda(x) \hat{b}_{E,k,m}^{M;\lambda} + 
 V_{E,k,m}^\lambda(x) \hat{d}_{E,k,m}^{M; \lambda\, \dagger}
 \right]\nonumber\\
 =& \sum_j \theta(E_j) (U_j \hat{b}^M_j + V_j \hat{d}^{M;\dagger}_j).
 \label{eq:vacM}
\end{align}
The above expansion is invertible since the set of solutions $\{U_j, V_j\}$ is complete, i.e.
\begin{equation}
 \sum_j \theta(E_j) [U_j(t,\bm{x}) \otimes U^\dagger_j(t, \bm{x}') + 
 V_j(t,\bm{x}) \otimes V^\dagger_j(t, \bm{x}')] = 
 \delta^3(\bm{x} - \bm{x}'),
 \label{eq:complete_M}
\end{equation}
where the sum over modes $\sum_j \theta(E_j)$ defined in Eq.~\eqref{eq:vacM} is such that $\sum_j f_j \delta(j,j') = f_{j'}$. 
Imposing the canonical anti-commutation relation 
$\{\widehat{\Psi}^\dagger(t, \bm{x}), \widehat{\Psi}(t, \bm{x}')\} 
= \delta^3(\bm{x} - \bm{x}')$, it can be seen that 
the particle creation and annihilation operators, 
$\hat{b}_j^{M;\dagger}$ and $\hat{b}_j^M$ and their 
antiparticle counterparts satisfy 
the following anti-commutation relations:
\begin{equation}
 \{\hat{b}^M_j, \hat{b}^{M;\dagger}_{j'}\} = 
 \{\hat{d}^M_j, \hat{d}^{M;\dagger}_{j'}\} = \delta(j,j').
\end{equation}

In the case of the rotating vacuum, by analogy to the RKT restriction on $\widetilde{p}^{\,t}$, given in Eq.~\eqref{eq:RKT_pt_positive}, the particle and anti-particle modes are split with respect to the sign of the co-rotating energy
\begin{equation}
 \widetilde{E}_j = E_j - \Omega m_j.
 \label{eq:E_rot}
\end{equation}
This is achieved at the level of the mode sum by introducing a step function $\theta(\widetilde{E})$, as follows \cite{iyer82,ambrus14plb}:
\begin{align}
 \widehat{\Psi} =& \sum_{\lambda = \pm \frac{1}{2}} 
 \sum_{m = -\infty}^\infty 
 \int_{|E| > M} dE \, |E|\,
 \int_{-p}^p dk \,\theta(\widetilde{E})
 \left[U_{E,k,m}^\lambda(x) \hat{b}_{E,k,m}^{\Omega;\lambda} + 
 V_{E,k,m}^\lambda(x) \hat{d}_{E,k,m}^{\Omega; \lambda\, \dagger}\right]\nonumber\\
 =& \sum_j \theta(\widetilde{E}_j) (U_j 
 \hat{b}^\Omega_j + V_j \hat{d}^{\Omega;\dagger}_j),
 \label{eq:vacr}
\end{align}
where the integral with respect to $E$ runs over both
negative and positive values, provided $|E| > M$.
Given the relation \eqref{eq:Vj} between the particle and anti-particle modes, it is not difficult to see that the completeness relation \eqref{eq:complete_M} can be recast for the rotating vacuum as follows:
\begin{equation}
 \sum_j \theta(\widetilde{E}_j) [U_j(t,\bm{x}) \otimes U^\dagger_j(t, \bm{x}') + 
 V_j(t,\bm{x}) \otimes V^\dagger_j(t, \bm{x}')] = 
 \delta^3(\bm{x} - \bm{x}'),
 \label{eq:complete_Omega}
\end{equation}
while the particle and anti-particle operators for the rotating vacuum satisfy the canonical anti-commutation relations:
\begin{equation}
 \{\hat{b}^\Omega_j, \hat{b}^{\Omega;\dagger}_{j'}\} = 
 \{\hat{d}^\Omega_j, \hat{d}^{\Omega;\dagger}_{j'}\} = \delta(j,j').
\end{equation}

Noting that both the Minkowski and the rotating one-particle operators are obtained by taking the inner product between $U_j$ and $V_j$ with the field operator, $\widehat{\Psi}$, simple Bogoliubov relations can be established between them, as follows \cite{ambrus14plb}:
\begin{align}
 b_j^M =&
 \begin{cases}
  b_j^\Omega, & \widetilde{E}_j > 0, \\
  i (-1)^{-m-\frac{1}{2}} d_{\overline{\jmath}}^{\Omega;\dagger}, & \widetilde{E}_j < 0,
 \end{cases} &
 d_j^{M,\dagger} =&
 \begin{cases}
  d_j^{\Omega,\dagger}, & \widetilde{E}_j > 0, \\
  i (-1)^{-m-\frac{1}{2}} b_{\overline{\jmath}}^{\Omega}, & \widetilde{E}_j < 0,
 \end{cases}\nonumber\\
 b_j^\Omega =&
 \begin{cases}
  b_j^M, & E_j > 0, \\
  i (-1)^{m+\frac{1}{2}} d_{\overline{\jmath}}^{M;\dagger}, & E_j < 0,
 \end{cases} &
 d_j^{\Omega,\dagger} =&
 \begin{cases}
  d_j^{M,\dagger}, & E_j > 0, \\
  i (-1)^{m+\frac{1}{2}} b_{\overline{\jmath}}^{M}, & E_j < 0.
 \end{cases}
\end{align}
It is not difficult to see that the expectation values of the products of two one-particle operators corresponding to the rotating vacuum with respect to the Minkowski vacuum is just
\cite{iyer82}:
\begin{equation}
 \braket{0_M| b^{\Omega,\dagger}_j b^\Omega_{j'}| 0_M} = 
 \braket{0_M| d^{\Omega,\dagger}_j d^\Omega_{j'}| 0_M} = 
 \begin{cases}
  0, & E_j > 0, \\
  \delta(j,j'), & E_j < 0.
 \end{cases}
 \label{eq:vac_bb}
\end{equation}

Let $\widehat{F}$ be an operator which is quadratic with respect to $\widehat{\Psi}$. Considering that at the classical level, $F$ can be written as:
\begin{equation}
 F = \mathcal{F}(\psi, \psi),
 \label{eq:F_bilinear_def}
\end{equation}
where $\mathcal{F}(\psi,\chi)$ is a bilinear form involving $\overline{\psi}$ and $\chi$ (being conjugate linear with respect to its first argument), at the quantum level, $\widehat{F}$ can be written as:
\begin{equation}
 \widehat{F} = \frac{1}{2} \sum_{j,j'} \theta(\widetilde{E}_j) \theta(\widetilde{E}_{j'})
 \left\{\mathcal{F}(U_j, U_{j'})[\hat{b}_j^{\Omega;\dagger}, \hat{b}^\Omega_{j'}] + \mathcal{F}(V_j, V_{j'})[\hat{d}^\Omega_j, \hat{d}_{j'}^{\Omega;\dagger}] + 
 \text{mixed terms}\right\}, 
 \label{eq:F_bilinear}
\end{equation}
where ``$\text{mixed terms}$'' refers to terms involving $\mathcal{F}(U_j,V_{j'})$ and $\mathcal{F}(V_j,U_{j'})$, which are proportional to $[\hat{b}_j^{\Omega;\dagger}, \hat{d}_{j'}^{\Omega;\dagger}]$ and $[\hat{d}^\Omega_j, \hat{b}^\Omega_{j'}]$, respectively. The field operators were expanded with respect to the rotating vacuum one-particle operators, for definiteness. 

The concept of normal ordering can be implemented for $\widehat{F}$ with 
respect to the rotating vacuum state, $\ket{0_\Omega}$, as follows:
\begin{equation}
 :\widehat{F}:_\Omega = \widehat{F} - \braket{0_\Omega|\widehat{F}|0_\Omega}.
 \label{eq:F_vacR}
\end{equation}
Taking into account that the state $\ket{0_\Omega}$ is annihilated 
by $b^\Omega_j$ and $d^\Omega_j$ and using the decomposition in 
Eq.~\eqref{eq:F_bilinear}, the following expression is obtained:
\begin{equation}
 :\widehat{F}:_\Omega = \sum_{j,j'} \theta(\widetilde{E}_j) \theta(\widetilde{E}_{j'})
 \left\{\mathcal{F}(U_j, U_{j'}) \hat{b}_j^\dagger  \hat{b}_{j'} - \mathcal{F}(V_j, V_{j'}) \hat{d}_{j'}^\dagger \hat{d}_j + 
 \text{mixed terms}\right\}.
 \label{eq:F_bilinear_rot}
\end{equation}

Considering Wick ordering with respect to the Minkowski vacuum, one can derive:
\begin{equation}
 :\widehat{F}:_M = :\widehat{F}:_\Omega - \braket{0_M| :\widehat{F}:_\Omega |0_M},\label{eq:F_vacM}
\end{equation}
where the expectation value of $:\widehat{F}:_\Omega$ computed in the Minkowsi vacuum state can be obtained using Eq.~\eqref{eq:vac_bb}:
\begin{align}
 \braket{0_M|:\widehat{F}:_\Omega|0_M} =& 
 \sum_j \theta(\widetilde{E}_j) \theta(-E_j) 
 [\mathcal{F}(U_j,U_j) - \mathcal{F}(V_j, V_j)] \nonumber\\
 =& \sum_j \theta(-\widetilde{E}_j) \theta(E_j) 
 [\mathcal{F}(U_{\overline{\jmath}},U_{\overline{\jmath}}) - \mathcal{F}(V_{\overline{\jmath}}, V_{\overline{\jmath}})], 
 \label{eq:vac_RM_aux}
\end{align}
where the flip $j \rightarrow \overline{\jmath}$ was considered on the second line. 
In general, the bilinear forms $\mathcal{F}(\psi,\chi)$ are scalars from the point 
of view of the spinor indices, involving the product of $\overline{\psi}$ and $\chi$ 
(as well as other spinor and/or differential operators acting on these spinors). 
By virtue of Eq.~\eqref{eq:Vj}, it can be assumed that
\begin{equation}
 \mathcal{F}(U_{\overline{\jmath}}, U_{\overline{\jmath}}) = \mathcal{F}(V_j, V_j), \qquad \mathcal{F}(V_{\overline{\jmath}}, V_{\overline{\jmath}}) = \mathcal{F}(U_j, U_j).
 \label{eq:Fbar_F}
\end{equation}
In this case, Eq.~\eqref{eq:vac_RM_aux} becomes:
\begin{align}
 \braket{0_M|:\widehat{F}:_\Omega|0_M} =& -\sum_j \theta(-\widetilde{E}_j) \theta(E_j) 
 [\mathcal{F}(U_j,U_j) - \mathcal{F}(V_j, V_j)]\nonumber\\
 =& -\sum_{\lambda = \pm \frac{1}{2}} 
 \sum_{m = -\infty}^\infty 
 \int_M^\infty dE \, E\,
 \int_{-p}^p dk \,\theta(-\widetilde{E})
 \left[\mathcal{F}(U_j, U_j) - \mathcal{F}(V_j, V_j)\right].
 \label{eq:vac_RM}
\end{align}

\subsection{Thermal expectation values} \label{sec:therm:tevs}


In this paper, the finite temperature expectation values are computed 
by taking the thermal average over the Fock space, as follows 
\cite{kapusta89,laine16}:
\begin{equation}
 \braket{\widehat{A}} = Z^{-1} {\rm Tr} (\hvrho \widehat{A}),
\end{equation}
where $Z = {\rm Tr} (\hvrho)$ is the partition function.
The density operator $\hvrho$, given in Eq.~\eqref{eq:rho},
can be obtained in a straightforward fashion, by promoting the 
microscopic momenta from RKT to quantum operators and taking into
account that the chemical potentials $\mu_V$ 
and $\mu_H$ are conjugate to the vector and helicity charge operators
$\widehat{Q}_V$ and $\widehat{Q}_H$ introduced in Eq.~\eqref{eq:QVH_hat}, respectively.
In particular, the t.e.v.s of $\hat{b}^\dagger_{j'} \hat{b}_j$ and $\hat{d}^\dagger_{j'} \hat{d}_j$ are \cite{vilenkin80,itzykson80,mallik16}:
\begin{equation}
 \braket{ \hat{b}^\dagger_j \hat{b}_{j'} } = 
 \frac{\delta(j,j')}{\exp[\beta_0(\widetilde{E}_j - 
 \mu_{\lambda_j;0})] + 1}, \qquad
 \braket{ \hat{d}^\dagger_j \hat{d}_{j'} } =
 \frac{\delta(j,j')}{\exp[\beta_0(\widetilde{E}_j + 
 \mu_{\lambda_j;0})] + 1},
 \label{eq:bblocks}
\end{equation}
where $\mu_{\lambda_j; 0} = \mu_{V;0} + 2 \lambda_j \mu_{H;0}$. 
Considering the limit $\beta_0 \rightarrow \infty$ and $\mu_{V;0}, \mu_{H;0} \rightarrow 0$, it can be seen that Eq.~\eqref{eq:bblocks} recovers the vacuum state only when $\widetilde{E}_j > 0$:
\begin{equation}
 \lim_{\substack{\beta_0 \rightarrow \infty\\ \mu_V, \mu_H \rightarrow 0}} \braket{ \hat{b}^\dagger_j \hat{b}_{j'} } = \begin{cases}
  0, & \widetilde{E}_j > 0,\\
  \delta(j,j), & \widetilde{E}_j < 0.
 \end{cases}
\end{equation}
It is therefore natural to discuss the construction of the thermal expectation values (t.e.v.s) at non-vanishing values of $\Omega$ by considering the normal ordering imposed via Eq.~\eqref{eq:vacr}, which corresponds to the rotating vacuum state. 

Focussing on normal ordering with respect to the rotating vacuum state, 
when $\widetilde{E}_j> 0$ is imposed through the step function 
$\theta(\widetilde{E}_j)$ in Eq.~\eqref{eq:vacr}, it can be seen that 
at vanishing temperatures, the t.e.v.s in Eq.~\eqref{eq:bblocks} vanish 
when $\widetilde{E}_j$ exceeds the Fermi level \eqref{eq:RKT_EF}, in 
agreement with the RKT prediction.

Taking the t.e.v. of Eq.~\eqref{eq:F_bilinear_rot} using Eq.~\eqref{eq:bblocks} gives:
\begin{equation}
 \braket{:\widehat{F}:_\Omega} = \sum_{j} \theta(\widetilde{E}_j) \left[\frac{\mathcal{F}(U_j, U_j)}{\exp[\beta_0(\widetilde{E}_j - 
 \mu_{\lambda_j;0})] + 1} - \frac{\mathcal{F}(V_j, V_j)}{\exp[\beta_0(\widetilde{E}_j + 
 \mu_{\lambda_j;0})] + 1}\right].
 \label{eq:F_aux1}
\end{equation}
The sum over $j$ is understood 
to stand for the summation and integration on the first line of 
Eq.~\eqref{eq:vacr}. 

Equation~\eqref{eq:F_aux1} can be written such
that the domain for the integration with respect to the energy $E$ 
spans only positive values, i.e. $M  < E < \infty$. To this end, the
notation $F_{m,E}$ is introduced through:
\begin{align}
 \braket{:\widehat{F}:_\Omega} =& \sum_{m = -\infty}^\infty \int_{|E| > M} dE\, |E| 
 \,\theta(\widetilde{E}) F_{m,E}(\widetilde{E}), \nonumber\\
 F_{m,E}(\widetilde{E}) =& \sum_{\lambda = \pm \frac{1}{2}} 
 \int_{-p}^p dk 
 \left[
 \frac{\mathcal{F}(U_{E,k,m}^\lambda, U_{E,k,m}^\lambda)}
 {e^{\beta_0(\widetilde{E} - \mu_{\lambda;0})} + 1} - 
 \frac{\mathcal{F}(V_{E,k,m}^\lambda, V_{E,k,m}^\lambda)}
 {e^{\beta_0(\widetilde{E} + \mu_{\lambda;0})} + 1}\right].
 \label{eq:F_aux2}
\end{align}
It can be seen that $F_{m,E}(\widetilde{E})$ depends explicitly on both $m$ 
and $E$ and the dependence on $\widetilde{E} = E - \Omega m$ is 
taken into account explicitly for future convenience.
The integration with respect to $E$ appearing in Eq.~\eqref{eq:F_aux2} can 
be broken into the corresponding positive and negative ranges. For the term 
containing the negative values of $E$, the transformation 
$(m,E) \rightarrow (-m, -E)$ is performed, yielding:
\begin{equation}
 \braket{:\widehat{F}:_\Omega} = \sum_{m = -\infty}^\infty \int_M^\infty dE\, E
 [\theta(\widetilde{E}) F_{m,E}(\widetilde{E}) + 
 \theta(-\widetilde{E}) F_{-m,-E}(-\widetilde{E})].
 \label{eq:F_aux3}
\end{equation}
In the expression for $F_{-m,-E}(-\widetilde{E})$, the flip $k \rightarrow -k$ can be performed, since the Fermi-Dirac weight factors in Eq.~\eqref{eq:F_aux2} are independent of $k$. Using Eq.~\eqref{eq:Fbar_F}, Eq.~\eqref{eq:F_aux3} can be written as:
\begin{multline}
 \braket{:\widehat{F}:_\Omega} = \frac{1}{2} \sum_{\lambda = \pm \frac{1}{2}}  \sum_{m = -\infty}^\infty \int_M^\infty dE\, E  \int_{-p}^p dk 
 \Bigg\{\\
 [\mathcal{F}(U_j, U_j) + \mathcal{F}(V_j, V_j)]
 \left[\frac{1}
 {e^{\beta_0(|\widetilde{E}| - \mu_{\lambda;0})} + 1} - 
 \frac{1}{e^{\beta_0(|\widetilde{E}| + \mu_{\lambda;0})} + 1}\right] \\
 + {\rm sgn}(\widetilde{E}) [\mathcal{F}(U_j, U_j) - \mathcal{F}(V_j, V_j)]
 \left[\frac{1}
 {e^{\beta_0(|\widetilde{E}| - \mu_{\lambda;0})} + 1} + 
 \frac{1}{e^{\beta_0(|\widetilde{E}| + \mu_{\lambda;0})} + 1}\right]
 \Bigg\}.
 \label{eq:F_R}
\end{multline}
For the particular operators considered in this paper, 
$\mathcal{F}(V_j,V_j) = \pm \mathcal{F}(U_j,U_j)$, such that only 
one of the terms above survives. Equation~\eqref{eq:F_R} forms the 
basis for the numerical analysis discussed in the following sections. 
However, further analysis of Eq.~\eqref{eq:F_R} is encumbered by 
the presence of the modulus $|\widetilde{E}|$ in the Fermi-Dirac 
factors, as well as by the sign function ${\rm sgn}(\widetilde{E})$. 
In order to simplify Eq.~\eqref{eq:F_R}, it is convenient to consider
the t.e.v. of $:\widehat{F}:_M$, Wick ordered with respect 
to the Minkowski vacuum, starting from Eq.~\eqref{eq:F_vacM}:
\begin{equation}
 \braket{:\widehat{F}:_M} = \braket{:\widehat{F}:_\Omega} - \braket{0_M|:\widehat{F}:_\Omega|0_M}.
\end{equation}
Using Eq.~\eqref{eq:vac_RM}, it can be shown that the t.e.v. of $:\widehat{F}:_M$ is given by
\begin{equation}
 \braket{:\widehat{F}:_M} = \sum_{\lambda = \pm \frac{1}{2}}  \sum_{m = -\infty}^\infty \int_M^\infty dE\, E  \int_{-p}^p dk 
 \left[\frac{\mathcal{F}(U_j, U_j)}
 {e^{\beta_0(\widetilde{E} - \mu_{\lambda;0})} + 1} - 
 \frac{\mathcal{F}(V_j, V_j)}{e^{\beta_0(\widetilde{E} + \mu_{\lambda;0})} + 1}\right].
 \label{eq:F_M}
\end{equation}
It is worth emphasising that the difference $\braket{0_M|:\widehat{F}:_\Omega|0_M}$
between Eq.~\eqref{eq:F_R} and \eqref{eq:F_M} is a purely vacuum quantity, depending only 
on the rotation parameter $\Omega$, which by virtue of Eq.~\eqref{eq:vac_RM}, vanishes 
whenever $\mathcal{F}(V_j, V_j) = \mathcal{F}(U_j,U_j)$. 
Previously, the t.e.v.s with respect to the Minkowski vacuum 
were computed by Vilenkin in Refs.~\cite{vilenkin79,vilenkin80} 
and indeed temperature-independent terms were revealed in the final results. 
It was argued in a previous publication \cite{ambrus14plb} that 
such temperature-independent terms are spurious and 
disappear when the t.e.v.s are computed with respect to 
the rotating vacuum.

\subsection{Small mass analysis}\label{sec:therm:tricks}

In the small mass limit, analytic closed form expressions 
can be obtained for the t.e.v.s of the form given in Eq.~\eqref{eq:F_M},
following the methodology introduced in Ref.~\cite{ambrus14plb}. 
The procedure will be illustrated in Subsecs.~\ref{sec:CC:M0}, 
\ref{sec:AC:M0}, \ref{sec:FC:M0} and \ref{sec:SET:M0} for the VCC and HCC, 
the ACC, the FC and the SET, respectively. The
basic idea is to expand the Fermi-Dirac distributions,
which depend on the co-rotating energy $\widetilde{E} = E - \Omega m$,
in a power series with respect to $\Omega$, following the prescription given below:
\begin{equation}
 \frac{1}{e^{\beta_0(E - \Omega m \mp \mu_{\lambda,0})} + 1} 
 = \sum_{n = 0}^\infty \frac{(-\Omega m)^n}{n!} \frac{d^n}{dE^n} 
 \left[\frac{1}{e^{\beta_0(E \mp \mu_{\lambda,0})} + 1}\right].
 \label{eq:M0_exp}
\end{equation}
When computing the small mass limit, a further expansion with respect to $M$ should be performed. Noting that a function $f(E)$ that depends only on the energy (no extra dependence on $p$) can be expanded as:
\begin{equation}
 f(E) = f(p) + \frac{df}{dp} \left(\frac{dE}{dM^2}\right)_{M\rightarrow 0} M^2 + O(M^4) = f(p) + \frac{M^2}{2p} \frac{df}{dp} + O(M^4),\label{eq:Etop}
\end{equation}
it can be seen that Eq.~\eqref{eq:M0_exp} can be written as:
\begin{multline}
 \frac{1}{e^{\beta_0(E - \Omega m \mp \mu_{\lambda,0})} + 1} 
 = \sum_{n = 0}^\infty \frac{(-\Omega m)^n}{n!} 
 \left\{
 \frac{d^n}{dp^n} \left[\frac{1}{e^{\beta_0(p \mp \mu_{\lambda,0})} + 1}
 \right] \right.\\ 
 \left. + \frac{M^2}{2p} \frac{d^{n+1}}{dp^{n+1}} \left[\frac{1}{e^{\beta_0(p \mp \mu_{\lambda,0})} + 1}
 \right] + O(M^4)\right\}.
 \label{eq:M0_exp2}
\end{multline}
After the above expansion, the sum over $m$ involves 
terms of the form $\sum_{m_j} m_j^n \mathcal{F}(U_j, U_j)$.
For the purpose of the t.e.v.s of interest in this paper,
it can be assumed that $\mathcal{F}(U_j, U_j)$ depends on $m$ through 
powers of $m$ multiplied by one of the following combinations of Bessel 
functions:
\begin{equation}
 J_m^\pm(q\rho) = J_{m - \frac{1}{2}}^2(q\rho) \pm 
 J_{m + \frac{1}{2}}^2(q\rho), \qquad 
 J_m^\times(q\rho) = 2 J_{m - \frac{1}{2}}(q\rho) 
 J_{m + \frac{1}{2}}(q\rho).
 \label{eq:Jstar_def}
\end{equation}
The following relations are useful for performing the sum over 
$m$ in Eq.~\eqref{eq:F_M} \cite{ambrus14plb}:\footnote{Note that in Ref.~\cite{ambrus14plb},
the parameter $m = 0, \pm 1, \pm2, \dots$ takes integer values, while 
in this paper, the convention that $m = \pm \frac{1}{2}, \pm \frac{3}{2}, \dots$ 
is an odd half-integer is employed.}
\begin{align}
 \sum_{m = -\infty}^\infty m^{2n} 
 \begin{pmatrix}
  J_m^+(z) \smallskip\\
  m J_m^-(z) \smallskip\\
  m J_m^\times(z) 
 \end{pmatrix} = \sum_{j = 0}^n 
 \frac{2}{\sqrt{\pi}}
 \begin{pmatrix}
  \Gamma(j + \frac{1}{2}) / j!  \smallskip\\
  \Gamma(j + \frac{3}{2}) / j! \smallskip\\
  z \Gamma(j + \frac{3}{2}) / (j+1)!
 \end{pmatrix} s_{n,j}^+ z^{2j},
 \label{eq:M0_summ}
\end{align}
where the coefficients $s_{n,j}^+$ are determined from:
\begin{equation}
 s_{n,j}^+ = \frac{1}{(2j+1)!} 
 \lim_{\alpha \rightarrow 0} 
 \frac{d^{2n+1}}{d\alpha^{2n+1}} 
 \left(2 \sinh\frac{\alpha}{2}\right)^{2j+1}.
\end{equation}
It can be seen that $s_{n,j}^+$ vanishes when $j > n$. For small values of 
$n - j \ge 0$, the first few coefficients are given by \cite{ambrus14plb}:
\begin{gather}
 s_{j,j}^+ = 1, \qquad s_{j+1,j} = \frac{1}{24}(2j+1)(2j+2)(2j+3),\nonumber\\
 s_{j+2,j}^+ = \frac{1}{5760} (2j+1)(2j+2)(2j+3)(2j+4)(2j+5)(10j+3).
 \label{eq:M0_summ_s}
\end{gather}
For general values of $n > j$, the following recurrences can be established:
\begin{equation}
 s_{n+1,j}^+  = s_{n,j-1}^+ + \left(j + \frac{1}{2}\right)^2 s_{n,j}^+, 
 \qquad 
 s_{n,j+1}^+ = \frac{1}{(j+1)(2j+3)} 
 \sum_{k = 1}^{n-j} \binom{2n+1}{2k} 
 s_{n-k,j}^+.
 \label{eq:M0_summ_rec}
\end{equation}
Finally, the following formula can be employed for the integration 
with respect to $k$:
\begin{equation}
 \int_0^p dk\, q^\nu = \frac{\Gamma(\frac{\nu}{2} + 1) \sqrt{\pi}}
 {2\Gamma(\frac{\nu + 1}{2} + 1)} 
 p^{\nu + 1}.
 \label{eq:M0_intk}
\end{equation}

\section{Vector and helicity charge currents}\label{sec:CC}

In this section, the thermal expectation values (t.e.v.s) 
of the vector charge current (VCC) and helicity charge current 
(HCC) operators are investigated. 
The axial charge current (ACC) operator will be considered in Sec.~\ref{sec:AC}. 
The general expressions that form the basis of the computation of the t.e.v.s 
are derived in Subsec.~\ref{sec:CC:gen}. Analytic expressions are derived in 
the small mass limit or on the rotation axis in Subsec.~\ref{sec:CC:M0}. 
The validity of these expressions at finite mass is considered using numerical 
integration in Subsec.~\ref{sec:CC:num}.

\subsection{General analysis}\label{sec:CC:gen}

In the notation of Sec.~\ref{sec:therm:tricks}, the bilinear forms 
introduced in Eq.~\eqref{eq:F_bilinear_def} which correspond to the 
HCC and VCC, introduced in Eq.~\eqref{eq:CC_hat}, are:
\begin{equation}
 \mathcal{J}^\halpha_V(\psi,\chi) = \overline{\psi} \gamma^\halpha \chi, \qquad 
 \mathcal{J}^\halpha_H(\psi,\chi) = \overline{\psi} \gamma^\halpha h \chi + 
 \overline{h\psi} \gamma^\halpha \chi. 
\end{equation}
Since $V_j = i \gamma^2 U_j^*$, is it easy to show that:
\begin{equation}
 \mathcal{J}_{V/H}^\halpha(V_j,V_j) =
 [\mathcal{J}_{V/H}^\halpha(U_j,U_j)]^* = 
 \mathcal{J}_{V/H}^\halpha(U_j,U_j),
 \label{eq:CC_VU_aux}
\end{equation}
where the final equality follows after noting that:
\begin{align}
 [\mathcal{J}^\halpha_V(U_j,U_j)]^* =& 
 U_j^\dagger (\gamma^0 \gamma^0) (\gamma^\halpha)^\dagger \gamma^0 U_j = 
 \mathcal{J}^\halpha_V(U_j,U_j), \nonumber\\
 [\mathcal{J}^\halpha_H(U_j,U_j)]^* =& 
 [\overline{U}_j \gamma^\halpha h U_j +
 \overline{(h U_j)} \gamma^\halpha U_j]^*
 = \mathcal{J}^\halpha_H(U_j,U_j).
\end{align}
Thus, the final line 
in Eq.~\eqref{eq:F_R} makes a vanishing contribution. This implies that 
the t.e.v.s computed with respect to the rotating and Minkowski vacua coincide,
i.e. $\braket{:\widehat{J}_{V/H}^\halpha:_\Omega} = \braket{:\widehat{J}_{V/H}^\halpha:_M}$, 
such that
\begin{multline}
 \braket{:\widehat{J}_{V/H}^\halpha:_\Omega} = \sum_{\lambda = \pm \frac{1}{2}} 
 \sum_{m = -\infty}^\infty \int_M^\infty dE\, E \left[
 \frac{1}{e^{\beta_0(\widetilde{E} - \mu_{\lambda;0})} + 1} - 
 \frac{1}{e^{\beta_0(\widetilde{E} + \mu_{\lambda;0})} + 1}\right]\\
 \times \int_{-p}^p dk \, \mathcal{J}^\halpha_{V/H}(U_j, U_j).
 \label{eq:CC_aux}
\end{multline}
Using the explicit expression for the modes, given 
in Eq.~\eqref{eq:U}, the following relations can 
be obtained:
\begin{align}
 \overline{U}_j \gamma^\hatt U_j =& 
 \frac{1}{8\pi^2}\left[J_{m_j}^+(q_j\rho) + 
 \frac{2\lambda_j k_j}{p_j} J_{m_j}^-(q_j\rho)\right],\nonumber\\
 \overline{U}_j \gamma^\hvarphi U_j =& 
 \frac{q_j}{8\pi^2 E_j} J_{m_j}^\times(q_j \rho),\nonumber\\
 \overline{U}_j \gamma^\hatz U_j =& 
 \frac{1}{8\pi^2}\left[\frac{k_j}{E_j} J_{m_j}^+(q_j\rho) + 
 \frac{2\lambda_j p_j}{E_j} J_{m_j}^-(q_j\rho)\right],
 \label{eq:CC_bblocks}
\end{align}
while $\overline{U}_j \gamma^\hrho U_j = 0$ for 
all $j$. The functions $J_m^*(q\rho)$ ($* \in \{+, -, \times\}$) 
were introduced in Eq.~\eqref{eq:Jstar_def}.

When substituting the expressions from Eq.~\eqref{eq:CC_bblocks} into
Eq.~\eqref{eq:CC_aux}, the terms proportional to 
$k$ can be discarded since they are odd with respect 
to the transformation $k \rightarrow -k$. 
The non-vanishing components of the charge currents 
can be summarised through:
\begin{multline}
 \begin{pmatrix}
  \braket{:\widehat{J}^\hatt_V:_\Omega} & 
  \braket{:\widehat{J}^\hvarphi_V:_\Omega} & 
  \braket{:\widehat{J}^\hatz_V:_\Omega} \\
  \braket{:\widehat{J}^\hatt_H:_\Omega} & 
  \braket{:\widehat{J}^\hvarphi_H:_\Omega} & 
  \braket{:\widehat{J}^\hatz_H:_\Omega}
 \end{pmatrix} = \frac{1}{4\pi^2} \sum_{\lambda = \pm \frac{1}{2}}
 \sum_{m = -\infty}^\infty \int_M^\infty dE \\
 \times \left[
 \frac{1}{e^{\beta_0(\widetilde{E} - \mu_{\lambda;0})} + 1} -
 \frac{1}{e^{\beta_0(\widetilde{E} + \mu_{\lambda;0})} + 1}
 \right] 
 \\ \times \int_{0}^p dk 
 \begin{pmatrix}
  E J_m^+(q \rho) & q J_m^\times(q\rho) & 2\lambda p J_m^-(q\rho)\\
  2\lambda E J_m^+(q \rho) & 2\lambda q J_m^\times(q\rho) & 
  p J_m^-(q\rho)
 \end{pmatrix}.
 \label{eq:CC}
\end{multline}
Looking at the terms involving the polarisation $\lambda$,
it is clear that when the helicity chemical potential 
vanishes ($\mu_{H;0} = 0$ and $\mu_{\lambda;0} = \mu_{V;0}$), the $z$ component 
of the VCC and the time and azimuthal components 
of the HCC vanish. However, the $z$ component of the HCC remains finite 
as long as the vector chemical potential $\mu_{V;0}$ is finite.

In the following, it is convenient to refer to the charge currents for the
right- and left-handed particles, 
$\widehat{J}^\halpha_\pm = \widehat{J}^\halpha_V \pm \widehat{J}^\halpha_H$, 
which can be computed as follows:
\begin{multline}
 \begin{pmatrix}
  \braket{:\widehat{J}^\hatt_\pm:_\Omega} \\
  \braket{:\widehat{J}^\hvarphi_\pm:_\Omega} \\
  \braket{:\widehat{J}^\hatz_\pm:_\Omega}
 \end{pmatrix} = \frac{1}{2\pi^2}
 \sum_{m = -\infty}^\infty \int_M^\infty dE \left[
 \frac{1}{e^{\beta_0(\widetilde{E} - \mu_{\pm;0})} + 1} -
 \frac{1}{e^{\beta_0(\widetilde{E} + \mu_{\pm;0})} + 1}
 \right] 
 \\ \times \int_{0}^p dk 
 \begin{pmatrix}
  E J_m^+(q \rho) \\
  q J_m^\times(q\rho) \\
  \pm p J_m^-(q\rho)
 \end{pmatrix},
 \label{eq:CC_pm}
\end{multline}
where $\mu_{\pm;0} = \mu_{V;0} \pm \mu_{H;0}$ [see Eq.~\eqref{eq:RKT_M0_Qpm}].

The t.e.v.s of the charge currents can be decomposed with respect 
to the orthogonal tetrad formed by the vectors $u^\halpha$, 
$a^\halpha$, $\omega^\halpha$ and $\tau^\halpha$,
introduced in Eqs.~\eqref{eq:RR_u}, \eqref{eq:RR_a}, 
\eqref{eq:RR_omega} and \eqref{eq:RR_tau}, as follows:
\begin{equation}
 \braket{:\widehat{J}_\pm^\halpha:_\Omega} = 
 Q_\pm u^\halpha + \mathcal{J}_\pm^\halpha, \qquad
 \mathcal{J}_\pm^\halpha = \sigma^\tau_\pm \tau^\halpha + \sigma^\omega_\pm \omega^\halpha,
 \label{eq:CC_dec}
\end{equation}
where $Q_\pm$ represents the charge density and 
the charge flow in the rest frame, $\mathcal{J}_\pm^\halpha$,
is by construction orthogonal to $u^\halpha$.
There is no term multiplying the acceleration, $a = -\rho \Omega^2 \Gamma^2 e_\hrho$, since 
$\braket{:\widehat{J}^\hrho_\pm:_\Omega} = 0$.
The charge densities $Q_\pm$ and the vortical and circular charge conductivities, 
$\sigma^\omega_\pm$ and $\sigma^\tau_\pm$, can be obtained via
\begin{align}
 Q_\pm =& u_\halpha \braket{:\widehat{J}_\pm^\halpha:_\Omega} = 
 \Gamma\left[\braket{:\widehat{J}_\pm^\hatt:_\Omega} - 
 \rho \Omega \braket{:\widehat{J}_\pm^\hvarphi:_\Omega}\right], \nonumber\\
 \sigma_\pm^\tau =& \frac{1}{\tau^2} \tau_\halpha
 \braket{:\widehat{J}_\pm^\halpha:_\Omega} =
 \frac{\rho \Omega \braket{:\widehat{J}_\pm^\hatt:_\Omega} -
 \braket{:\widehat{J}_\pm^\hvarphi:_\Omega}}
 {\rho \Omega^3 \Gamma^3}, \nonumber\\
 \sigma_\pm^\omega =& \frac{1}{\omega^2} \omega_\halpha
 \braket{:\widehat{J}_\pm^\halpha:_\Omega} = 
 \frac{\braket{:J^\hatz_\pm:_\Omega}}{\Omega \Gamma^2}.
 \label{eq:CC_dec_inv}
\end{align}
Comparing Eq.~\eqref{eq:CC_dec} to the RKT decomposition in Eq.~\eqref{eq:RKT_CC_SET},
it can be seen that $\mathcal{J}_\pm^\halpha$ has no classical correspondent and 
therefore it describes anomalous transport due to vortical effects.

\subsection{Small mass limit}\label{sec:CC:M0}

The algorithm described in Sec.~\ref{sec:therm:tricks} is applied 
to the t.e.v.s $\braket{:\widehat{J}^\halpha_\pm:_\Omega} = 
\braket{:\widehat{J}^\halpha_V:_\Omega} \pm \braket{:\widehat{J}^\halpha_H:_\Omega}$.
Expanding the Fermi-Dirac factors with respect to $\Omega$ using Eq.~\eqref{eq:M0_exp}, 
performing the summation over $m$ using Eq.~\eqref{eq:M0_summ} and the integrating over $k$ 
using Eq.~\eqref{eq:M0_intk} yields:
\begin{multline}
 \begin{pmatrix}
  \braket{:\widehat{J}^\hatt_\pm:_\Omega}\\ 
  \braket{:\widehat{J}^\hvarphi_\pm:_\Omega} \\
  \braket{:\widehat{J}^\hatz_\pm:_\Omega}
 \end{pmatrix} = \frac{1}{\pi^2} 
 \sum_{j = 0}^\infty (\rho \Omega)^{2j}
 \sum_{n = 0}^\infty \frac{\Omega^{2n} s_{n+j,j}^+}{(2n+2j)!}
 \int_0^\infty dp\, p^{2j+1} \\
 \times \frac{d^{2n+2j}}{dE^{2n+2j}} 
 \left[\frac{1}{e^{\beta_0(E - \mu_{\pm,0})} + 1} - 
 \frac{1}{e^{\beta_0(E + \mu_{\pm,0})} + 1}\right] \begin{pmatrix}
  p/(2j+1) \\
  \rho \Omega p / (2n + 2j+1) \\
  \pm \Omega (j+1) / (2n + 2j + 1)
 \end{pmatrix},\label{eq:CC_M0_aux}
\end{multline}
where the summations over $j$ and $n$ were interchanged and 
the transformation $n \rightarrow n + j$ was subsequently performed 
in the sum over $n$, in order to shift the summation range 
from $j \le n < \infty$ to $0 \le n < \infty$. 
The integration variable was changed from $E$ to $p$.
For the $\hvarphi$ and $\hatz$ components, an integration by parts 
was performed prior to this change of variable.

The analysis for the small mass regime can be performed starting from Eq.~\eqref{eq:M0_exp2}.
Due to the nature of the integrands, it is convenient to discuss first $\braket{:J^\hatt_\pm:_\Omega}$ and $\braket{:J^\hvarphi_\pm:_\Omega}$. Integration by parts can be performed $2j$ times for the leading order (massless) term and $2j+1$ times for the $O(M^2)$ term, allowing Eq.~\eqref{eq:CC_M0_aux} to be written as follows:
\begin{multline}
 \begin{pmatrix}
  \braket{:\widehat{J}^\hatt_\pm:_\Omega}\\ 
  \braket{:\widehat{J}^\hvarphi_\pm:_\Omega}
 \end{pmatrix} = \frac{1}{2\pi^2} 
 \sum_{j = 0}^\infty (\rho \Omega)^{2j}
 \sum_{n = 0}^\infty \frac{\Omega^{2n} s_{n+j,j}^+ (2j)!}{(2n+2j+1)!}
 \int_0^\infty dp\, [(2j+2)p^2 - M^2] 
 \begin{pmatrix}
 2n+2j+1 \\ \rho \Omega(2j+1)
 \end{pmatrix} \\
 \times \frac{d^{2n}}{dp^{2n}} 
 \left[\frac{1}{e^{\beta_0(p - \mu_{\pm,0})} + 1} - 
 \frac{1}{e^{\beta_0(p + \mu_{\pm,0})} + 1}\right].\label{eq:CC_M0_aux2}
\end{multline}
Focussing on the integration with respect to $p$ and noting that:
\begin{equation}
 \frac{1}{2}\int_0^\infty dx \, x^2 \frac{d^{2n}}{dx^{2n}} 
 \left[\frac{1}{e^{x - a} + 1} - 
 \frac{1}{e^{x + a} + 1}\right] = 
 \begin{cases}
  I_1^-, & n = 0, \\
  2I_0^-, & n = 1, \\
  0, & n > 1,
 \end{cases}
\end{equation}
where the functions $I_0^-$ and $I_1^-$ are given in Eq.~\eqref{eq:FD},
it can be seen that the series with respect to $n$ appearing in 
Eq.~\eqref{eq:CC_M0_aux2} terminates after a finite number of terms. 
For each value of $n$, the summation over $j$ can be performed by noting that:
\begin{equation}
 \sum_{j = 0}^\infty \frac{(j + n)!}{j!} z^j = \frac{d^n}{dz^n} 
 \left(\frac{1}{1 - z}\right) = \frac{n!}{(1 - z)^{n+1}}.
 \label{eq:sumj}
\end{equation}
This leads to the following exact results:
\begin{align}
 \braket{:\widehat{J}^\hatt_\pm:_\Omega} =& 
 \Gamma \mu_\pm \left[\frac{T^2}{3} + \frac{\mu_\pm^2}{3\pi^2} - \frac{M^2}{2\pi^2} + 
 \frac{\Omega^2 \Gamma^2}{12\pi^2} 
 (4\Gamma^2 - 1) + O(M^4)\right],\nonumber\\
 \braket{:\widehat{J}^\hvarphi_\pm:_\Omega} =& 
 \rho \Omega \Gamma \mu_\pm \left[\frac{T^2}{3} + \frac{\mu_\pm^2}{3\pi^2} - \frac{M^2}{2\pi^2} + 
 \frac{\Omega^2 \Gamma^2}{12\pi^2} 
 (4\Gamma^2 - 3) + O(M^4)\right].
 \label{eq:M0_CC_tphi}
\end{align}
The results in Eq.~\eqref{eq:M0_CC_tphi} can be written in 
terms of $Q_\pm$ and $\sigma_\pm^\tau$ as follows:
\begin{equation}
 Q_\pm = Q^{\rm RKT}_\pm + \Delta Q_\pm, \qquad 
 \Delta Q = \frac{\mu_\pm}{4\pi^2}
  (\bm{\omega}^2 + \bm{a}^2) + O(M^4), \qquad 
 \sigma^\tau_\pm = 
 \frac{\mu_\pm}{6\pi^2} + O(M^4),\label{eq:CC_Qst}
\end{equation}
where the acceleration $a^\halpha$ and vorticity $\omega^\halpha$ vectors are defined in 
Eqs.~\eqref{eq:RR_a} and \eqref{eq:RR_omega}, respectively, while the squares 
$\bm{a}^2 = -a^2 \ge 0$ and $\bm{\omega}^2 = -\omega^2 \ge 0$ of their spatial 
parts are given in Eq.~\eqref{eq:vecs_squares}. In the above, the term $Q^{\rm RKT}_\pm$ 
corresponds to the relativistic kinetic theory (RKT) prediction, given up to $O(M^4)$ in 
Eq.~\eqref{eq:RKT_M_corr}, while $\Delta Q_\pm$ and $\sigma^\tau_\pm$ represent 
quantum corrections, which are independent of the particle mass up to $O(M^4)$.

Turning to $\braket{:\widehat{J}^\hatz_\pm:_\Omega}$, the 
summation with respect to $m$ and integration with respect to $k$ 
can be performed as before, yielding:
\begin{multline}
 \braket{:\widehat{J}^\hatz_\pm:_\Omega} = \pm \frac{\Omega}{2\pi^2} 
 \sum_{j = 0}^\infty (\rho \Omega)^{2j}
 \sum_{n = 0}^\infty \frac{\Omega^{2n} s_{n+j,j}^+ (2j+2)!}{(2n+2j+1)!}
 \int_0^\infty dp\, \left[p + \frac{M^2}{2(2j+1)} \frac{d}{dp}\right]  \\
 \times \frac{d^{2n}}{dp^{2n}} 
 \left[\frac{1}{e^{\beta_0(p - \mu_{\pm,0})} + 1} - 
 \frac{1}{e^{\beta_0(p + \mu_{\pm,0})} + 1}\right].\label{eq:CC_Jz_aux1}
\end{multline}
After performing the integration with respect to $p$, the following result is obtained:
\begin{multline}
 \braket{:\widehat{J}^\hatz_\pm:_\Omega} = \pm \frac{\Omega}{2\pi^2} \sum_{j = 0}^\infty (\rho\Omega)^{2j} \left\{\frac{4(j+1)}{\beta_0^2} I_{1/2}^-(\beta\mu_{\pm}) \right.\\
 \left. + \sum_{n = 0}^\infty \frac{(2j+2)!\Omega^{2n}}{(2n+2j+1)!} \left[\frac{\Omega^2 s_{n+j+1,j}^+}{(2n+2j+2)(2n+2j+3)} - \frac{M^2 s_{n+j,j}^+}{2(2j+1)}\right] \frac{d^{2n}}{d\mu_{\pm}^{2n}} \tanh\frac{\beta \mu_{\pm}}{2}\right\},
 \label{eq:CC_Jz_aux2}
\end{multline}
where the expression for $I_{1/2}^-(a)$ can be found in
Eq.~\eqref{eq:FD_Half}. It can be seen that the series over $n$
does not terminate, indicating that the dependence on $\Gamma$ is not polynomial. In the following, only the term on the first line together with the term corresponding to $n = 0$ on the second line of Eq.~\eqref{eq:CC_Jz_aux2} are considered. 
Performing the sum over $j$ using Eq.~\eqref{eq:sumj}, together with the relation:
\begin{equation}
 \sum_{j = 0}^\infty \frac{(\rho \Omega)^{2j}}{2j + 1} = \frac{{\rm arcsinh} (\rho \Omega \Gamma)}{\rho \Omega},\label{eq:sumj_2jp1}
\end{equation}
the following expression can be obtained for $\sigma^\omega_\pm$:
\begin{multline}
 \sigma^\omega_\pm = 
 \pm \frac{T^2}{\pi^2}
 \left[{\rm Li}_2(-e^{-\mu_\pm/T}) - 
 {\rm Li}_2(-e^{\mu_\pm/T})\right]\\
 \mp \left\{\frac{M^2}{4\pi^2} 
 \left[1 + \frac{{\rm arcsinh}(\rho \Omega \Gamma)}{\rho \Omega \Gamma^2}\right]
 - \frac{\bm{\omega}^2 + 3\bm{a}^2}{24\pi^2}\right\} 
 \tanh \frac{\mu_\pm}{2T}
 + O(\Omega^4,\Omega^2 M^2,M^4), \label{eq:CC_so_Omega}
\end{multline}
where it is understood that $T = T_0 \Gamma$ and
$\mu_\pm = \mu_{\pm;0} \Gamma$, while the acceleration $a = -\rho \Omega^2 \Gamma^3 e_\hrho$ and kinematic vorticity $\omega = \Omega \Gamma^2 e_\hatz$ are given in Eqs.~\eqref{eq:RR_a} and \eqref{eq:RR_omega}, respectively. The high temperature limit of Eq.~\eqref{eq:CC_so_Omega} can be computed by noting that:
\begin{equation}
 {\rm Li}_2(-e^{-z}) = -\frac{\pi^2}{12} + z \ln 2 - 
 \frac{z^2}{4} + \frac{z^3}{24} + O(z^5).\label{eq:Li2_small}
\end{equation}
The following result is obtained:
\begin{multline}
 \sigma^\omega_\pm = 
 \pm \frac{2 \mu_\pm T}{\pi^2} \ln 2 
 \pm \frac{\mu_\pm}{12\pi^2 T} 
 \left[\mu_\pm^2 - \frac{3M^2}{2} \left(1 + \frac{{\rm arcsh}(\rho \Omega \Gamma)}{\rho \Omega \Gamma^2} \right) + \frac{\bm{\omega}^2 + 3\bm{a}^2}{4}\right] \\
 + O(\Omega^4, \Omega^2 M^2, M^4, T^{-2}). 
 \label{eq:CC_so_HighT}
\end{multline}

The result derived in Eq.~\eqref{eq:CC_so_Omega} is valid only for small 
values of the rotation parameter. An expression which is exact on the rotation 
axis can be derived starting from Eq.~\eqref{eq:CC}, by noting that 
$J_{\pm 1/2}^-(0) = \pm 1$, such that,
\begin{equation}
 \braket{:\widehat{J}^\hatz_\pm:_\Omega}_{\rho=0} = 
 \pm \frac{1}{\pi^2} \int_M^\infty dE\, p^2 \left[f^+_{\beta_0}\left(\mu_{\pm; 0} + \frac{\Omega}{2}\right) -
 f^+_{\beta_0}\left(\mu_{\pm; 0} - \frac{\Omega}{2}\right)\right],\label{eq:CC_Jz_axis_aux1}
\end{equation}
where the following notation was employed:
\begin{equation}
 f^\pm_{\beta_0}\left(\frac{a}{\beta_0}\right) = \frac{1}{2}\left( \frac{1}{e^{\beta_0 E - a} + 1} \pm \frac{1}{e^{\beta_0 E + a} + 1}\right).
 \label{eq:fpm_def}
\end{equation}
The techniques introduced for the analysis of the case of non-vanishing mass in the RKT formulation, in Sec.~\ref{sec:RKT}, can now be employed. First, the integration variable is changed to $x = E / M$, as indicated in Eq.~\eqref{eq:RKT_x}. Then, the formula on the second line of Eq.~\eqref{eq:RKT_exp_aux3} can be used to transform the Fermi-Dirac factors, allowing Eq.~\eqref{eq:CC_Jz_axis_aux1} to be written as:
\begin{equation}
 \braket{:\widehat{J}^\hatz_\pm:_\Omega}_{\rho=0} = 
 \pm \frac{2M^3}{\pi^2} \sum_{\ell = 1}^\infty 
 (-1)^{\ell+1} \sinh(\ell\beta \mu_\pm) \sinh\left(\frac{\ell \beta \Omega}{2}\right) 
 \int_1^\infty dx\,(x^2 - 1)e^{-\ell \beta M x}.
\end{equation}
It is not difficult to perform the integration with respect to $x$, giving:
\begin{equation}
 \braket{:\widehat{J}^\hatz_\pm:_\Omega}_{\rho=0} = 
 \pm \frac{4}{\pi^2 \beta^3} \sum_{\ell = 1}^\infty 
 \frac{(-1)^{\ell+1}}{\ell^3} e^{-\ell \beta M}
 (1 + \ell \beta M)
 \sinh(\ell\beta \mu_\pm) \sinh\left(\frac{\ell \beta \Omega}{2}\right).
\end{equation}
The sum over $\ell$ can be performed in terms of the polylogarithm function, allowing 
$\sigma^\omega_\pm = \braket{:\widehat{J}^\hatz_\pm:_\Omega} / \Omega \Gamma^2$ 
to be expressed on the rotation axis as:
\begin{equation}
 \left.\sigma^\omega_\pm\right\rfloor_{\rho=0} = \mp \frac{T^3}{\pi^2 \Omega} 
 \sum_{\varsigma_\pm = \pm 1} \varsigma_\pm 
 \sum_{\varsigma_\Omega = \pm 1} \varsigma_\Omega 
 \left[{\rm Li}_3(-e^{\zeta}) + 
 \frac{M}{T} {\rm Li}_2(-e^\zeta)\right],
 \label{eq:CC_so_axis}
\end{equation}
where $\zeta = (\varsigma_\pm \mu_\pm + 
 \varsigma_\Omega\frac{\Omega}{2} - M) / T$.
The above expression is exact for any value of the mass. 
The large temperature limit can be extracted using 
Eq.~\eqref{eq:Li2_small} together with:
\begin{equation}
 {\rm Li}_3(-e^{-z}) = -\frac{3}{4} \zeta(3) + \frac{\pi^2 z}{12} 
 - \frac{z^2}{2} \ln 2 + \frac{z^3}{12} - 
 \frac{z^4}{96} + O(z^6).\label{eq:Li3_small}
\end{equation}
The result is:
\begin{equation}
 \left.\sigma^\omega_\pm\right\rfloor_{\rho = 0} = \pm \mu_\pm\left[\frac{2T}{\pi^2} \ln 2 +
 \frac{1}{12\pi^2 T} 
 \left(\mu_{\pm}^2 + \frac{\Omega^2}{4} - 3M^2\right) + 
 O(T^{-3},\rho^2) \right],
 \label{eq:CC_so_axis_T}
\end{equation}
which is consistent with the expression derived in Eq.~\eqref{eq:CC_so_HighT}.
Both Eqs.~\eqref{eq:CC_so_HighT} and \eqref{eq:CC_so_axis_T} show that at large temperature, $\sigma^\omega_\pm$ scales 
linearly with the temperature $T$ and 
the local chemical potential $\mu_\pm$. It is noteworthy that 
$\sigma^\omega_H = \frac{1}{2} (\sigma^\omega_+ - \sigma^\omega_-) 
\simeq \frac{2}{\pi^2} \mu_V T \ln 2$ is proportional to $\mu_V$, while
$\sigma^\omega_V = \frac{1}{2}(\sigma^\omega_+ + \sigma^\omega_-) 
\simeq \frac{2}{\pi^2} \mu_H T \ln 2$ is proportional to $\mu_H$.


The result is summarised below:
\begin{align}
 \braket{:\widehat{J}^\halpha_\pm:_\Omega} =& 
 Q_\pm u^\halpha + 
 \sigma^\tau_\pm \tau^\halpha + 
 \sigma^\omega_\pm \omega^\halpha, &
 Q_\pm =& Q^{\rm RKT}_\pm + \Delta Q_\pm, \nonumber\\
 Q^{\rm RKT}_\pm =& \mu_\pm
 \left( \frac{T^2}{3} + \frac{\mu_\pm^2}{3\pi^2} 
 - \frac{M^2}{2\pi^2} \right) + O(M^4), &
 \Delta Q_\pm =& \frac{\mu_{\pm}}{4\pi^2}
 (\bm{\omega}^2 + \bm{a}^2) + O(M^4),\nonumber\\
 \sigma^\tau_\pm =& \frac{\mu_{\pm}}{6\pi^2} + O(M^4), &
 \sigma^\omega_\pm =& \pm \frac{2\mu_{\pm} T}{\pi^2} \ln 2 + O(T^{-1}).\label{eq:CC_summary}
\end{align}
The quantities referring to the vector and helicity charge currents can be obtained 
by adding or subtracting the above expressions, e.g. $Q_V = \frac{1}{2}(Q_+ + Q_-)$ 
and $Q_H = \frac{1}{2}(Q_+ - Q_-)$:
\begin{align}
 Q^{\rm RKT}_V =& \mu_V
 \left(\frac{T^2}{3} + \frac{\mu_V^2 + 3 \mu_H^2}{3\pi^2} 
 - \frac{M^2}{2\pi^2} \right) + O(M^4), \nonumber\\
 Q^{\rm RKT}_H =& \mu_H
 \left(\frac{T^2}{3} + \frac{\mu_H^2 + 3 \mu_V^2}{3\pi^2} 
 - \frac{M^2}{2\pi^2} \right) + O(M^4), \nonumber\\
 \Delta Q_{V/H} =& \frac{\mu_{V/H}}{4\pi^2}(\bm{\omega}^2 + \bm{a}^2) + O(M^4), 
 \qquad 
 \sigma^\tau_{V/H} = \frac{\mu_{V/H}}{6\pi^2} + O(M^4), \nonumber\\
 \sigma^\omega_V =& \frac{2\mu_H T}{\pi^2} \ln 2 + O(T^{-1}), \qquad\qquad
 \sigma^\omega_H = \frac{2\mu_V T}{\pi^2} \ln 2 + O(T^{-1}).
 \label{eq:CC_summary_VH}
\end{align}
At vanishing mass and helicity chemical potential, the results in Eq.~\eqref{eq:CC_summary} coincide with those reported in Eq.~(4.9) and Table~2 of Ref.~\cite{buzzegoli18} when the axial chemical potential is assumed to vanish.

\subsection{Numerical analysis}\label{sec:CC:num}

In this section, the validity of the constitutive equations derived 
in the previous subsection for the quantum corrections are analysed 
at finite mass. The explorations presented in this section are based 
on the ``HIC'' conditions described in the introduction.
For simplicity, the discussion in this subsection is kept
at the level of $J^\mu_\pm$, which depends only on the chemical 
potential $\mu \equiv \mu_{\pm} = \mu_V \pm \mu_H$. The HIC conditions are 
therefore enforced as $\mu_\pm = 30\ {\rm MeV}$.
The convention when presenting results 
for different parameters is to start from the HIC values and change 
only one parameter per curve. The parameter change is shown in the plot 
legend in the form \texttt{parameter} $\times$ \texttt{multiplier}.  
E.g., a curve corresponding to a temperature which is twice that 
corresponding to the HIC parameters ($T = 300\ {\rm MeV}$) is labelled 
as $T \times 2$.

\begin{figure}
    \centering
\begin{tabular}{cc}
    \includegraphics[width=0.45\linewidth]{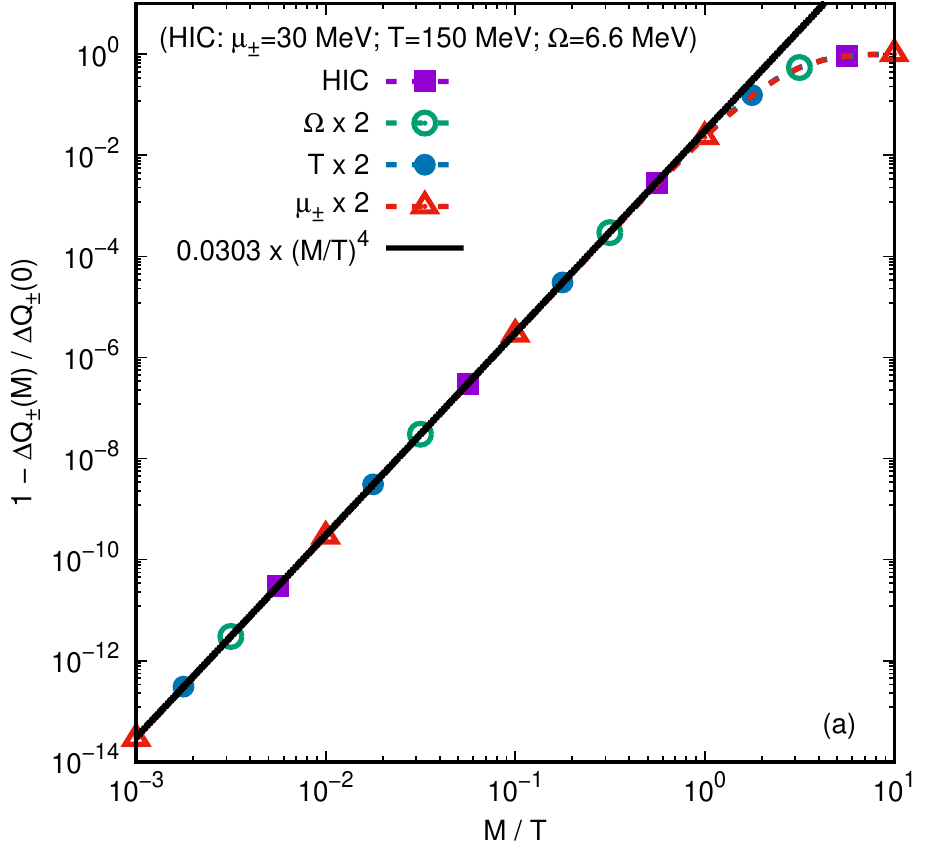} & 
    \includegraphics[width=0.45\linewidth]{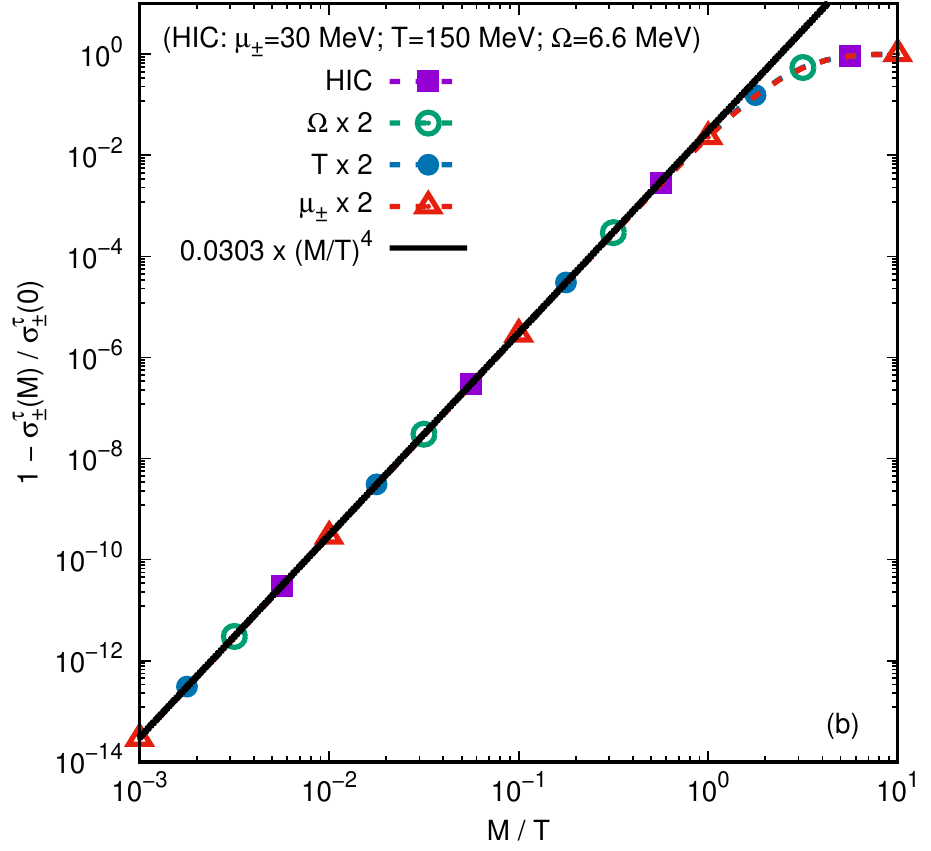} \\
    \includegraphics[width=0.45\linewidth]{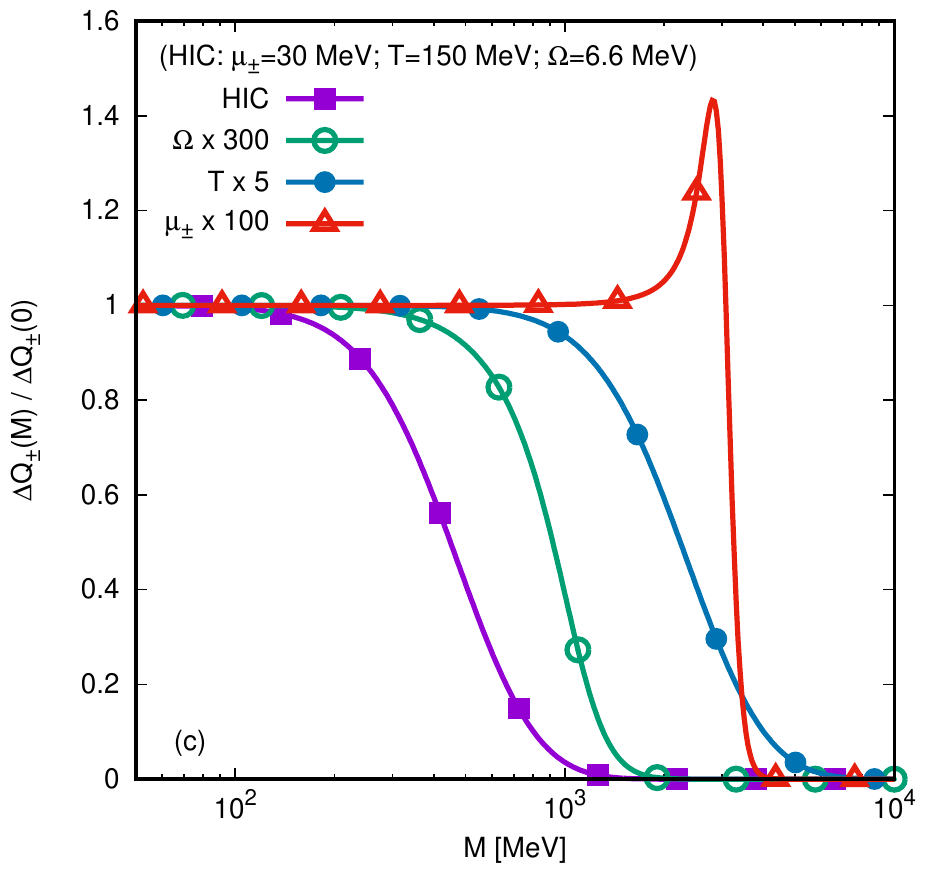} & 
    \includegraphics[width=0.45\linewidth]{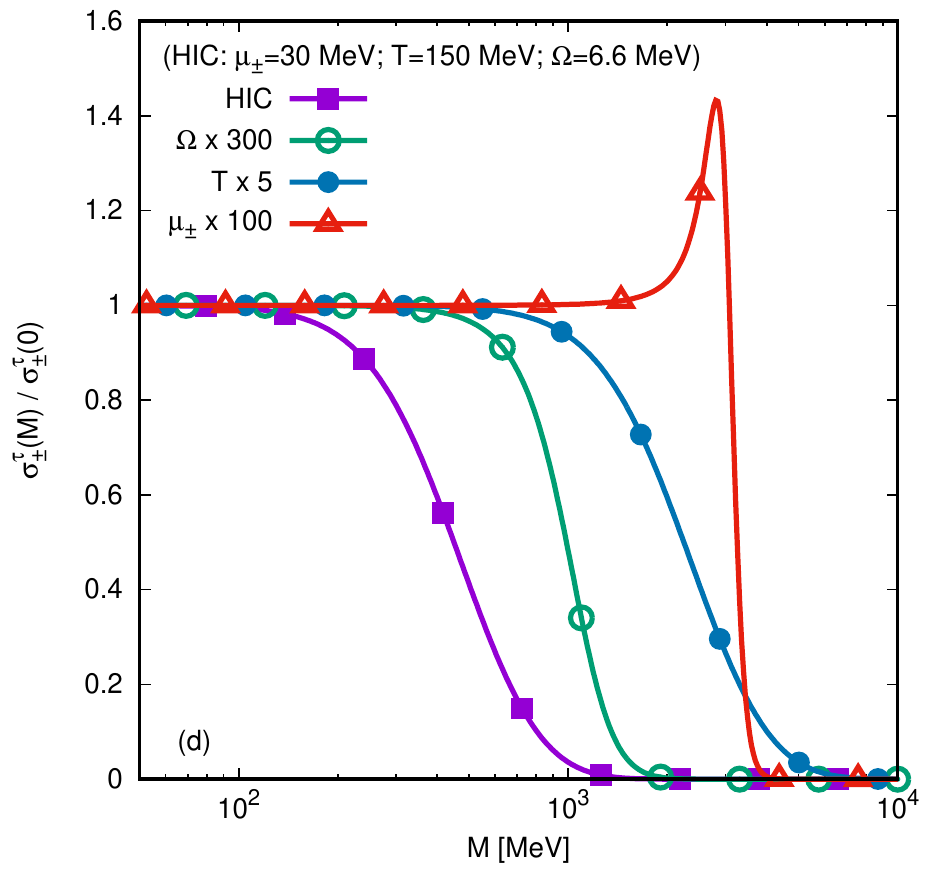}
\end{tabular}
\caption{
(a-b) Relative mass corrections 
(a) $1 - \Delta Q_{\pm}(M) / \Delta Q_\pm(0)$ and (b) $1 -\sigma^\tau_\pm(M) / \sigma^\tau_\pm(0)$,
represented with respect to $M$, for various values of $\Omega$, $T$ and $\mu_\pm$. The black curves represent the best fit of the function $\alpha (M/T)^4$ to the numerical data ($\alpha \simeq 0.0303$ in both cases).
(c-d) Ratios (c) $\Delta Q_{\pm}(M) / \Delta Q_\pm(0)$ and (d) $\sigma^\tau_\pm(M) / \sigma^\tau_\pm(0)$,
represented with respect to $M$, for various values of $\Omega$, $T$ and $\mu_\pm$. The set of parameters indicated at the top of each image corresponds to the values typical of Heavy Ion Collisions (HIC) and the corresponding data set is represented using the purple curve and filled squares. For each subsequent curve, only the parameter indicated in the legend is changed, compared to the HIC set.}
\label{fig:CC_Qst}
\end{figure}

The discussion in this section begins by considering the behaviour 
of the correction $\Delta Q_\pm$ to the charge density and the circular
conductivity $\sigma_\pm^\tau$ as the mass is increased. The analysis 
in the previous subsection showed that there are no contributions to 
$\Delta Q_\pm$ and $\sigma^\tau_\pm$ from the mass term up to $O(M^4)$.
Figures~\ref{fig:CC_Qst}(a) and \ref{fig:CC_Qst}(b) validate this 
prediction by considering the dependence of the relative mass corrections, 
$1 - \Delta Q_\pm(M) / \Delta Q_\pm(0)$ and $1 - \sigma^\tau_\pm(M) / \sigma^\tau_\pm(0)$, 
with respect to $M / T$. It can be seen that, close to the parameters 
relevant to heavy ion collisions (labelled HIC and shown with dashed 
purple lines and squares), the relative mass corrections have a universal 
behaviour of the type $\alpha (M / T)^4$, where $\alpha \simeq 0.0303$ gives 
the best fit of this power law to the numerical data. Figures~\ref{fig:CC_Qst}(c) 
and \ref{fig:CC_Qst}(d) show the ratios $\Delta Q_\pm(M) / \Delta Q_\pm(0)$ 
and $\sigma^\tau_\pm(M) / \sigma^\tau_\pm(0)$ with respect to $M$. The massless 
limits $\Delta Q_\pm(0)$ and $\sigma_\pm^\tau(0)$ are taken from 
Eq.~\eqref{eq:CC_summary}, while $\Delta Q_\pm(M)$ and $\sigma^\tau_\pm(M)$ 
are obtained by directly integrating Eq.~\eqref{eq:CC_pm}. For the HIC 
parameters, it can be seen that the constitutive relations derived in 
Eq.~\eqref{eq:CC_summary} are valid for $M \lesssim T = 150\ {\rm MeV}$. 
Increasing the mass has the expected effect of suppressing the t.e.v.s with 
respect to their values obtained in the massless case. To further test the 
robustness of the constitutive equations, three more curves are represented. 
For each curve, only one parameter from the original list is increased. 
This parameter is indicated in the legend, together with the corresponding 
multiplier. The second curve (green and empty circles), corresponding to 
$\Omega \simeq 2\ {\rm GeV}$, shows that the validity domain of 
Eq.~\eqref{eq:CC_summary} is enhanced at higher values of $\Omega$, with 
deviations occurring at $M \gtrsim \Omega / 4$. The third curve (blue and 
filled circles) shows deviations from the massless prediction also when 
$M \gtrsim T = 750\ {\rm MeV}$. Finally, the last curve (red line and empty 
triangles) curresponds to $\mu_\pm = 3\ {\rm GeV} \gg T = 150\ {\rm MeV}$. 
In this case, the fermion fluid is strongly degenerate, such that a strong suppression 
can be seen when $M$ exceeds $\mu_\pm$. Also, for $Mc^2 < \mu_\pm$, the constitutive 
relations for the massless case retain their validity, except in the vicinity 
$M \simeq \mu_\pm$. Here, a strong deviation can be seen, indicating an unexpected 
resonance. While this regime does not seem to be of immediate relevance to 
the field of relativistic heavy ion collisions, it is worth exploring its origin.

\begin{figure}
    \centering
\begin{tabular}{cc}
    \includegraphics[width=0.45\linewidth]{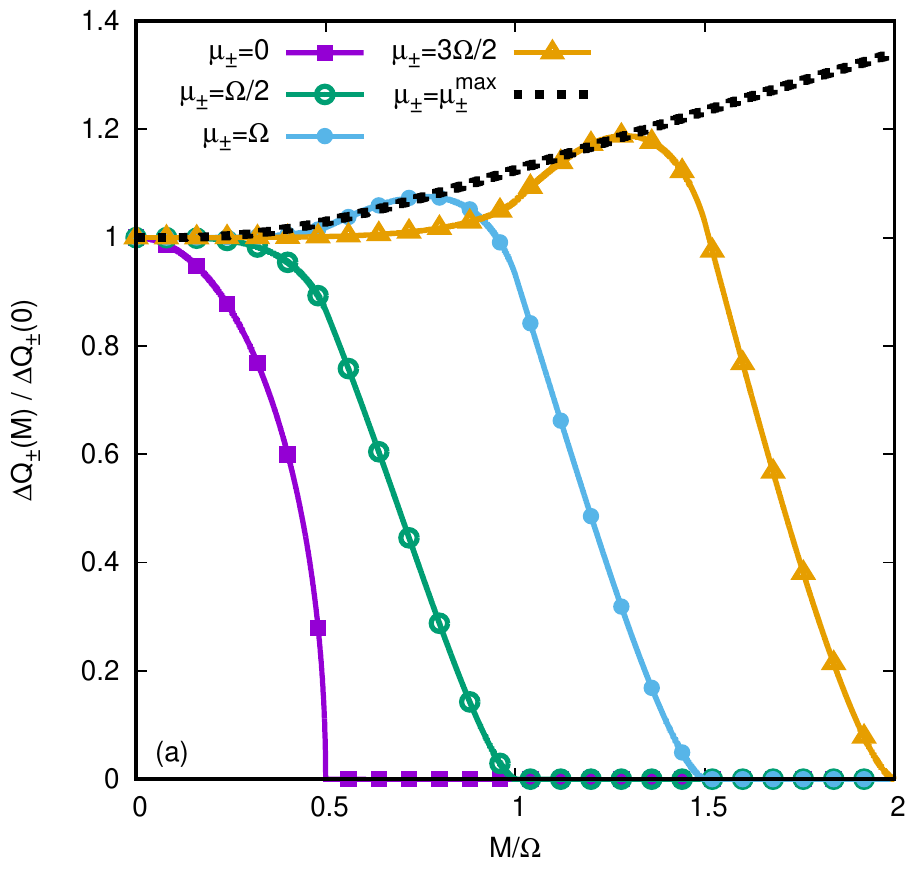} & 
    \includegraphics[width=0.45\linewidth]{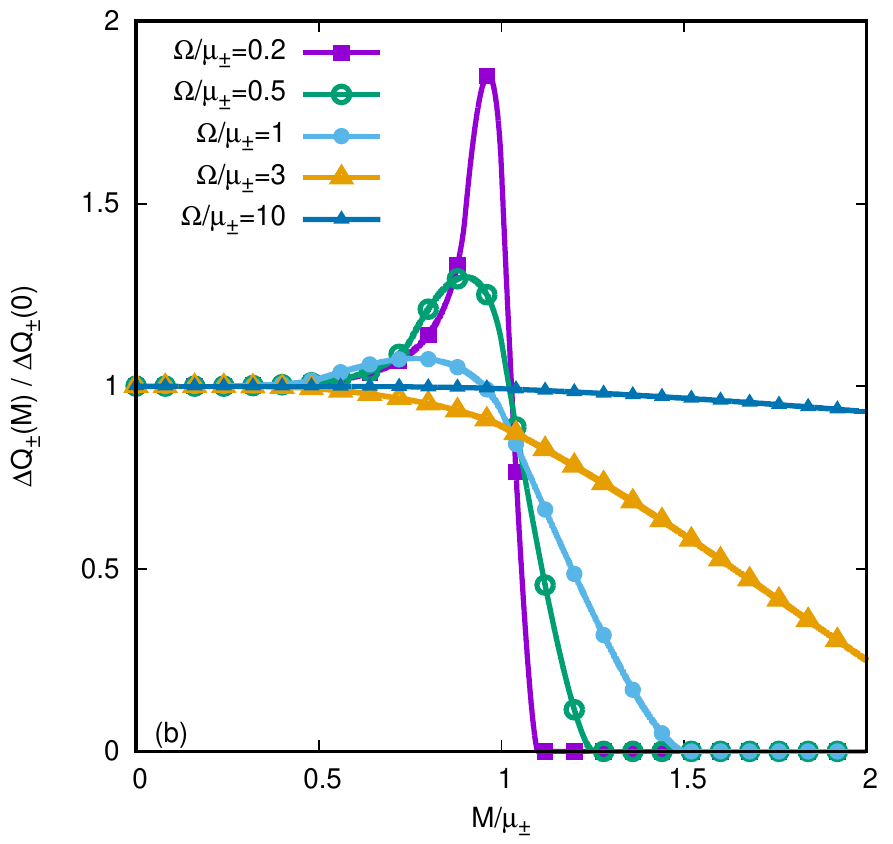}
\end{tabular}
\caption{Ratio $\Delta Q_{\pm}(M) / \Delta Q_\pm(0)$ at vanishing temperature
represented with respect to (a) $M / \Omega$ at fixed $\Omega$, for various 
values of $\mu_\pm / \Omega$; and (b) $M / \mu_\pm$ at fixed $\mu_\pm$, 
for various values of $\Omega / \mu_\pm$.}
\label{fig:CC_Q}
\end{figure}

The starting point for the analysis of the vanishing temperature regime 
is to take the $\beta_0 \rightarrow \infty$ limit of Eq.~\eqref{eq:CC_pm}. 
Restricting the analysis to the axis of rotation, the following result is 
obtained for the quantum correction to the charge density:
\begin{multline}
 \Delta Q_\pm\rfloor_{T\rightarrow 0} = \frac{{\rm sgn}(\mu_\pm)}{6\pi^2}
 \left\{\left[\left(|\mu_\pm| + \frac{|\Omega|}{2}\right)^2 - M^2\right]^{3/2}
 \theta\left(|\mu_\pm| + \frac{|\Omega|}{2} - M\right) \right.\\
 +
 \left[\left(|\mu_\pm| - \frac{|\Omega|}{2}\right)^2 - M^2\right]^{3/2}
 \left[\theta\left(|\mu_\pm| - \frac{|\Omega|}{2} - M\right) -
 \theta\left(\frac{|\Omega|}{2} -|\mu_\pm| - M\right) \right] \\
 \left.-  2(\mu_\pm^2 - M^2)^{3/2} \theta(|\mu_\pm| - M)\right\},
 \label{eq:CC_deltaQ_T0}
\end{multline}
where the last line corresponds to the RKT contribution $Q^{\rm RKT}_\pm$, 
which can be obtained by taking the vanishing temperature limit of 
Eq.~\eqref{eq:RKT_gen}:
\begin{equation}
 Q^{\rm RKT}_\pm\rfloor_{T\rightarrow 0} = \frac{{\rm sgn}(\mu_\pm)}{3\pi^2} (\mu_\pm^2 - M^2)^{3/2} \theta(|\mu_\pm| - M).
 \label{eq:RKT_Q_T0}
\end{equation}
When $|\mu_\pm| - \frac{|\Omega|}{2} < M < |\mu_\pm|$ and the term on 
the second line in Eq.~\eqref{eq:CC_deltaQ_T0} cancels, $\Delta Q_\pm$ 
admits an extremum with respect to $M$, which is obtained by 
solving $d\Delta Q_\pm / dM = 0$:
\begin{align}
 M_{\rm max} =& \sqrt{\left(|\mu_\pm| + \frac{|\Omega|}{6}\right) 
 \left(|\mu_\pm| - \frac{|\Omega|}{2}\right)}, \nonumber\\
 \Delta Q_\pm(M_{\rm max}) =& \frac{\mu_\pm^3}{3\pi^2 \sqrt{3}} 
 \left[\left(1 + \frac{|\Omega|}{2|\mu_\pm|}\right)^2 - 1\right]^{3/2}.
\end{align}
It is understood that this maximum occurs only when $|\mu_\pm| > |\Omega| / 2$. 
This behaviour is confirmed in Fig.~\ref{fig:CC_Q}(a), where the ratio 
$\Delta Q_{\pm}(M) / \Delta Q_{\pm}(0)$ is represented at fixed $\Omega$ 
with respect to the ratio $M / \Omega$, for various values of $\mu_\pm$. 
The dotted black line represents the value of this ratio when 
$|\mu_\pm| = \mu_\pm^{\rm max}$, where
\begin{equation}
 \mu_\pm^{\rm max} = \frac{|\Omega|}{6} + \sqrt{M^2 + \frac{\Omega^2}{9}}.
\end{equation}
When $|\Omega / \mu_\pm| \ll 1$ and $M_{\rm max} \rightarrow |\mu_\pm|$,
$\Delta Q_{\pm}(M_{\rm max}) / \Delta Q_{\pm}(0)$ peaks sharply 
around $M = M_{\rm max}$:
\begin{equation}
 \frac{\Delta Q_{\pm}(M_{\rm max})}{\Delta Q_{\pm}(0)}
 \simeq \frac{4}{3} \sqrt{\frac{|\mu_{\pm}|}{3|\Omega|}} + 
 O(|\Omega / \mu_\pm|^{1/2}).
\label{eq:CC_DQ_T0_peak}
\end{equation}
The development of this peak is highlighted in Fig.~\ref{fig:CC_Q}(b). 
For the case considered in Fig.~\ref{fig:CC_Qst}(c), the ratio between the 
chemical potential and the rotation parameter is $\mu_\pm / \Omega \simeq 450$. 
In this case, Eq.~\eqref{eq:CC_DQ_T0_peak} predicts that 
$\Delta Q_{\pm}(M_{\rm max})/\Delta Q_{\pm}(0) \simeq 16.4$, compared to 
$\simeq 1.5$ observed in Fig.~\ref{fig:CC_Qst}(c). This discrepancy is an 
indication of the thermal suppression of this effect, which is already 
significant for the values in Fig.~\ref{fig:CC_Qst}(c), when $T_0 / \mu_\pm = 0.05$.

\begin{figure}
    \centering
\begin{tabular}{cc}
    \includegraphics[width=0.45\linewidth]{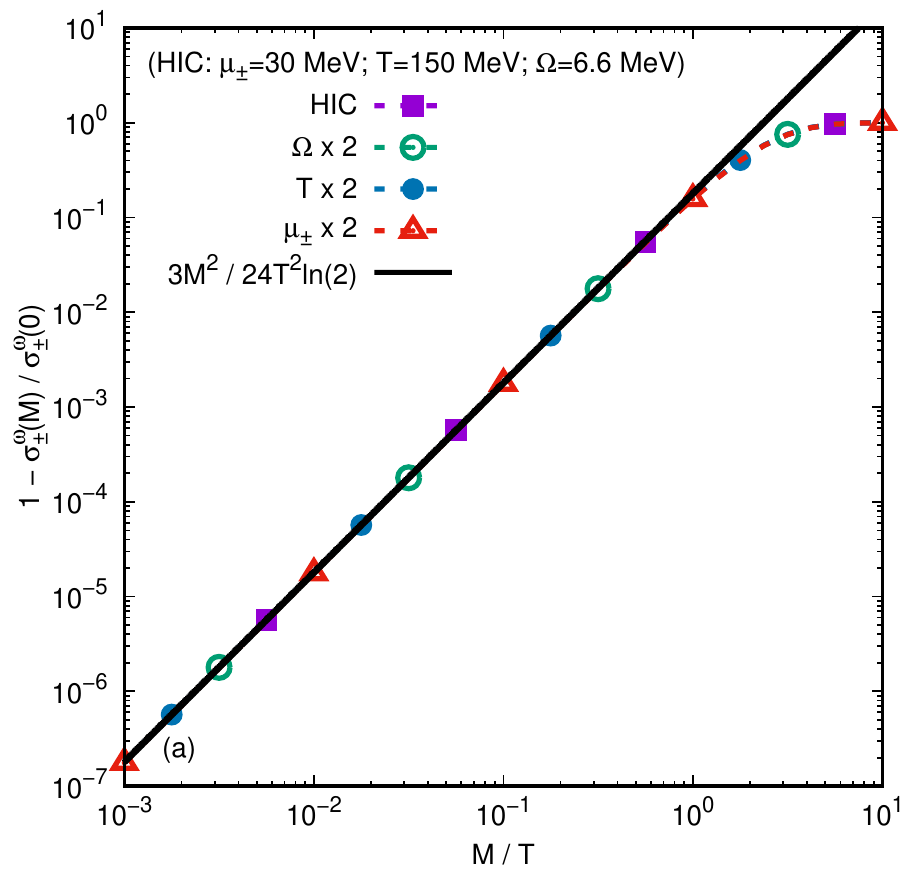} &
    \includegraphics[width=0.45\linewidth]{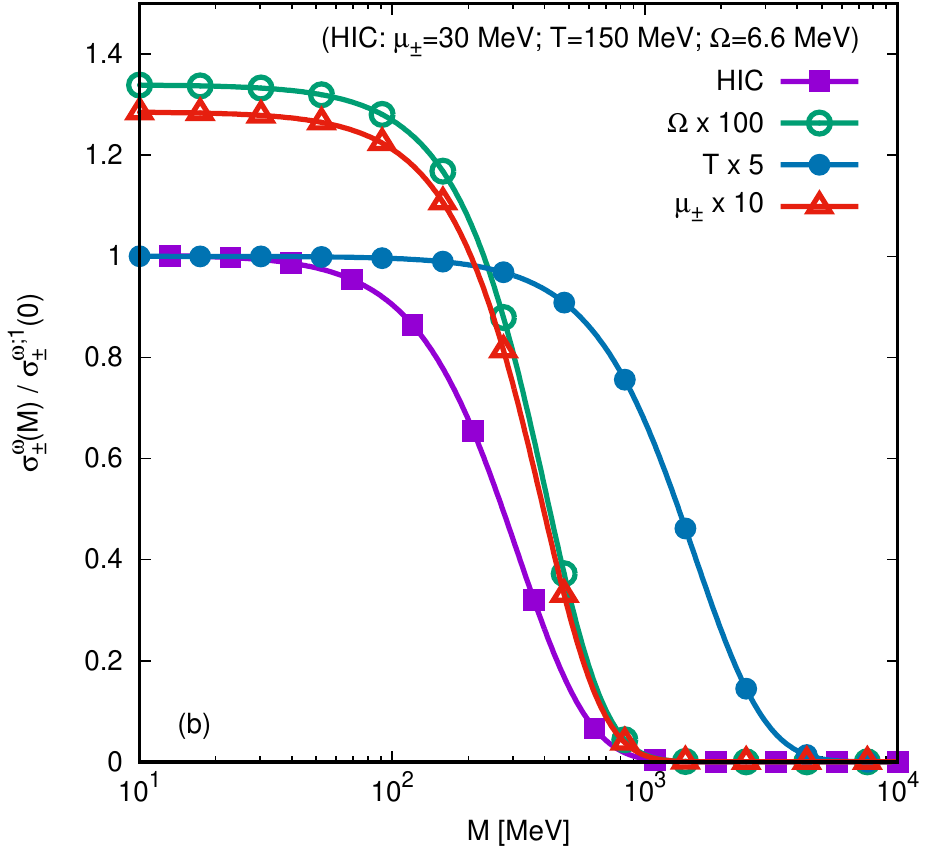} \\ 
    \includegraphics[width=0.45\linewidth]{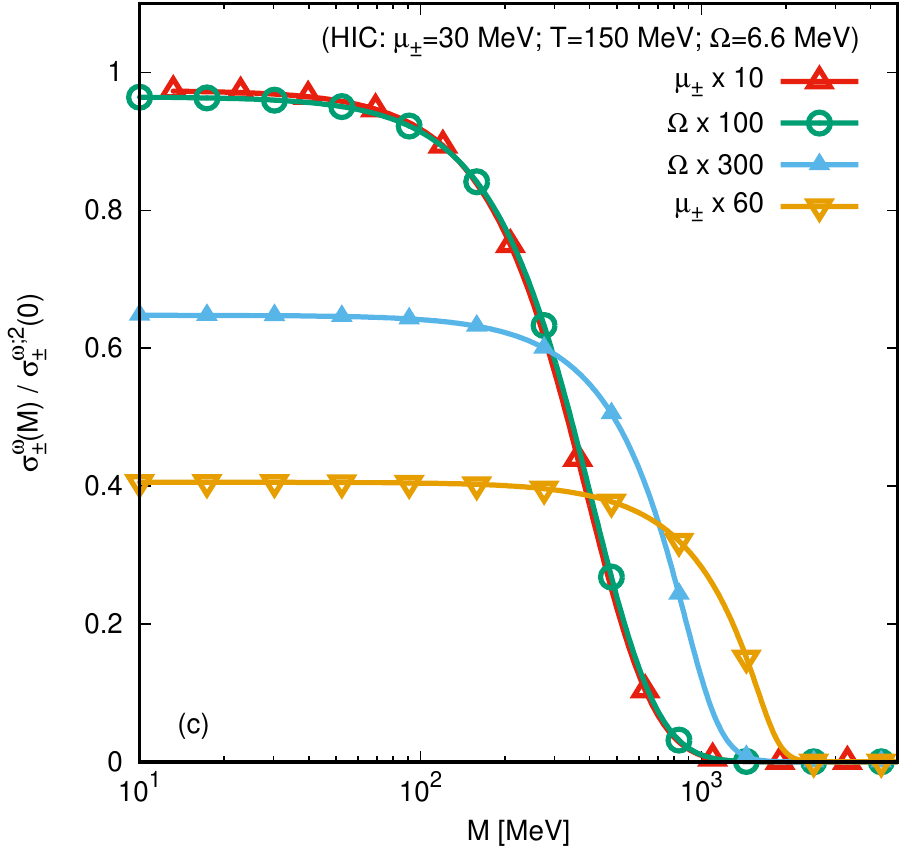} &
    \includegraphics[width=0.45\linewidth]{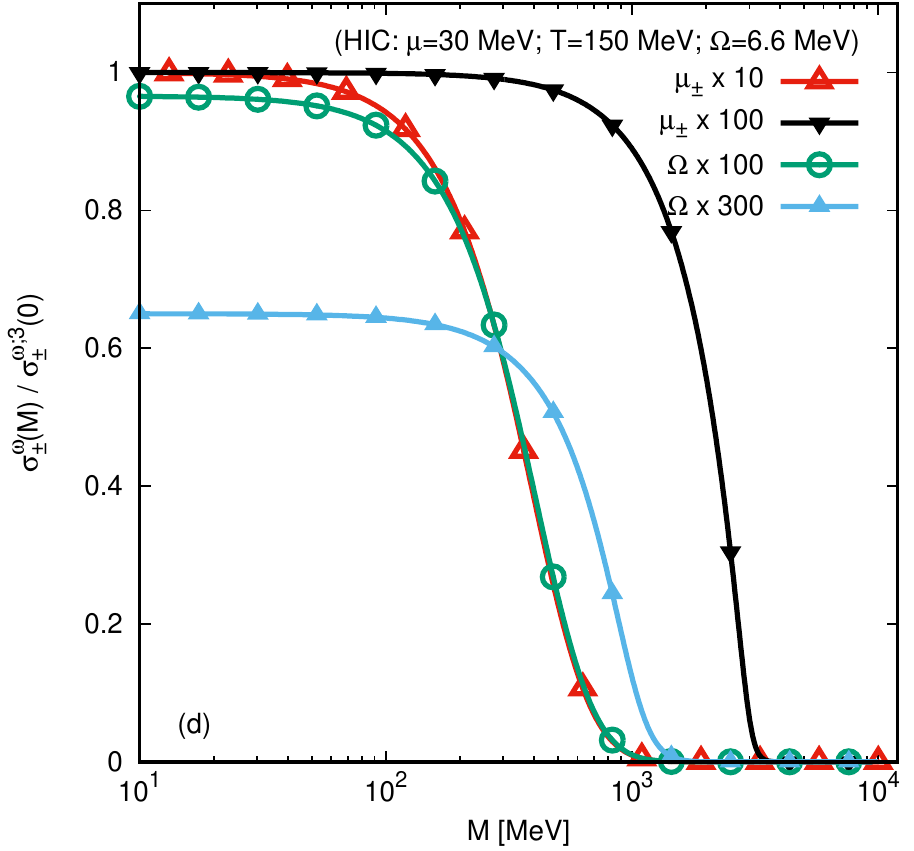}
\end{tabular}
\caption{
(a) Relative mass correction $1 - \sigma^\omega_\pm(M) / \sigma^\omega_\pm(0)$, 
with both $\sigma^\omega_\pm(M)$ and $\sigma^\omega_\pm(0)$ computed from 
Eq.~\eqref{eq:CC_so_axis}, represented with respect to $M/T$ with dotted lines 
and symbols. The solid black line represents the analytic prediction 
$3M^2/24 T^2 \ln(2)$, derived from Eqs.~\eqref{eq:CC_so_HighT} and \eqref{eq:CC_so_axis_T}.
(b)--(d) Ratio $\sigma^\omega_{\pm}(M) / \sigma^{\omega;n}_\pm(0)$, evaluated on 
the rotation axis, with $\sigma^\omega_\pm(M)$ computed using 
Eq.~\eqref{eq:CC_so_axis} and the approximations $\sigma^{\omega;n}_\pm(0)$ are 
given for (b) $n = 1$, (c) $n = 2$ and (d) $n = 3$ in Eqs.~\eqref{eq:CC_so_O1}, 
\eqref{eq:CC_so_O2} and \eqref{eq:CC_so_O3}, respectively.}
\label{fig:CC_so}
\end{figure}

The validity of the results derived for the vortical charge current conductivity 
$\sigma^\omega_\pm$ is now considered. An exact result was obtained in 
Eq.~\eqref{eq:CC_so_axis} for this quantity evaluated on the rotation axis, 
which is valid at any mass. Figure~\ref{fig:CC_so}(a) shows that, in the 
vicinity of the HIC parameters, the relative mass correction 
$1 - \sigma^\omega(M) / \sigma^\omega(0)$ is well represented by the 
quadratic term $3M^2 / 24T^2 \ln(2)$ highlighted in 
Eqs.~\eqref{eq:CC_so_HighT} and \eqref{eq:CC_so_axis_T}.

Away from the rotation axis, the result for $\sigma^\omega_\pm$ was obtained 
only up to O($\Omega^4$) in Eq.~\eqref{eq:CC_so_Omega}. This expression, 
involving the polylogarithm function, is approximate even on the rotation axis. 
To test its validity for relativistic heavy ion collisions, various levels of 
approximation $\sigma^{\omega;n}_\pm$ are considered based on 
Eq.~\eqref{eq:CC_so_Omega}, distinguished using the index $n$. The $n = 1$ 
approximation corresponds to the leading order with respect to the 
temperature, highlighted in Eq.~\eqref{eq:CC_so_HighT} and reproduced below:
\begin{equation}
 \sigma^{\omega;1}_\pm = \pm \frac{2\mu_\pm T}{\pi^2} \ln 2.
 \label{eq:CC_so_O1}
\end{equation}
As can be seen from Fig.~\ref{fig:CC_so}(b), $\sigma^{\omega;1}_\pm$ offers 
a very good approximation under the HIC conditions. 
The mass corrections are important only when $M \gtrsim T$, 
even for $T= 750\ {\rm MeV}$. However, when $\mu_\pm$ or $\Omega$ become comparable 
with $T$, deviations can be seen. In order to understand the nature of these 
deviations, the next to leading order is considered below, which can also be 
read from Eq.~\eqref{eq:CC_so_HighT}:
\begin{equation}
 \sigma^{\omega;2}_\pm = \pm \frac{2\mu_\pm T}{\pi^2} \ln 2 
 \pm \frac{\mu_\pm}{12\pi^2 T} \left[\mu_\pm^2 +
 \frac{\bm{\omega}^2 + 3\bm{a}^2}{4}\right].\label{eq:CC_so_O2}
\end{equation}
In the above, the mass contribution was not taken into account.
As can be seen from Fig.~\ref{fig:CC_so}(c), this approximation improves 
the small $M$ results at moderate values for $\Omega$ and $\mu_\pm$. However, 
increasing $\mu_\pm$ and $\Omega$ further brings about new discrepancies. 
The final approximation, $\sigma^{\omega;3}_\pm$, corresponds to the 
result in Eq.~\eqref{eq:CC_so_Omega}, written without the mass term:
\begin{equation}
 \sigma^{\omega;3}_\pm = \pm \frac{T^2}{\pi^2} \left[{\rm Li}_2(-e^{-\mu_\pm / T}) - {\rm Li}_2(-e^{\mu_\pm/ T})\right] \pm \frac{\bm{\omega}^2 + 3\bm{a}^2}{24\pi^2} 
 \tanh\frac{\mu_\pm}{2T}.\label{eq:CC_so_O3}
\end{equation}
Figure~\ref{fig:CC_so}(d) shows that the above approximation correctly 
accounts for the chemical potential, even when $\mu_\pm / T \gg 1$. However, 
no improvement can be seen for the case of large values of $\Omega$, which is 
to be expected since Eq.~\eqref{eq:CC_so_Omega} 
was obtained following a truncation at order $O(\Omega^4)$.

\section{Axial current}\label{sec:AC}

Initially shown to be so by Vilenkin \cite{vilenkin78}, it is known that 
the local vorticity induces an axial current parallel to the vorticity 
vector through the chiral vortical effect \cite{kharzeev16}. This section 
presents an analysis of the effect of helicity imbalance induced by the 
helicity chemical potential $\mu_H$ on the thermal expectation value 
(t.e.v.) of the axial charge current (ACC) operator $\widehat{J}_A^\halpha$,
introduced in Eq.~\eqref{eq:CC_hat}.
The analysis begins with a general discussion of the t.e.v. of the ACC, presented 
in Subsec.~\ref{sec:AC:gen}, followed by analytical and numerical analyses 
of the finite mass regime in Subsecs.~\ref{sec:AC:M0} and \ref{sec:AC:num}.

\subsection{General analysis}\label{sec:AC:gen}

The bilinear form introduced in Eq.~\eqref{eq:F_bilinear_def} corresponding to 
the ACC, defined in Eq.~\eqref{eq:CC_hat}, is
\begin{equation}
 \mathcal{J}^\halpha_A(\psi, \chi) = \overline{\psi} \gamma^\halpha \gamma^5 \chi.
\end{equation}
Noting that $\mathcal{J}^\halpha_A(V_j,V_j) = 
-[\mathcal{J}^\halpha_A(U_j,U_j)]^* = -\mathcal{J}^\halpha_A(U_j,U_j)$, 
it can be seen that the second line in Eq.~\eqref{eq:F_R} does not 
contribute to the t.e.v. $\braket{:J^\halpha_A:_\Omega}$, however,
there will be an extra vacuum contribution to the t.e.v. computed with 
respect to the Minkowski vacuum, $\braket{:J^\halpha_A:_M}$.
Using the following relations:
\begin{align}
 \mathcal{J}^\hatt_A(U_j, U_j) =& \frac{p_j}{8\pi^2 E_j} \left[2 \lambda_j J_{m_j}^+(q_j \rho) + \frac{k_j}{p_j} J_{m_j}^-(q_j \rho) \right], \nonumber\\
 \mathcal{J}^\hvarphi_A(U_j, U_j) =& \frac{\lambda_j q_j}{4\pi^2 p_j} J_{m_j}^\times(q_j \rho), \nonumber\\
 \mathcal{J}^\hatz_A(U_j, U_j) =& \frac{1}{8\pi^2} \left[J_{m_j}^-(q_j \rho) + \frac{2\lambda_j k_j}{p_j} J_{m_j}^+(q_j \rho) \right],
\end{align}
the t.e.v.s of the components of the ACC taken with respect to the rotating vacuum 
can be written as follows:
\begin{multline}
 \begin{pmatrix}
  \braket{:\widehat{J}^{\hatt}_A:_\Omega} \\
  \braket{:\widehat{J}^{\hvarphi}_A:_\Omega} \\
  \braket{:\widehat{J}^{\hatz}_A:_\Omega}
 \end{pmatrix} = 
 \frac{1}{4\pi^2} \sum_{\lambda = \pm\frac{1}{2}} 
 \sum_{m = -\infty}^\infty \int_M^\infty dE \, {\rm sgn}(\widetilde{E})\\
 \times \left[
 \frac{1}{e^{\beta_0(|\widetilde{E}| - \mu_{\lambda;0})} + 1} +
 \frac{1}{e^{\beta_0(|\widetilde{E}| + \mu_{\lambda;0})} + 1}
 \right] \int_{0}^p dk 
 \begin{pmatrix}
   2\lambda p \, J_m^+(q \rho) \\
  2\lambda \frac{q E}{p} J_m^\times(q\rho) \\
  E \, J_m^-(q\rho)
 \end{pmatrix}.\label{eq:AC}
\end{multline}
The t.e.v.s expressed with respect to the Minkowski vacuum are:
\begin{multline}
 \begin{pmatrix}
  \braket{:\widehat{J}^{\hatt}_A:_M} \\
  \braket{:\widehat{J}^{\hvarphi}_A:_M} \\
  \braket{:\widehat{J}^{\hatz}_A:_M}
 \end{pmatrix} = 
 \frac{1}{4\pi^2} \sum_{\lambda = \pm\frac{1}{2}} 
 \sum_{m = -\infty}^\infty \int_M^\infty dE \\
 \times \left[
 \frac{1}{e^{\beta_0(\widetilde{E} - \mu_{\lambda;0})} + 1} +
 \frac{1}{e^{\beta_0(\widetilde{E} + \mu_{\lambda;0})} + 1}
 \right] \int_{0}^p dk 
 \begin{pmatrix}
  2\lambda p \, J_m^+(q \rho) \\
  2\lambda \frac{q E}{p} J_m^\times(q\rho) \\
  E \, J_m^-(q\rho)
 \end{pmatrix}.\label{eq:AC_simp}
\end{multline}
Due to the factors of $2\lambda$ in $\braket{:\widehat{J}^{\hatt}_A:_M}$ and 
$\braket{:\widehat{J}^{\hvarphi}_A:_M}$, it is clear that these components vanish 
when $\mu_{H;0} = 0$. Thus, the t.e.v.s of these two components computed with 
respect to the Minkowski and rotating vacua coincide, i.e.
\begin{equation}
 \braket{:\widehat{J}^{\hatt}_A:_M} =  \braket{:\widehat{J}^{\hatt}_A:_\Omega}, \qquad
 \braket{:\widehat{J}^{\hvarphi}_A:_M} =  \braket{:\widehat{J}^{\hvarphi}_A:_\Omega}.
\end{equation}

As was the case for the VCC and HCC, the t.e.v. of the ACC can be decomposed as:
\begin{equation}
 \braket{:\widehat{J}^\halpha_A:_\Omega} = 
 Q_A u^\halpha + \mathcal{J}_A^\halpha, \qquad
 \mathcal{J}_A^\halpha = \sigma_A^\tau \tau^\halpha
 + \sigma_A^\omega \omega^\halpha.
 \label{eq:ACC_dec}
\end{equation}

\subsection{Small mass limit}\label{sec:AC:M0}

The t.e.v.s of the components of the ACC taken with 
respect to the Minkowski vacuum can be put in the form:
\begin{multline}
 \begin{pmatrix}
  \braket{:\widehat{J}^\hatt_A:_M} \\
  \braket{:\widehat{J}^\hvarphi_A:_M} \\
  \braket{:\widehat{J}^\hatz_A:_M}
 \end{pmatrix}
 = \frac{1}{2\pi^2} \sum_{\lambda = \pm \frac{1}{2}}
 \sum_{j = 0}^\infty (\rho \Omega)^{2j}
 \sum_{n = 0}^\infty \frac{\Omega^{2n} s_{n+j,j}^+}{(2n+2j+1)!}
 \int_M^\infty dE\, p^{2j} \\
 \times \frac{d^{2n+2j}}{dE^{2n+2j}} 
 \left[\frac{1}{e^{\beta_0(E - \mu_{\lambda,0})} + 1} + 
 \frac{1}{e^{\beta_0(E + \mu_{\lambda,0})} + 1}\right] 
 \begin{pmatrix}
 {\displaystyle \frac{2n + 2j + 1}{2j+1} 2 \lambda p^2 }\\
 {\displaystyle \frac{2\lambda \rho \Omega}{2j + 3} 
 [p^2 + (2j+2) E^2]} \\
 {\displaystyle \frac{\Omega}{2p} [p^2 + (2j+1) E^2]}
 \end{pmatrix}.
 \label{eq:AC_M0_aux}
\end{multline}

For the $z$ component of the t.e.v. of the AC, the small mass
limit can be derived in closed form. After 
changing the integration variable to $dp = \frac{E}{p} dE$, 
the procedure introduced in Eq.~\eqref{eq:M0_exp} can be used 
to obtain:
\begin{multline}
 \braket{:\widehat{J}^\hatz_A:_M} = \frac{\Omega}{4\pi^2} 
 \sum_{\lambda = \pm\frac{1}{2}} 
 \sum_{j = 0}^\infty (\rho \Omega)^{2j}
 \sum_{n = 0}^\infty \frac{\Omega^{2n} s_{n+j,j}^+}{(2n+2j+1)!} \int_0^\infty dp \left[(2j+2)p^{2j+1} - 
 j(2j+1) M^2 p^{2j-1}\right] \\\times
 \frac{d^{2n+2j}}{dp^{2n+2j}} \left[\frac{1}{e^{\beta_0(E - \mu_{\lambda,0})} + 1} + 
 \frac{1}{e^{\beta_0(E + \mu_{\lambda,0})} + 1}\right],
\end{multline}
where the coefficient of $M^2$ was obtained after noting 
that $E \simeq p + \frac{M^2}{2p}$ and $\frac{1}{E} \simeq \frac{1}{p} - \frac{M^2}{2p^3}$.
In the term proportional to $M$, the $j = 0$ term is infrared divergent and must be treated separately.
After performing integration by parts $2j$ times, the following formula can 
be used for the integration with respect to $p$:
\begin{equation}
 \int_0^\infty dp\, p \frac{d^{2n}}{dp^{2n}} \left[\frac{1}{e^{\beta_0(E - \mu_{\lambda,0})} + 1} + 
 \frac{1}{e^{\beta_0(E + \mu_{\lambda,0})} + 1}\right] = 
 \begin{cases}
  2I_0^+(\beta \mu_\lambda) / \beta_0^2, & n = 0,\\
  1, & n = 1, \\
  0, & n > 1,
 \end{cases}
\end{equation}
where $I_0^+(a)$ is given in Eq.~\eqref{eq:FD}.
It is not difficult to see that the sum over $n$ terminates 
at $n = 1$ for the first term and at $n = 0$ for the second
term. The exact result is:
\begin{align}
 \braket{:\widehat{J}^\hatz_A:_M} =& 
 \left[\sigma_A^\omega + \frac{\bm{\omega}^2 + 3\bm{a}^2}
 {24\pi^2} - \frac{M^2}{4\pi^2} + O(M^4)\right]\omega^\hatz, \nonumber\\
 \sigma_A^\omega =& \frac{T^2}{6} + 
 \frac{\mu_V^2 + \mu_H^2}{2\pi^2} + O(M^4),
 \label{eq:AC_so_M}
\end{align}
where $a = -\rho \Omega^2 \Gamma^2 e_\hrho$ and 
$\omega = \Omega \Gamma^2 e_\hatz$ are the 
acceleration and vorticity vectors, introduced 
in Eqs.~\eqref{eq:RR_a} and \eqref{eq:RR_omega},
respectively, while their squares $\bm{\omega}^2 = -\omega^2 \ge 0$ and 
$\bm{a}^2 = -a^2 \ge 0$ are given in Eq.~\eqref{eq:vecs_squares}.
The second and third terms inside the square brackets in Eq.~\eqref{eq:AC_so_M} 
are purely vacuum contributions which are not taken into account 
in the definition of $\sigma_A^\omega$. 

It is worth comparing the result in Eq.~\eqref{eq:AC_so_M} with 
the one obtained in Ref.~\cite{prokhorov18} using the Wigner 
function proposed in Ref.~\cite{becattini13}:
\begin{equation}
 \braket{j^5_\mu} = \left(\frac{1}{6} \left[T^2 + 
 \frac{a^2 - \omega^2}{4\pi^2}\right] + \frac{\mu^2}{2\pi^2}\right) \omega_\mu.
 \label{eq:prokh_j5}
\end{equation}
Noting that $\omega^2 = -\bm{\omega}^2$ and $a^2 = -\bm{a}^2$, it 
can be seen that the parts which depend on $T$ and $\mu = \mu_V$ 
reproduced in Eq.~\eqref{eq:prokh_j5} agree with those obtained 
in Eq.~\eqref{eq:AC_so_M}, when $\mu_H = 0$. There is a discrepancy 
in the vacuum term, equal to 
\begin{equation}
 \braket{:\widehat{J}^\hatz_A:_M} - \braket{j_5^z} = \frac{\bm{a}^2}{6\pi^2} \omega^\hatz,
\end{equation}
which may be due to a fundamental difference between the formulation 
based on the Wigner function (employed in Ref.~\cite{prokhorov18}) 
and the QFT formulation employed in this work.

As opposed to the case of the $\hatz$ component, the sum over $n$ in 
Eq.~\eqref{eq:AC_M0_aux} no longer terminates when considering the 
$\hatt$ and $\hvarphi$ components, which is an indication that their 
dependence on $\Gamma$ and $\Omega$ is non-polynomial (see also the 
discussion in Subsec.~\ref{sec:CC:M0}). Performing the change of 
integration variable from $E$ to $p$ ($dp = E dE / p$) in 
Eq.~\eqref{eq:AC_M0_aux} and using Eq.~\eqref{eq:M0_exp2} gives:
\begin{align}
 \braket{:\widehat{J}^\hatt_A:_\Omega} =& \frac{1}{4\pi^2} \sum_{\lambda = \pm \frac{1}{2}} 2\lambda 
 \sum_{j = 0}^\infty (\rho \Omega)^{2j}
 \sum_{n = 0}^\infty \frac{\Omega^{2n} (2j)!(2j+2)}{(2n+2j)!} s_{n+j,j}^+
 \int_0^\infty dp\, \left(p^2 - \frac{M^2}{2j+1}\right)
 \nonumber\\
 &\qquad\qquad \times \frac{d^{2n}}{dp^{2n}} 
 \left[\frac{1}{e^{\beta_0(p - \mu_{\lambda,0})} + 1} + 
 \frac{1}{e^{\beta_0(p + \mu_{\lambda,0})} + 1}\right],\nonumber\\
 \braket{:\widehat{J}^\hvarphi_A:_\Omega} =& \frac{\rho \Omega}{4\pi^2} \sum_{\lambda = \pm \frac{1}{2}} 2\lambda 
 \sum_{j = 0}^\infty (\rho \Omega)^{2j}
 \sum_{n = 0}^\infty \frac{\Omega^{2n} (2j+2)!}{(2n+2j+1)!} s_{n+j,j}^+
 \int_0^\infty dp\, \left(p^2 - \frac{M^2}{2j+3}\right)
 \nonumber\\
 &\qquad\qquad \times \frac{d^{2n}}{dp^{2n}} 
 \left[\frac{1}{e^{\beta_0(p - \mu_{\lambda,0})} + 1} + 
 \frac{1}{e^{\beta_0(p + \mu_{\lambda,0})} + 1}\right].
\end{align}
Taking into account the $n = 0$ and $n = 1$ terms gives:
\begin{align}
 \braket{:\widehat{J}^\hatt_A:_\Omega} =& 
 \frac{\Gamma}{4\pi^2} 
 \sum_{\lambda = \pm \frac{1}{2}} 2\lambda \left\{ 
 \frac{4}{\beta^3} I_{1/2}^+(\beta \mu_\lambda) + 
 \frac{\Omega^2 \Gamma^2(4\Gamma^2 - 1)}{3\beta} I_{-1/2}^+(\beta \mu_{\lambda})\right.\nonumber\\
 & \left.- \frac{2M^2}{\beta}\left[1 + \frac{{\rm arcsinh}(\rho\Omega\Gamma)}{\rho\Omega\Gamma^2}\right] 
 I_{-1/2}^+(\beta \mu_\lambda)
 + O(\Omega^4,M^4,\Omega^2 M^2)\right\},\nonumber\\
 \braket{:\widehat{J}^\hvarphi_A:_\Omega} =& \frac{\rho\Omega\Gamma}{4\pi^2} 
 \sum_{\lambda = \pm \frac{1}{2}} 2\lambda \left\{ 
 \frac{4}{\beta^3} I_{1/2}^+(\beta \mu_\lambda) + 
 \frac{\Omega^2 \Gamma^2(4 \Gamma^2 - 3)}{3\beta} I_{-1/2}^+(\beta \mu_{\lambda})\right.\nonumber\\
 & \left.
 - \frac{2M^2}{\beta \rho^2 \Omega^2}\left[1 - \frac{{\rm arcsinh}(\rho\Omega\Gamma)}{\rho\Omega\Gamma^2}\right] 
 I_{-1/2}^+(\beta \mu_\lambda) + O(\Omega^4,M^4,\Omega^2 M^2)\right\}.
\end{align}
The charge $Q_A$ and the circular conductivity $\sigma_A^\tau$ are 
\begin{align}
 Q_A =& \frac{T^3}{\pi^2} \sum_{\lambda = \pm\frac{1}{2}} 2\lambda I_{1/2}^+\left(\frac{\mu_\lambda}{T}\right) +
 \frac{T}{\pi^2} \left[
 \frac{\bm{\omega}^2 + \bm{a}^2}{4} - M^2
 \frac{{\rm arcsinh}(\rho\Omega \Gamma)}{\rho\Omega}\right]\sum_{\lambda = \pm\frac{1}{2}} 2\lambda I_{-1/2}^+\left(\frac{\mu_\lambda}{T}\right) \nonumber\\
 & + O(\Omega^4,M^4,\Omega^2 M^2),\nonumber\\ 
 \sigma_A^\tau =& \frac{T}{6\pi^2}
 \left\{1 -
 \frac{3M^2/\Omega^2}{\Gamma^2(\Gamma^2 - 1)} \left[(2\Gamma^2 - 1) \frac{{\rm arcsinh}(\rho\Omega \Gamma)}{\rho\Omega\Gamma^2} - 1\right]\right\} \sum_{\lambda = \pm \frac{1}{2}} 2\lambda I_{-1/2}^+\left(\frac{\mu_\lambda}{T}\right) \nonumber\\
 &+ O(\Omega^4,M^4, \Omega^2 M^2).
\end{align}
The sum over $\lambda$ can be performed by taking into account 
the expressions for $I_{\pm1/2}^+$ given in Eq.~\eqref{eq:FD_Half}:
\begin{align}
 Q_A =& \frac{\mu_H}{3} \left(T^2 + \frac{\mu_H^2 + 3 \mu_V^2}{\pi^2}\right)
 - \frac{2T^3}{\pi^2} \left[
 {\rm Li}_3(-e^{-\mu_+/T}) - 
 {\rm Li}_3(-e^{-\mu_-/T})\right] \nonumber\\
 &+\frac{T}{\pi^2} \left[\frac{\bm{\omega}^2 + \bm{a}^2}{4} - \frac{M^2}{\rho\Omega} {\rm arcsinh}(\rho\Omega \Gamma) \right] 
 \ln \left[\frac{\cosh(\mu_+/2T)}
 {\cosh(\mu_-/2T)}\right] + O(\Omega^4,M^4,\Omega^2 M^2),\nonumber\\
 \sigma_A^\tau =& 
 \frac{T}{6\pi^2} \left\{1 - 
 \frac{3M^2/\Omega^2}{\Gamma^2(\Gamma^2 - 1)} \left[(2\Gamma^2 - 1) \frac{{\rm arcsinh}(\rho\Omega \Gamma)}{\rho\Omega\Gamma^2} - 1\right]\right\} \ln \left[\frac{\cosh(\mu_+/2T)}
 {\cosh(\mu_-/2T)}\right] \nonumber\\
 &+ O(\Omega^4,M^4,\Omega^2 M^2).
 \label{eq:AC_Q_st}
\end{align}
The high temperature limit of the above expressions is:
\begin{align}
 Q_A =& 
 \frac{4 \mu_V \mu_H T}{\pi^2} \ln 2 + 
 \frac{\mu_V \mu_H}{2\pi^2T} \left[
 \frac{\mu_V^2 + \mu_H^2}{3} +
 \frac{\bm{\omega}^2 + \bm{a}^2}{4} - \frac{M^2}{\rho\Omega} {\rm arcsinh}(\rho\Omega \Gamma) \right] + O(T^{-2}),\nonumber\\
 \sigma_A^\tau =& 
 \frac{\mu_V \mu_H}{12\pi^2 T} \left\{1 - 
 \frac{3M^2/\Omega^2}{\Gamma^2(\Gamma^2 - 1)} \left[(2\Gamma^2 - 1) \frac{{\rm arcsinh}(\rho\Omega \Gamma)}{\rho\Omega\Gamma^2} - 1\right]\right\} + O(T^{-2}).
 \label{eq:AC_Q_st_T}
\end{align}
It can be seen that the mass term makes a significant 
contribution to $\sigma^\tau_A$ when $\Omega$ is sufficiently small. 
This is an indication that the finite mass correction to $\sigma_A^\tau$ behaves 
formally different from $\sigma_A^\tau(M = 0)$. Indeed, multiplying $\delta \sigma_A^\tau = \sigma_A^\tau(M) - \sigma_A^\tau(0)$ by the circular vector $\tau^\halpha$ gives:
\begin{align}
 \delta \sigma_A^\tau\, \tau =& \frac{\mu_V \mu_H M^2 \Gamma}{4\pi^2 T \rho \Omega} 
 \left[\frac{{\rm arcsinh}(\rho \Omega \Gamma)}{\rho \Omega} - 1 + 
 \rho \Omega {\rm arcsinh}(\rho \Omega \Gamma)\right](\rho \Omega e_\hatt + e_\hvarphi) \nonumber\\
 \simeq& \frac{\mu_V \mu_H M^2}{3\pi^2 T} [u^\hvarphi e_\hvarphi + O(\rho^2 \Omega^2)],
\end{align}
where $u^\hvarphi = \rho \Omega \Gamma$, indicating that $\delta \sigma_A^\tau \tau^\hvarphi$ 
contributes to $\braket{:J^\hvarphi_A:_\Omega}$ on the same footing as $Q_A u^\hvarphi$.

The results in Eq.~\eqref{eq:AC_Q_st} are valid only at small values 
of $\Omega$. It is possible to obtain exact expressions for 
$Q_A$ and $\sigma^\tau_A$ on the rotation axis, where 
\begin{equation}
 Q_A\rfloor_{\rho = 0} = \braket{:\widehat{J}_A^\hatt:_\Omega}\rfloor_{\rho = 0}, \qquad 
 \sigma_A^\tau = \left(\braket{:\widehat{J}_A^\hatt:_\Omega} - 
 \frac{\braket{:\widehat{J}_A^\hvarphi:_\Omega}}
 {\rho \Omega^3}\right)_{\rho = 0}.
\end{equation}
Noting that 
\begin{equation}
 \lim_{\rho \rightarrow 0} J_{m}^+(q\rho) = 
 \begin{cases}
  1, & m = \pm \frac{1}{2}, \\
  0, & \text{otherwise},
 \end{cases} \qquad 
 \lim_{\rho \rightarrow 0} \frac{J_{m}^\times(q\rho)}{q\rho} = 
 \begin{cases}
  \pm 1, & m = \pm \frac{1}{2}, \\
  0, & \text{otherwise},
 \end{cases}
\end{equation}
it is not difficult to obtain the following expressions:
\begin{align}
 \braket{:\widehat{J}^\hatt_A:_\Omega} =&
 \frac{2M^3}{\pi^2} \sum_{\ell = 1}^\infty (-1)^{\ell+1} 
 \cosh\frac{\ell \beta \Omega}{2} \sinh(\ell \beta \mu_V) 
 \sinh(\ell \beta \mu_H) \int_1^\infty dx(x^2 - 1) e^{-\ell \beta x M},\nonumber\\
 \frac{\braket{:\widehat{J}^\hvarphi_A:_\Omega}}{\rho\Omega} =&
 \frac{4M^4}{3\pi^2 \Omega} \sum_{\ell = 1}^\infty (-1)^{\ell+1} 
 \sinh\frac{\ell \beta \Omega}{2} \sinh(\ell \beta \mu_V) 
 \sinh(\ell \beta \mu_H) \int_1^\infty dx\, x (x^2 - 1) e^{-\ell \beta x M},
\end{align}
where $x= E / M$ and Eq.~\eqref{eq:RKT_exp_aux3} was used to 
expand the Fermi-Dirac factors. The integration with respect 
to $x$ can be readily performed and the sums over $\ell$ 
yield polylogarithm functions. The result can be summarised as:
\begin{align}
 Q_A\rfloor_{\rho \rightarrow 0} =& 
 -\frac{T^3}{2\pi^2} \sum_{\sigma_\Omega = \pm 1} 
 \sum_{\sigma_V = \pm 1} \sigma_V
 \sum_{\sigma_H = \pm 1} \sigma_H \left[{\rm Li}_3(-e^\zeta) + \frac{M}{T} {\rm Li}_2(-e^{\zeta})\right],\nonumber\\
 \sigma_A^\tau\rfloor_{\rho \rightarrow 0} =& 
 -\frac{T^4}{\pi^2 \Omega^3} \sum_{\sigma_\Omega = \pm 1} 
 \sum_{\sigma_V = \pm 1} \sigma_V
 \sum_{\sigma_H = \pm 1} \sigma_H \left\{
 \frac{\Omega}{2T} \left[{\rm Li}_3(-e^\zeta) + \frac{M}{T} {\rm Li}_2(-e^{\zeta})\right]\right.\nonumber\\
 &\left. -\sigma_\Omega \left[{\rm Li}_4(-e^\zeta) + \frac{M}{T} {\rm Li}_3(-e^{\zeta}) + \frac{M^2}{3T^2} {\rm Li}_2(-e^{\zeta})\right]\right\},
 \label{eq:AC_Q_st_axis}
\end{align}
where $\zeta = (\sigma_\Omega\frac{\Omega}{2} + \sigma_V \mu_V + \sigma_H \mu_H - M) / T$.
The above expressions are exact for any particle mass. Extracting the large temperature limit yields:
\begin{align}
 Q_A\rfloor_{\rho = 0} =& \frac{4\mu_H \mu_V T}{\pi^2} \ln 2 + \frac{\mu_H \mu_V}{2\pi^2 T} \left(\frac{\mu_H^2 + \mu_V^2}{3} + \frac{\Omega^2}{4} - M^2\right) + O(T^{-3}),\nonumber\\
 \sigma_A^\tau\rfloor_{\rho = 0} =& \frac{\mu_H \mu_V}{12\pi^2 T}\left(1 - \frac{4M^2}{\Omega^2}\right) + O(T^{-3}).
 \label{eq:AC_Q_st_axis_T}
\end{align}
The above results are in agreement with Eq.~\eqref{eq:AC_Q_st_T}.

The high temperature limit of the results obtained in 
this section can be summarised as follows:
\begin{align}
 \braket{:\widehat{J}^\halpha_A:_\Omega} =& Q_A u^\halpha + \sigma_A^\tau \tau^\halpha + \sigma_A^\omega \omega^\halpha,\nonumber\\ 
 Q_A =& \frac{4\mu_V \mu_H T}{\pi^2} \ln 2 + O(T^{-1}),\nonumber\\
 \sigma_A^\tau =& \frac{\mu_V\mu_H}{12\pi^2 T}\left(1 - \frac{4M^2}{\Omega^2} \right) + O(T^{-3}),\nonumber\\ 
 \sigma_A^\omega =& \frac{T^2}{6} + \frac{\mu_V^2 + \mu_H^2}{2\pi^2} + O(M^4).
 \label{eq:AC_summary}
\end{align}
The mass term was retained in the coefficient of $\tau^\halpha$ due to its unusual effect.
It is worth noting that, in the limit $\mu_H = M = 0$, the results in Eq.~\eqref{eq:AC_summary} reproduce those reported in Eq.~(4.14) and Table~2 of Ref.~\cite{buzzegoli18} for the case of a vanishing axial chemical pontential ($\mu_A = 0$).

\subsection{Numerical analysis}\label{sec:AC:num}

\begin{figure}
    \centering
\begin{tabular}{cc}
    \includegraphics[width=0.45\linewidth]{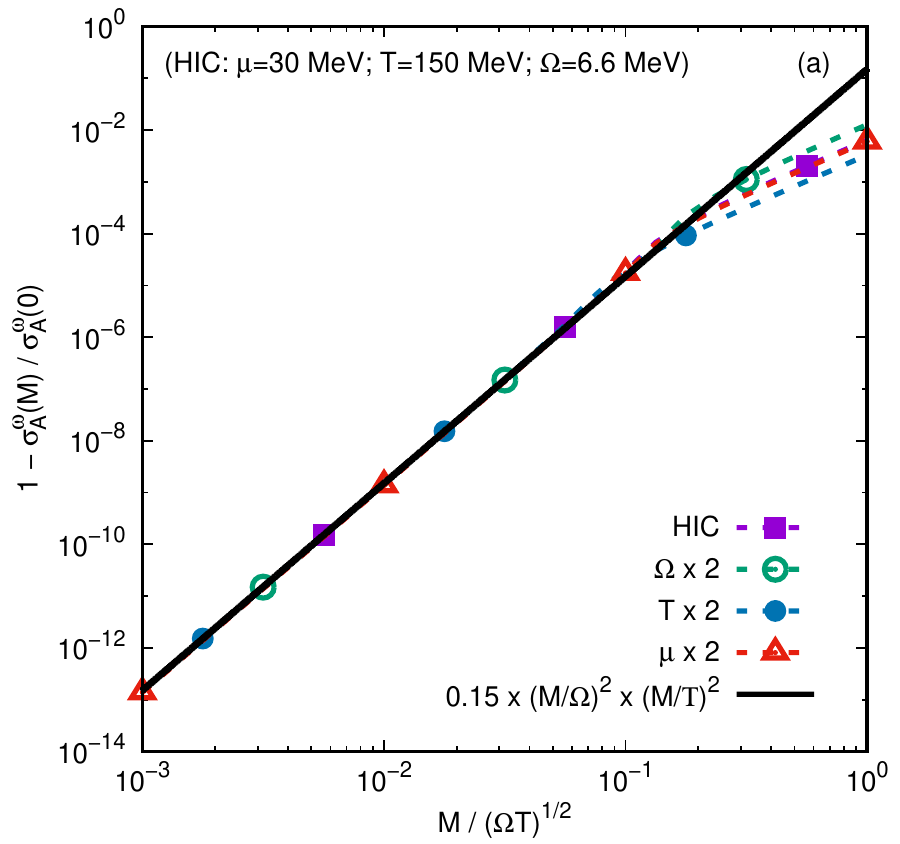} &
    \includegraphics[width=0.45\linewidth]{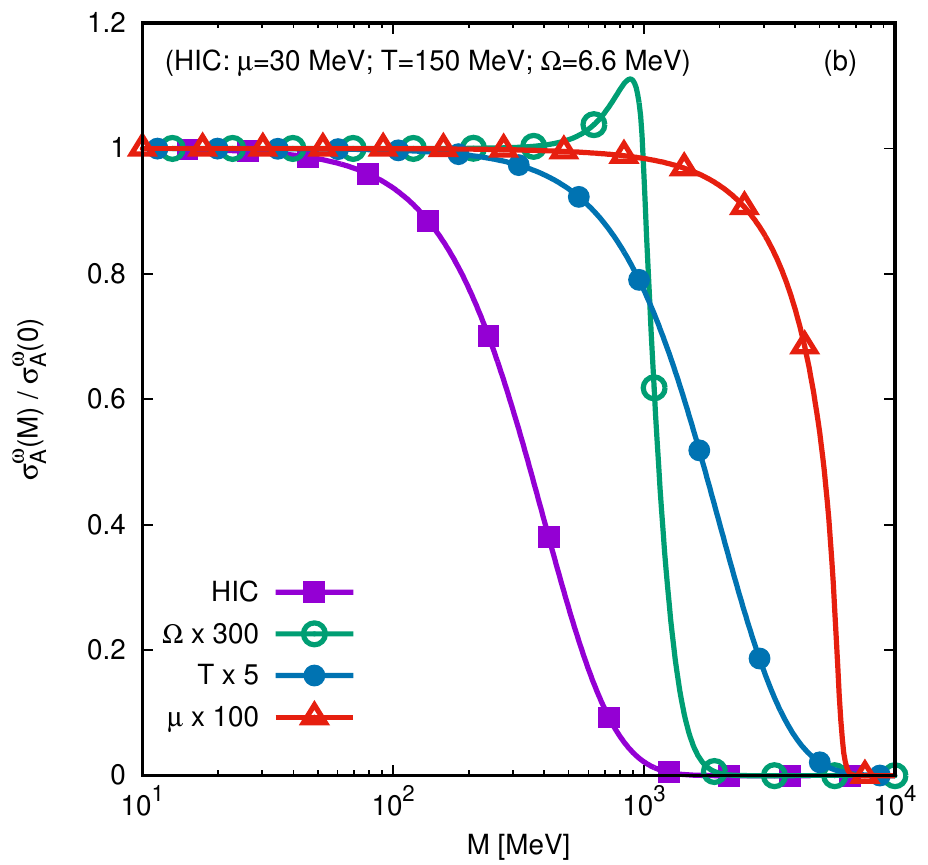}
\end{tabular}
\caption{(a) Relative mass correction 
$1 - \sigma^\omega_A(M) / \sigma^\omega_A(0)$ and (b) ratio $\sigma^\omega_A(M) / \sigma^{\omega}_A(0)$, evaluated on the rotation axis, where $\sigma^\omega_A(M)$ is computed numerically using Eq.~\eqref{eq:AC} and $\sigma^{\omega}_A(0)$ is taken from Eq.~\eqref{eq:AC_summary}. The solid line in (a) represents the best fit of the function $\alpha M^4 / \Omega^2 T^2$ to the numerical data.}
\label{fig:AC_so}
\end{figure}

In this section, the validity of the results summarised for the 
small mass limit in Eq.~\eqref{eq:AC_summary} is investigated as 
the mass is increased. As mentioned in the introduction, the 
numerical analysis is focussed on the HIC parameters. 
In the case of the axial current, the vector and 
helical chemical potentials play equivalent roles. Thus, for simplicity,
the convention $\mu_V = \mu_H = \mu$ is used throughout this section,
which is set to the ``HIC'' value of $\mu = 30\ {\rm MeV}$.

First, the validity of the constitutive equation for $\sigma^\omega_A$
derived in the small mass limit in Eq.~\eqref{eq:AC_summary} is considered 
as the mass is increased. At nonvanishing $M$, $\sigma^\omega_A(M)$ is 
evaluated numerically starting from Eq.~\eqref{eq:AC}. For simplicity, 
the analysis is limited to the case of the rotation axis. Figure~\ref{fig:AC_so}(a) 
confirms that the relative mass correction, 
$1 - \sigma_A^\omega(M) / \sigma_A^\omega(0)$, is of fourth order with 
respect to $M$. Around the HIC parameters, the mass term makes a relative 
contribution of the form 
$\alpha M^4 / \Omega^2 T^2$, where $\alpha \simeq 0.150$ is a 
dimensionless number. Figure~\ref{fig:AC_so}(b) presents the ratio 
$\sigma^\omega_A(M) / \sigma^\omega_A(0)$. 
It can be seen that the constitutive relation holds for 
$M \lesssim 100\ {\rm MeV}$, which is below the thermal energy 
($T = 150\ {\rm MeV}$). Increasing the temperature by a factor of 
$5$ (blue line with filled circles) increases the validity up to 
$M \simeq 500\ {\rm MeV}$, i.e. by a factor of $5$. At higher chemical 
potentials ($\mu_V = \mu_H = \mu = 3\ {\rm GeV}$), it can be seen that 
the constitutive relation remains valid until the mass approaches the 
Fermi level, which is given by $\sim \mu_V + \mu_H = 6\ {\rm GeV}$. 
The ratio $\sigma^\omega_A(M) / \sigma^\omega_A(0)$ seems to have a 
monotonic behaviour, as also observed for the ratio 
$\sigma^\omega_\pm(M) / \sigma^\omega_\pm(0)$, shown in Fig.~\ref{fig:CC_so}. 
An interesting effect can be observed when the rotation parameter is 
increased by a factor of $300$, to $\Omega \simeq 2\ {\rm GeV}$. At 
such a high value, $\Omega / 2$ acts like a Fermi level and the 
maximum observed in Fig.~\ref{fig:AC_so} resembles the one highlighted 
for the ratios $\Delta Q_\pm(M) / \Delta Q_\pm(0)$ and 
$\sigma^\tau_\pm(M) / \sigma^\tau_\pm(0)$ in Fig.~\ref{fig:CC_Q}. 
However, in the latter case, the maximum was observed at large 
chemical potential. 

\begin{figure}
    \centering
\begin{tabular}{cc}
    \includegraphics[width=0.45\linewidth]{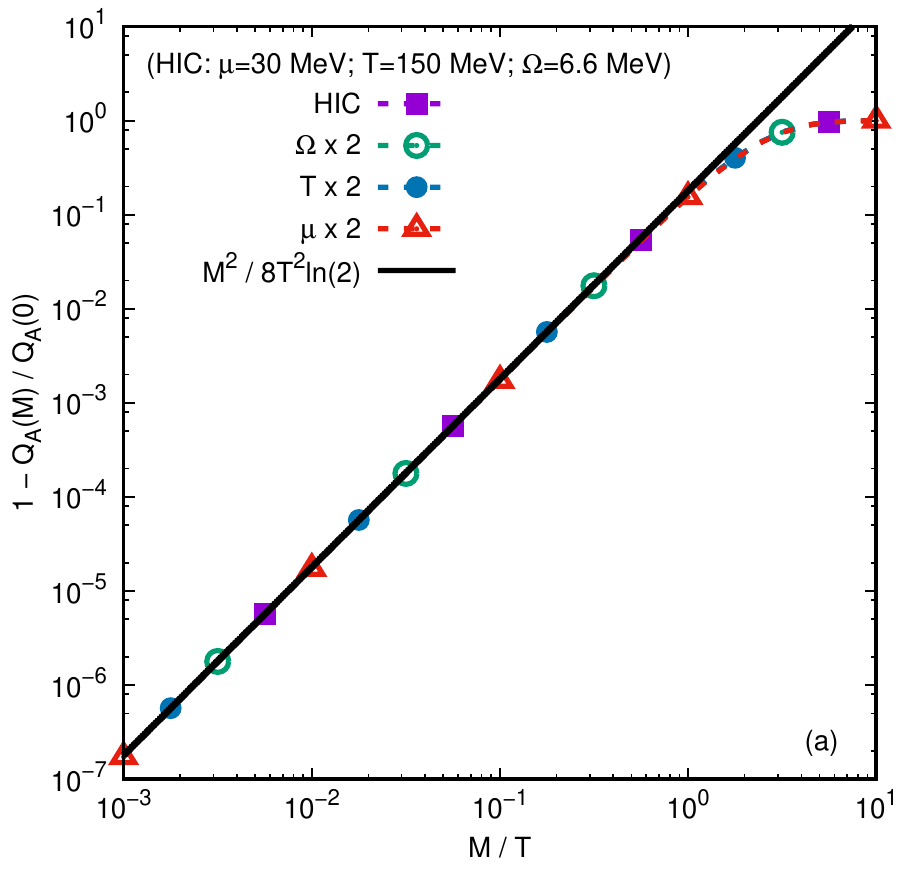} &
    \includegraphics[width=0.45\linewidth]{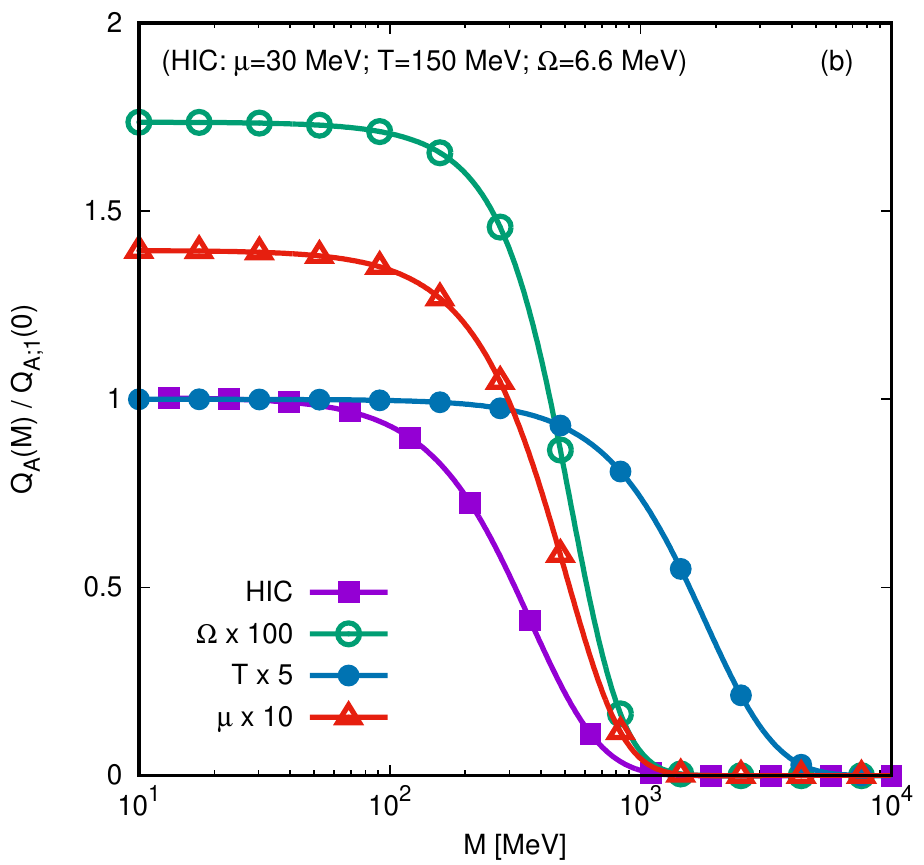} \\ 
    \includegraphics[width=0.45\linewidth]{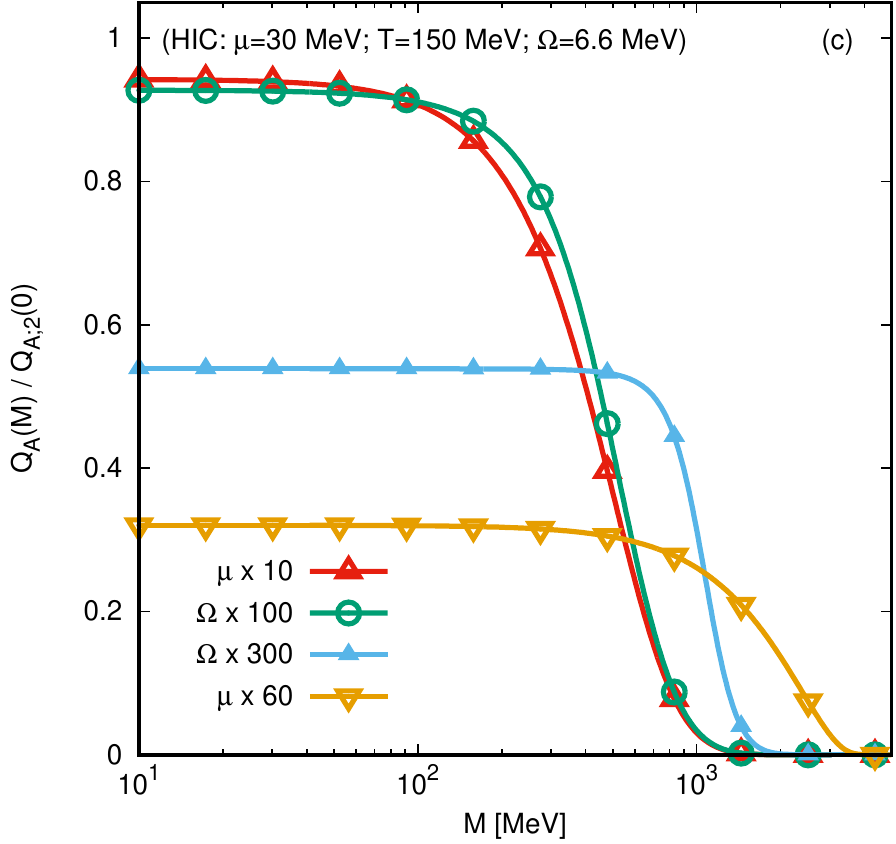} &
    \includegraphics[width=0.45\linewidth]{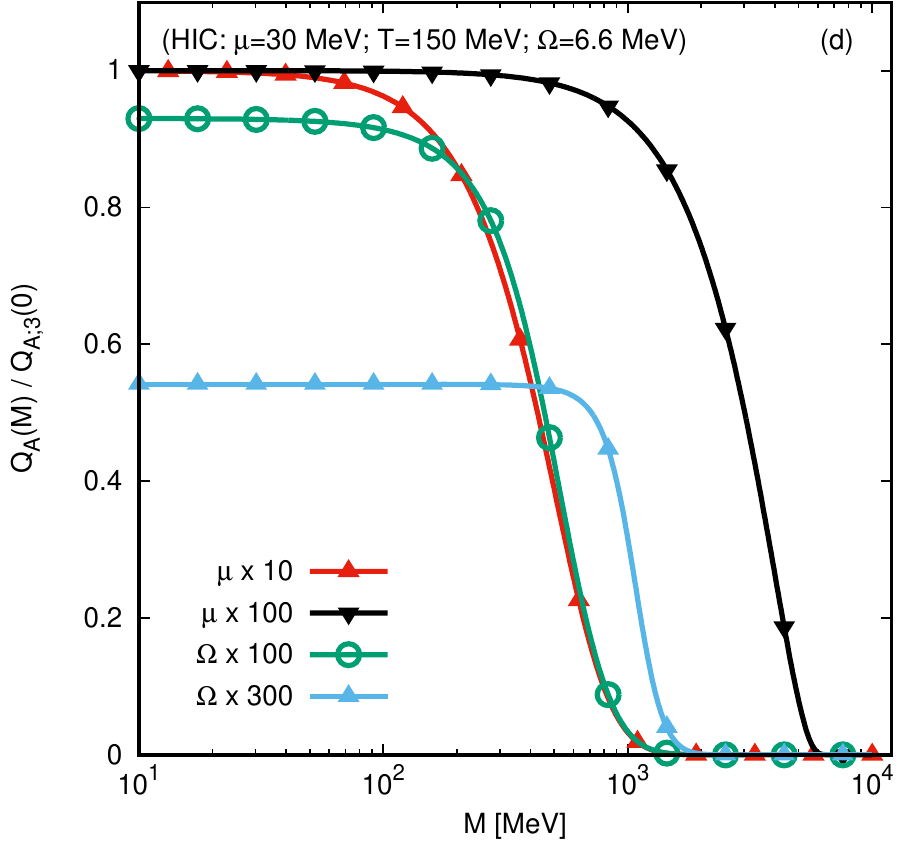}
\end{tabular}
\caption{(a) Relative mass correction $1 - Q_A(M) / Q_A(0)$ represented with respect to $M$. The solid black line shows the second order prediction in Eq.~\eqref{eq:AC_Q_st_axis_T}. (b)-(d) Ratio $Q_A(M) / Q_{A;n}(0)$ with respect to $M$, where $Q_A(M)$ is computed using Eq.~\eqref{eq:AC_Q_st_axis}, while the approximations $Q_{A;n}(0)$ are given for
(b) $n = 1$ (c) $n = 2$ and (d) $n = 3$ in Eqs.~\eqref{eq:AC_Q_O1}, \eqref{eq:AC_Q_O2} and \eqref{eq:AC_Q_O3}, respectively.}
\label{fig:AC_Q}
\end{figure}

Now, the axial charge density $Q_A$ and the conductivity 
$\sigma^\tau_A$ along $\tau$ are discussed. As opposed to 
$\sigma^\omega_A$, it is not possible to obtain the radial 
profiles of these quantities analytically, even in the massless 
limit. Instead, Eq.~\eqref{eq:AC_Q_st} gives $Q_A$ and 
$\sigma_A^\tau$ up to order $O(\Omega^4, M^4, \Omega^2 M^2)$ 
at any distance $\rho$ from the rotation axis, while 
Eq.~\eqref{eq:AC_Q_st_axis} 
gives exact expressions for their values on the rotation axis. 
The latter expressions involve the polylogarithm function. 
More insightful expressions can be obtained by considering 
the high temperature expansions \eqref{eq:AC_Q_st_T} and 
\eqref{eq:AC_Q_st_axis_T}. 

In the case of $Q_A$, according to Eq.~\eqref{eq:AC_Q_st_T}, 
the mass term makes second order contributions of the form 
$O(M^2/T^2)$. This is confirmed in Fig.~\ref{fig:AC_Q}(a), 
where the relative mass correction $1 - Q_A(M)/Q_A(0)$ is 
represented with respect to $M/T$. Both $Q_A(M)$ and $Q_A(0)$ 
are evaluated using Eq.~\eqref{eq:AC_Q_st_axis}, which is 
valid on the rotation axis for any mass. In the vicinity of 
the HIC parameters, $M^2 / 8T^2 \ln 2$ provides a good 
approximation for the relative mass corrections. Away from 
the rotation axis, Eq.~\eqref{eq:AC_Q_st} provides an 
approximation which is obtained using a small $\Omega$ 
expansion. Eq.~\eqref{eq:AC_Q_st} involves the polylogarithm 
function, and is thus less insightful compared to its high 
temperature expansion presented in  Eq.~\eqref{eq:AC_Q_st_T}. 
In what follows, three levels of approximation are considered, 
denoted using $Q_{A;n}$ ($1 \le n \le 3$), which are based on 
Eq.~\eqref{eq:AC_Q_st}. Their validity is tested compared with 
the exact solution in Eq.~\eqref{eq:AC_Q_st_axis}. The first 
two approximations take into account the leading and next-to-leading 
order terms in the high temperature expansion, given in 
Eq.~\eqref{eq:AC_Q_st_T}:
\begin{align}
 Q_{A;1} =&  \frac{4 \mu_H \mu_V T}{\pi^2} \ln 2,\label{eq:AC_Q_O1}\\
 Q_{A;2} =& \frac{4 \mu_H \mu_V T}{\pi^2} \ln 2 + 
 \frac{\mu_H \mu_V}{2\pi^2T} \left(
 \frac{\mu_H^2 + \mu_V^2}{3} +
 \frac{\bm{\omega}^2 + \bm{a}^2}{4}\right).\label{eq:AC_Q_O2}
\end{align}
The third expression is the massless limit of Eq.~\eqref{eq:AC_Q_st}:
\begin{multline}
  Q_{A;3} = \frac{\mu_H}{3} \left(T^2 + \frac{\mu_H^2 + 3 \mu_V^2}{\pi^2}\right)
 - \frac{2T^3}{\pi^2} \left[
 {\rm Li}_3(-e^{-\mu_+/T}) - 
 {\rm Li}_3(-e^{-\mu_-/T})\right] \\
 + \frac{T}{\pi^2} \frac{\bm{\omega}^2 + \bm{a}^2}{4} 
 \ln \left[\frac{\cosh(\mu_+/2T)}
 {\cosh(\mu_-/2T)}\right].\label{eq:AC_Q_O3}
\end{multline}

Figures~\ref{fig:AC_Q}(b), \ref{fig:AC_Q}(c) and \ref{fig:AC_Q}(d) 
show the ratio $Q_A(M) / Q_{A;n}(M)$ for $n = 1$, $2$ and $3$, 
respectively. It can be seen in Fig.~\ref{fig:AC_Q}(b) that 
$Q_{A;1}(0)$ is a good approximation for $Q_A$ on the axis in the 
case of the HIC parameters for $M \lesssim T / 3$. This inequality 
is confirmed also when the temperature is increased by a factor of $5$ 
(red curve and filled circles). At higher values of the 
rotation parameter ($\Omega \simeq 660\ {\rm MeV}$) or of the 
chemical potentials ($\mu_V = \mu_H = 300\ {\rm MeV}$), $Q_{A;1}$ is 
no longer a good approximation for $Q_A(0)$. Panel (c) shows that $Q_{A;2}$ 
is much better at these values, though it still presents a relative error 
of about $10\%$. The validity of $Q_{A;2}$ worsens as $\Omega$ and $\mu$ 
are further increased. Finally, panel (d) shows that considering the 
polylogarithms in Eq.~\eqref{eq:AC_Q_O3} is sufficient to correctly account 
for any value of the chemical potential. The agreement between 
$Q_{A;3}(0)$ and $Q_A(0)$ is very good even at 
$\mu_V = \mu_H = \mu = 3\ {\rm GeV}$. However, the discrepancies at 
high $\Omega$ persist, since Eq.~\eqref{eq:AC_Q_st_axis} is valid only 
up to second order in $\Omega$. For all parameters studied, it seems 
that the variations with respect to the mass appear only at 
$M \gtrsim 100\ {\rm MeV}$.

\begin{figure}
    \centering
\begin{tabular}{cc}
    \includegraphics[width=0.45\linewidth]{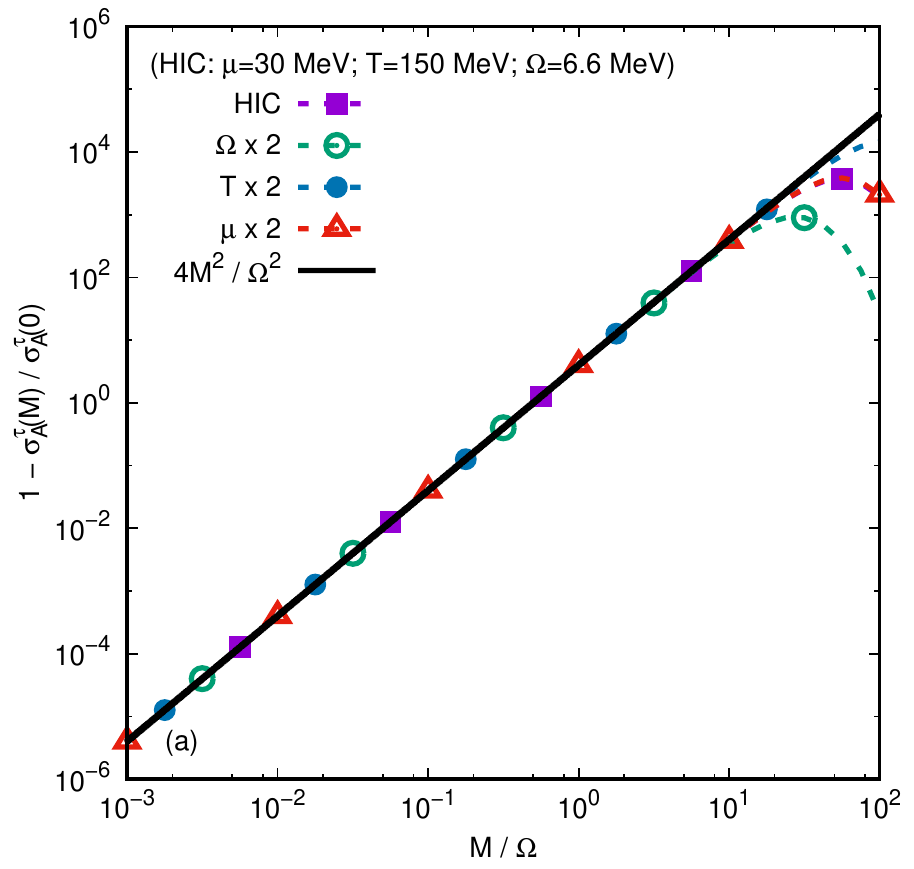} &
    \includegraphics[width=0.45\linewidth]{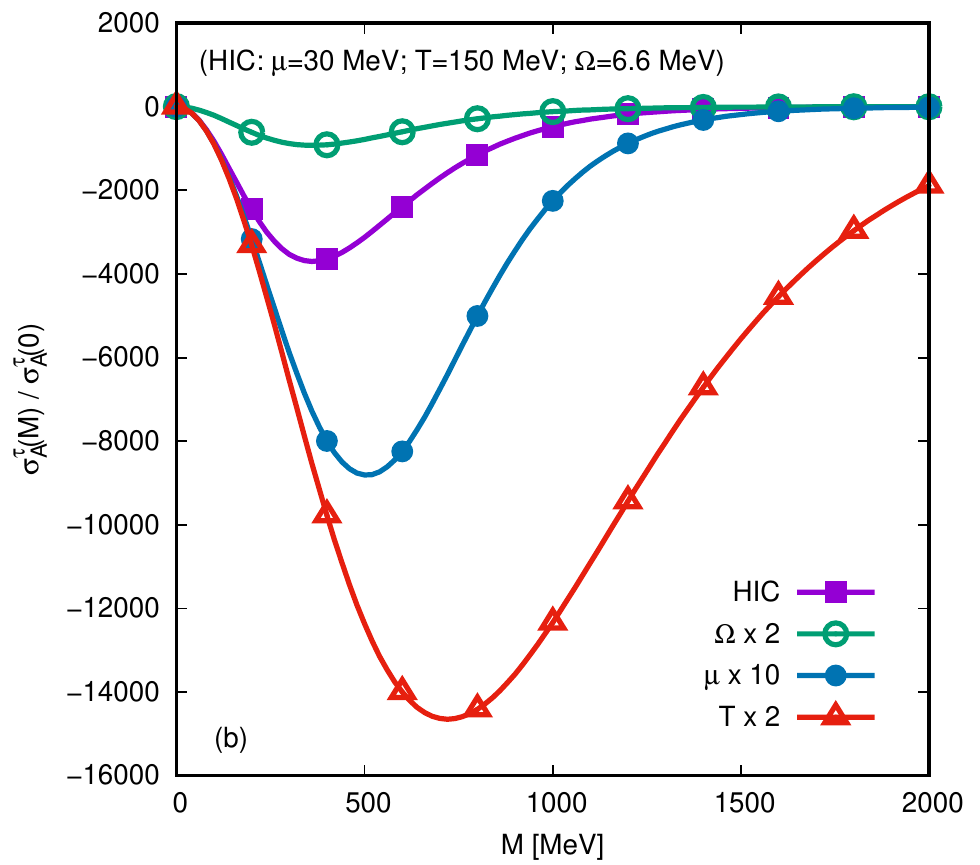}
\end{tabular}
\caption{(a) Relative mass correction 
$1 - \sigma^\tau_A(M) / \sigma^\tau_A(0)$ and (b) ratio 
$\sigma^\tau_A(M) / \sigma^{\tau}_A(0)$, where 
$\sigma^\tau_A(M)$ and $\sigma^\tau_A(0)$ are computed using 
the exact expression in Eq.~\eqref{eq:AC_Q_st_axis}. The 
solid black line in (a) corresponds to the analytic prediction
for the $M^2$ correction given in the high temperature expansion 
of $\sigma^\tau_A(M)$ in Eq.~\eqref{eq:AC_Q_st_axis_T}.}
\label{fig:AC_st}
\end{figure}

Considering now the properties of $\sigma^\tau_A$, 
Eq.~\eqref{eq:AC_Q_st_T} indicates that it receives 
corrections due to the mass term which are proportional 
to $M^2 / \Omega^2$. This is confirmed in Fig.~\ref{fig:AC_st}(a), 
which shows the relative mass correction 
$1 - \sigma^\tau_A(M) / \sigma^\tau_A(0)$ with respect 
to the ratio $M / \Omega$. Surprisingly, the analytic 
prediction $4M^2 / \Omega^2$ is dominant for parameters 
in the vicinity of the HIC values at high values of 
$M / \Omega$, where the mass term dominates by a few orders 
of magnitude over the massless limit. This behaviour is also 
confirmed in Fig.~\ref{fig:AC_st}(b), where the ratio 
$\sigma^\tau_A(M) / \sigma^\tau_A(0)$ is shown with respect to 
$M$. It can be seen that this ratio achieves negative values 
for the HIC parameters. Furthermore, $|\sigma^\tau_A(M)|$ 
increases with respect to the massless prediction 
$\sigma^\tau_A(0)$ by a few orders of magnitude as $M$ is increased, 
decreasing to $0$ after reaching a maximum. The amplitude of this 
maximum decreases when $\Omega$ is increased, but it increases 
when either $\mu_V = \mu_H = \mu$ or $T$ are increased (the two 
chemical potentials play an equivalent role). Furthermore, 
the maximum shifts to higher values of $M$ when the chemical 
potentials or $T$ are increased. 

\section{Fermion condensate}\label{sec:FC}

In this section, the t.e.v. of the fermion condensate (FC), 
$\frac{1}{2}[\widehat{\overline{\Psi}}, \widehat{\Psi}]$, 
is considered. Subsection~\ref{sec:FC:gen} presents the 
general expression for the t.e.v. of the FC, while 
Subsections~\ref{sec:FC:M0} and \ref{sec:FC:num} are 
dedicated to its analytical and numerical analyses.

\subsection{General analysis}\label{sec:FC:gen}

The bilinear form introduced in Eq.~\eqref{eq:F_bilinear_def} which 
corresponds to the FC is:
\begin{equation}
 {\rm FC}(\psi, \chi) = \overline{\psi} \chi.
\end{equation}
After noting that $\overline{V}_j V_j = -(\overline{U}_j U_j)^*$, 
it is easy to see that the term on the second line of 
Eq.~\eqref{eq:F_R} makes a vanishing contribution to the t.e.v. 
of the FC. Furthermore, the expressions in Eq.~\eqref{eq:U} for 
the particle modes allow the following result to be obtained:
\begin{equation}
 {\rm FC}(U_j, U_j) = \frac{M}{8\pi^2 E_j} \left[ 
 J_{m_j}^+(q_j\rho) + \frac{2\lambda_j k_j}{p_j} 
 J_{m_j}^-(q_j\rho)\right],
\end{equation}
allowing the t.e.v. of the FC to be put in the following form:
\begin{multline}
 \braket{:\frac{1}{2}[\widehat{\overline{\Psi}}, \widehat{\Psi}]:_\Omega} 
 = \frac{M}{4\pi^2} \sum_{\lambda = \pm\frac{1}{2}} 
 \sum_{m = -\infty}^\infty \int_M^\infty dE \, 
 {\rm sgn}(\widetilde{E})\\
 \times\left[
 \frac{1}{e^{\beta_0(|\widetilde{E}| - \mu_{\lambda;0})} + 1} +
 \frac{1}{e^{\beta_0(|\widetilde{E}| + \mu_{\lambda;0})} + 1}
 \right] \int_{0}^p dk \, J_m^+(q\rho).\label{eq:FC}
\end{multline}
Taken with respect to the Minkowski vacuum, the t.e.v. 
of the FC becomes:
\begin{multline}
 \braket{:\frac{1}{2}[\widehat{\overline{\Psi}}, \widehat{\Psi}]:_M}
 = \frac{M}{4\pi^2} \sum_{\lambda = \pm\frac{1}{2}} 
 \sum_{m = -\infty}^\infty \int_M^\infty dE \, 
 \left[
 \frac{1}{e^{\beta_0(\widetilde{E} - \mu_{\lambda;0})} + 1} +
 \frac{1}{e^{\beta_0(\widetilde{E} + \mu_{\lambda;0})} + 1}
 \right] \\\times \int_{0}^p dk \, J_m^+(q\rho),\label{eq:FC_simp}
\end{multline}
which differs from the t.e.v. in Eq.~\eqref{eq:FC} by a vacuum term.

\subsection{Small mass limit}\label{sec:FC:M0}

Employing the same method as in Subsec.~\ref{sec:CC:M0}, the 
t.e.v. of the FC taken with respect to the Minkowski vacuum 
can be put in the form:
\begin{multline}
 \braket{:\frac{1}{2}[\widehat{\overline{\Psi}}, \widehat{\Psi}]:_M}
 = \frac{M}{2\pi^2} \sum_{\lambda = \pm \frac{1}{2}}
 \sum_{j = 0}^\infty \frac{(\rho \Omega)^{2j}}{2j+1}
 \sum_{n = 0}^\infty \frac{\Omega^{2n} s_{n+j,j}^+}{(2n+2j)!}
 \int_0^\infty dp\, \frac{p^{2j+2}}{E}\\\times
 \frac{d^{2n+2j}}{dE^{2n+2j}} 
 \left[\frac{1}{e^{\beta_0(E - \mu_{\lambda,0})} + 1} + 
 \frac{1}{e^{\beta_0(E + \mu_{\lambda,0})} + 1}\right].
 \label{eq:FC_M0_aux}
\end{multline}
The correction due to the mass cannot be obtained using the 
methodology from Eq.~\eqref{eq:M0_exp2}, due to an infrared 
divergence of the $j = 0$ term. 
Dividing Eq.~\eqref{eq:FC_M0_aux} by $M$ and taking the 
massless limit leads to:
\begin{equation}
 \braket{:\frac{1}{2}[\widehat{\overline{\Psi}}, \widehat{\Psi}]:_M}
 = M \left[\frac{T^2}{6} + \frac{\mu_V^2 + \mu_H^2}{2\pi^2} 
 + \frac{3\bm{\omega}^2 - \bm{a}^2}{24\pi^2}\right] + O(M^2),
\end{equation}
where $T = T_0 \Gamma$, $\mu_{V/H} = \mu_{V/h,0} \Gamma$,
$\omega = \Omega \Gamma^2 e_\hatz$ and $a = \rho \Omega^2 \Gamma^2 e_\hrho$, 
while $\bm{\omega}^2 = -\omega^2$ and $\bm{a}^2 = -a^2$.
The last term is the vacuum contribution corresponding to the 
difference between the t.e.v.s taken with respect to the rotation 
and Minkowski vacua, respectively. This contribution must be subtracted 
in order to obtain the t.e.v. of the FC with respect to the rotating 
vacuum:
\begin{equation}
 \braket{:\frac{1}{2}[\widehat{\overline{\Psi}}, \widehat{\Psi}]:_\Omega}
 = M\left(\frac{T^2}{6} + \frac{\mu_V^2 + \mu_H^2}{2\pi^2}\right) + O(M^2). 
 \label{eq:FC_M0}
\end{equation}
It is surprising that Eq.~\eqref{eq:FC_M0} coincides with the 
RKT prediction $(E_{\rm RKT} - 3 P_{\rm RKT})/M$
given in Eq.~\eqref{eq:RKT_M0}.

\subsection{Numerical analysis}\label{sec:FC:num}

\begin{figure}
    \centering
\begin{tabular}{cc}
    \includegraphics[width=0.48\linewidth]{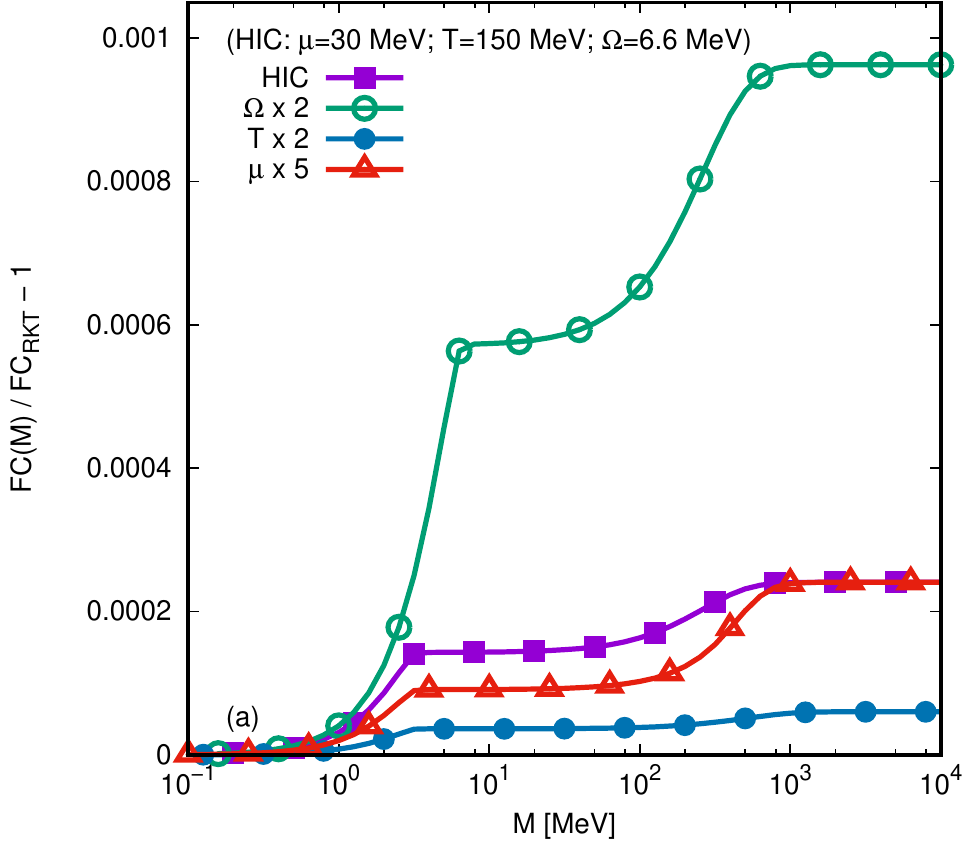} & 
    \includegraphics[width=0.45\linewidth]{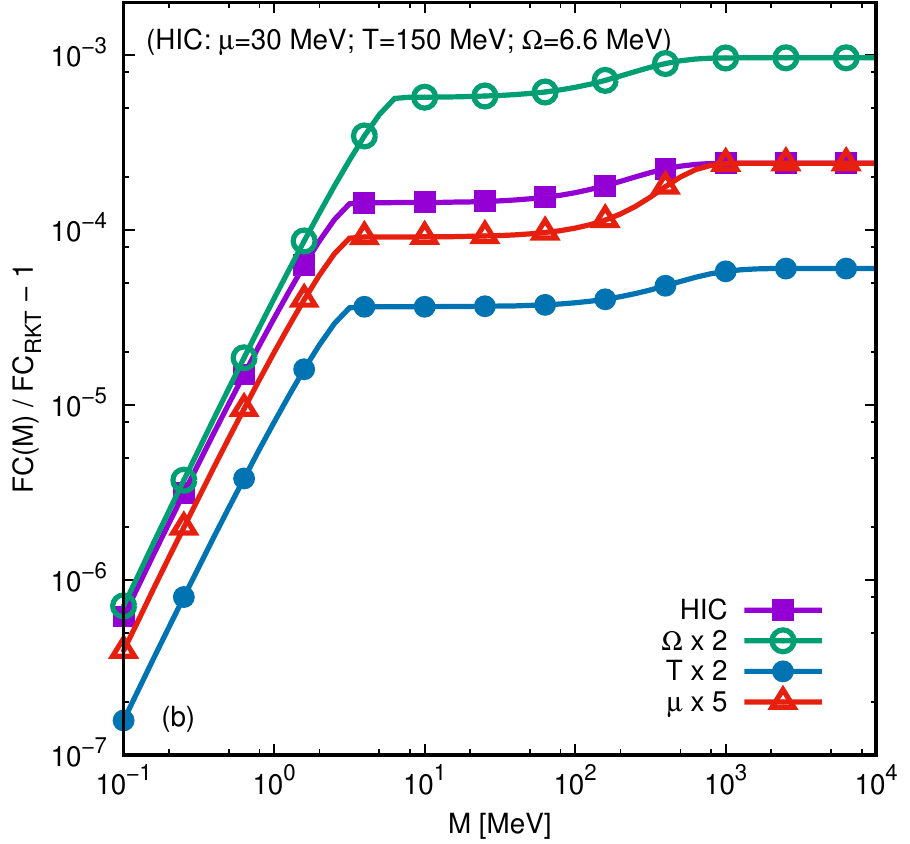}
\end{tabular}
\caption{Relative difference of the quantum and kinetic theory 
fermion condensates, on the rotation axis, computed by numerically 
integrating Eqs.~\eqref{eq:FC} and \eqref{eq:RKT_CC_SET}, respectively. 
The HIC parameters are indicated at the top of the images and the 
corresponding curve is represented using purple and filled squares. 
For each subsequent curve, only the parameter indicated in the 
legend is changed, compared to the original set. The results are 
represented in log-linear (a) and log-log (b) scale.}
\label{fig:FC_axis}
\end{figure}

At vanishing mass, a comparison between Eqs.~\eqref{eq:FC_M0} 
and \eqref{eq:RKT_M0} indicates that the t.e.v. of the FC 
(divided by $M$) is identical with the RKT prediction for the 
trace of the SET (divided by $M^2$). At finite mass, it can be 
expected that this equality holds only approximately. 
Figure~\ref{fig:FC_axis} presents the relative difference between 
the quantum and RKT predictions for the FC, evaluated on the rotation axis. 
It can be seen that there is a rapid increase from $0$ in the massless 
case to some plateau values. The first plateau occurs quite early, 
for $M \gtrsim 5\ {\rm MeV}$. The second plateau appears when 
$M \simeq T$. The values on these plateaus depend on the parameters 
of the system. Doubling $\Omega$ increases the plateaus by a factor 
of $4$, while doubling the temperature decreases the value, again by 
a factor of $4$, as can be seen from Fig.~\ref{fig:FC_axis}(a). Increasing 
the chemical potentials ($\mu_V = \mu_H$ is considered here) seems to 
lower the plateau in the central region, however, for 
$M \gtrsim 1\ {\rm GeV}$, it gives the same result as in the case of the HIC 
parameters. The increase from the massless limit to the first plateau 
seems to follow a power law of the form $\sim M^{3/2}$, as can be seen 
from Fig.~\ref{fig:FC_axis}(b). 

\section{Stress-energy tensor}\label{sec:SET}

The stress-energy tensor (SET) operator is defined as:
\begin{equation}
 \widehat{T}_{\halpha\hsigma} = \frac{i}{4} 
 e_{\halpha}^\mu e_{\hsigma}^\nu
 \left\{[\widehat{\overline{\Psi}}, \gamma_{(\mu} \partial_{\nu)} \widehat{\Psi}] - 
 [\partial_{(\mu} \widehat{\overline{\Psi}} \gamma_{\nu)}, \widehat{\Psi}]\right\}.
 \label{eq:SET_def}
\end{equation}
The general expressions for the computation of the thermal expectation values (t.e.v.s) of the SET operator are presented in Subsec.~\ref{sec:SET:gen}. The analytical and numerical analyses of these expressions are presented in Subsections~\ref{sec:SET:M0} and \ref{sec:SET:num}.

\subsection{General analysis}\label{sec:SET:gen}

The bilinear form defined in Eq.~\eqref{eq:F_bilinear_def} which corresponds 
to the SET operator is
\begin{equation}
 \mathcal{T}_{\halpha\hsigma}(\psi,\chi) = 
 \frac{i}{2} e_\halpha^\mu e_\hsigma^\nu
 \left[\,\overline{\psi} \gamma_{(\mu} \partial_{\nu)} \chi - 
 \partial_{(\mu} \overline{\psi} \gamma_{\nu)} \chi\right].
\end{equation}
Substituting $\psi = \chi = V_j$ in the above 
equation and using the relations in Eq.~\eqref{eq:Vj},
it can be shown that
\begin{equation}
 \mathcal{T}_{\halpha\hsigma}(V_j,V_j) = 
 \mathcal{T}_{\halpha\hsigma}
 (U_{\overline{\jmath}},U_{\overline{\jmath}}) = 
 -[\mathcal{T}_{\halpha\hsigma}(U_j,U_j)]^*.
\end{equation}
As in Sections~\ref{sec:AC} and \ref{sec:FC}, 
the second line of Eq.~\eqref{eq:F_R} does not
contribute to the t.e.v. of the SET, which can be 
summarised as follows:
\begin{multline}
 \braket{:\widehat{T}_{\halpha\hsigma}:_\Omega} = \sum_{\lambda = \pm \frac{1}{2}}  \sum_{m = -\infty}^\infty \int_M^\infty dE\, E\,
 {\rm sgn}(\widetilde{E})
 \left[\frac{1}
 {e^{\beta_0(|\widetilde{E}| - \mu_{\lambda;0})} + 1} + 
 \frac{1}{e^{\beta_0(|\widetilde{E}| + \mu_{\lambda;0})} + 1}\right]\\\times
 \int_{-p}^p dk 
 \,\mathcal{T}_{\halpha\hsigma}(U_j, U_j).
 \label{eq:SET_aux}
\end{multline}
Using the explicit expression for the modes 
$U_j$, given in Eq.~\eqref{eq:U}, the following 
auxiliary expressions can be computed:
\begin{align}
 i \overline{U}_j \gamma_{\halpha} e_{\hatt}^\nu \partial_\nu U_j 
 =& E_j \overline{U}_j \gamma_{\halpha} U_j, \qquad 
 i \overline{U}_j \gamma_{\halpha} e_{\hatz}^\nu \partial_\nu U_j 
 = -k_j \overline{U}_j \gamma_{\halpha} U_j, \nonumber\\
 i \overline{U}_j \gamma_{\hatt} e_{\hrho}^\nu \partial_\nu U_j 
 =& -\frac{i \lambda_j k_j q_j}{4\pi^2 p_j} J_{m_j}^\times(q_j \rho)\nonumber\\
 &+ \frac{i}{8\pi^2\rho} \left[ 
 \left(\frac{2\lambda_j k_j}{p_j} m_j - \frac{1}{2}\right) 
 J_{m_j}^+(q_j \rho) + 
 \left(m_j - \frac{\lambda_j k_j}{p_j}\right)
 J_{m_j}^-(q_j \rho)\right],\nonumber\\
 i \overline{U}_j \gamma_{\hrho} e_{\hrho}^\nu \partial_\nu U_j 
 =& \frac{q_j^2}{8\pi^2 E_j} \left[J_{m_j}^+(q_j \rho) - \frac{m_j}{q_j \rho} J_{m_j}^\times(q_j \rho)\right],\nonumber\\
 i \overline{U}_j \gamma_{\hvarphi} e_{\hrho}^\nu \partial_\nu U_j 
 =& -\frac{i q_j}{8\pi^2 E_j \rho} \left[q_j\rho\, J_{m_j}^-(q_j \rho) - 
 \frac{1}{2} J_{m_j}^\times(q_j \rho)\right], \nonumber\\
 i \overline{U}_j \gamma_{\hatz} e_{\hrho}^\nu \partial_\nu U_j 
 =& \frac{i \lambda_j q_j p_j}{4\pi^2 E_j} J_{m_j}^\times(q_j \rho) \nonumber\\
 &- \frac{i \lambda_j p_j}{4\pi^2 E_j \rho} \left[
 \left(m_j - \frac{\lambda_j k_j}{p_j}\right) J_{m_j}^+(q_j \rho) + 
 \left(\frac{2\lambda_j k_j}{p_j} m_j - \frac{1}{2}\right) J_{m_j}^-(q_j\rho)\right],\nonumber\\
 i \overline{U}_j \gamma_{\hatt} e_{\hvarphi}^\nu \partial_\nu U_j 
 =& -\frac{1}{8\pi^2 \rho} \left[\left(m_j - \frac{\lambda_j k_j}{p_j}\right) J_{m_j}^+(q_j\rho) + 
 \left(\frac{2\lambda_j k_j}{p_j} m_j - \frac{1}{2}\right) 
 J_{m_j}^-(q_j \rho)\right],\nonumber\\
 i \overline{U}_j \gamma_{\hrho} e_{\hvarphi}^\nu \partial_\nu U_j 
 =& \frac{i q_j}{16\pi^2 E_j \rho} J_{m_j}^\times(q_j\rho),\nonumber\\
 i \overline{U}_j \gamma_{\hvarphi} e_{\hvarphi}^\nu \partial_\nu U_j 
 =& \frac{m_j q_j}{8\pi^2 E_j \rho} J_{m_j}^\times(q_j\rho),\nonumber\\
 i \overline{U}_j \gamma_{\hatz} e_{\hvarphi}^\nu \partial_\nu U_j 
 =& \frac{\lambda_j p_j}{4\pi^2 E_j \rho} \left[
 \left(\frac{2\lambda_j k_j}{p_j} m_j - \frac{1}{2}\right) 
 J_{m_j}^+(q_j\rho) + 
 \left(m_j - \frac{\lambda_j k_j}{p_j}\right) J_{m_j}^-(q_j\rho)\right].
 \label{eq:SET_aux2}
\end{align}
The derivatives of the Bessel functions were replaced using 
the following identities:
\begin{align}
 J_{m - \frac{1}{2}}'(q\rho) =& -J_{m+\frac{1}{2}}(q\rho) + 
 \frac{m - \frac{1}{2}}{q\rho} J_{m-\frac{1}{2}}(q\rho), \nonumber\\
 J_{m + \frac{1}{2}}'(q\rho) =& J_{m-\frac{1}{2}}(q\rho) - 
 \frac{m + \frac{1}{2}}{q\rho} J_{m+\frac{1}{2}}(q\rho).
 \label{eq:Jderiv}
\end{align}

Since $\mathcal{T}_{\halpha\hsigma}(U_j, U_j) = 
{\rm Re}[i\overline{U}_j 
\gamma_{(\halpha} e_{\hsigma)}^\mu \partial_\mu U_j]$,
the terms proportional to the imaginary unit $i$ appearing in 
Eq.~\eqref{eq:SET_aux2} can be ignored.
The non-vanishing components of the bilinear form 
$\mathcal{T}_{\halpha\hsigma}(U_j, U_j)$ are:
\begin{align}
 \mathcal{T}_{\hatt\hatt}(U_j, U_j) =& \frac{E_j}{8\pi^2} 
 \left[J_{m_j}^+(q_j\rho) + \frac{2\lambda_j k_j}{p_j} 
 J_{m_j}^-(q_j\rho)\right],\nonumber\\
 \mathcal{T}_{\hatt\hvarphi}(U_j, U_j) =& 
 -\frac{1}{16\pi^2 \rho} 
 \left[\left(m_j - \frac{\lambda_j k_j}{p_j}\right) J_{m_j}^+(q_j \rho) +
 \left(\frac{2 \lambda_j k_j m_j}{p_j} - \frac{1}{2}\right) J_{m_j}^-(q_j \rho)\right] \nonumber\\
 &-\frac{q_j}{16\pi^2} J_{m_j}^\times(q_j\rho),\nonumber\\
 \mathcal{T}_{\hatt\hatz}(U_j, U_j) =& 
 -\frac{\lambda_j (p^2_j + k_j^2)}{8\pi^2 p_j} J_{m_j}^-(q_j \rho)
 - \frac{k_j}{8\pi^2} J_{m_j}^+(q_j \rho),\nonumber\\
 \mathcal{T}_{\hrho\hrho}(U_j, U_j) =& \frac{q_j^2}{8\pi^2 E_j} 
 \left[J_{m_j}^+(q_j\rho) - \frac{m_j}{q_j \rho} 
 J_{m_j}^\times(q_j\rho)\right],\nonumber\\
 \mathcal{T}_{\hvarphi\hvarphi}(U_j, U_j) =& 
 \frac{q_j m_j}{8\pi^2 \rho E_j} 
 J_{m_j}^\times(q_j\rho),\nonumber\\
 \mathcal{T}_{\hvarphi\hatz}(U_j, U_j) =& 
 \frac{\lambda_j p_j}{8\pi^2 \rho E_j} 
 \left[m_j J_{m_j}^-(q_j\rho) - \frac{1}{2} J_{m_j}^+(q_j\rho)\right]
 \nonumber\\
 & + \frac{k_j}{16\pi^2 \rho E_j} 
 \left[q_j \rho J_{m_j}^\times(q_j\rho) - 
 \frac{1}{2} J_{m_j}^-(q_j \rho) +
 m_j J_{m_j}^+(q_j \rho)\right],\nonumber\\
 \mathcal{T}_{\hatz\hatz}(U_j, U_j) =& 
 \frac{k_j^2}{8\pi^2 E_j} J_{m_j}^+(q_j\rho) + 
 \frac{\lambda_j k_j p_j}{4\pi^2 E_j}  
 J_{m_j}^-(q_j\rho).
 \label{eq:SET_aux3}
\end{align}
The terms appearing in Eq.~\eqref{eq:SET_aux3} which are odd 
with respect to $k_j \rightarrow -k_j$ 
do not contribute to the t.e.v. of the SET. Starting from the 
above expressions, it is possible to show that
the t.e.v.s $\braket{:\widehat{T}_{\hrho\hrho}:}$ and 
$\braket{:\widehat{T}_{\hatz\hatz}:}$ are exactly equal. 
This can be seen by first noting that
\begin{equation}
 J_m^-(z) = \frac{1}{2m} \frac{d}{dz}[z J_m^+(z)] 
 = \frac{1}{2z} \frac{d}{dz} [z J_m^\times(z)].
 \label{eq:Jstarderiv}
\end{equation}
The above relations can be derived using the properties of 
the derivatives of the Bessel functions $J_{m-1/2}(z)$ and 
$J_{m+1/2}(z)$ given in Eq.~\eqref{eq:Jderiv}. Next, $I_m(p)$ 
is introduced as follows:
\begin{equation}
 I_m(p) = \frac{1}{2} \int_{-p}^p  dk \, k^2 J_m^-(q\rho) 
 = \int_{0}^p  dk \, k^2 J_m^-(q\rho),
\end{equation}
where the last equality follows after noting that the 
integrand is even with respect to $k$.
Replacing now $J_m^-(q\rho)$ using the two expressions
in Eq.~\eqref{eq:Jstarderiv} and using integration by 
parts to remove the derivative term, it can be shown that
\begin{equation}
 I_m(p) = \frac{1}{2m} \int_0^p dk(q^2-k^2) J_m^+(q\rho) 
 = \frac{1}{2\rho} \int_0^p dk\, q\, J_m^\times(q\rho).
\end{equation}
Rearranging the above expressions, it can be seen that
\begin{equation}
 \int_{-p}^p dk\, \mathcal{T}_{\hrho\hrho}(U_j,U_j) = 
 \int_{-p}^p dk\, \mathcal{T}_{\hatz\hatz}(U_j,U_j).
\end{equation}
Thus, it is easy to conclude that 
\begin{equation}
 \braket{:\widehat{T}_{\hrho\hrho}:_\Omega} = \braket{:\widehat{T}_{\hatz\hatz}:_\Omega},
 \label{eq:SET_equal}
\end{equation}
the above relation being valid for all values of 
$\rho$, $\Omega$, $M$, $\beta_0$, $\mu_{V,0}$ and $\mu_{H,0}$.

For a given velocity field $u^\halpha$, the SET can be decomposed as follows:
\begin{equation}
 \braket{:\widehat{T}^{\halpha\hsigma}:_\Omega} = 
 E\, u^\halpha u^\hsigma - (P + \overline{\omega})\Delta^{\halpha\hsigma} + 
 \Pi^{\halpha\hsigma} + u^\halpha W^\hsigma + u^\hsigma W^\halpha,
 \label{eq:SET_dec}
\end{equation}
where $E$ and $P$ are the usual energy density and pressure,
$W^\halpha$ represents the heat flux in the local rest frame,
$\overline{\omega}$ is the dynamic pressure
and $\Pi^{\halpha\hsigma}$ is the pressure deviator, which is traceless
by construction. The tensor 
$\Delta^{\halpha\hsigma} = \eta^{\halpha\hsigma} - u^\halpha u^\hsigma$ 
is a projector on the hypersurface orthogonal to $u^\halpha$. 
The anomalous contributions $\Pi^{\halpha\hsigma}$ and $W^\halpha$ 
are also orthogonal to $u^\halpha$, by construction. The isotropic 
pressure $P + \overline{\omega}$ is given as the sum of the hydrostatic
pressure $P$, which is computed using the equation of state of the fluid,
and of the dynamic pressure $\overline{\omega}$, which in general depenends
on the divergence of the velocity. With this convention, the pressure 
deviator is considered to be traceless. It should be noted that for
ultrarelativistic fluids, the SET is traceless (this is true in 
the quantum case as well since the Dirac field is conformally coupled)
and $E = 3P$. Since $T^{\halpha}{}_\halpha = E - 3(P + \overline{\omega})$,
it can be seen that $\overline{\omega} = 0$ for ultrarelativistic
(massless) fluid constituents. Furthermore, for massive fluid constituents,
$\overline{\omega}$ is usually related to the expansion of the fluid \cite{rezzolla13},
which vanishes in the case of rigid rotation. Thus, for the rest of this section,
the relation $\overline{\omega} = 0$ is assumed to hold true.

The macroscopic quantities can be extracted from $T^{\halpha\hsigma}$ 
as follows:
\begin{align}
 E =& \braket{:\widehat{T}_{\halpha\hsigma}:_\Omega} u^\halpha u^\hsigma, &
 P + \overline{\omega} =& -\frac{1}{3} \Delta^{\halpha\hsigma} 
 \braket{:\widehat{T}_{\halpha\hsigma}:_\Omega}, \nonumber\\
 W^\halpha =& \Delta^{\halpha\hsigma} u^\hlambda 
 \braket{:\widehat{T}_{\hsigma\hlambda}:_\Omega}, &
 \Pi_{\halpha\hsigma} =& \braket{:\widehat{T}_{\braket{\halpha\hsigma}}:_\Omega},
 \label{eq:SET_dec_inv}
\end{align}
where the notation $A_{\braket{\halpha\hsigma}}$ for a general two-tensor refers to:
\begin{equation}
 A_{\braket{\halpha\hsigma}} = \frac{1}{2} \left[\left(
 \Delta_{\halpha\hbeta} \Delta_{\hsigma\hgamma} + 
 \Delta_{\hsigma\hbeta} \Delta_{\halpha\hgamma}\right)-
 \frac{2}{3} \Delta_{\halpha\hsigma} \Delta_{\hbeta\hgamma}\right] 
 A^{\hbeta\hgamma}.
 \label{eq:SET_shear}
\end{equation}

The anomalous terms $W^\halpha$ and $\Pi^{\halpha\hsigma}$ can be 
further decomposed with respect to the tetrad formed by $u$, $a$, 
$\omega$ and $\tau$, introduced in Sec.~\ref{sec:RR}. Noting that 
$W\cdot u = 0$ and that $\braket{:\widehat{T}_{\hrho\halpha}:_\Omega} = 0$ when 
$\alpha \neq \rho$, the most general decomposition for $W$ is
\begin{align}
 W^\halpha =& \sigma_\varepsilon^\tau \tau^\halpha + 
 \sigma_\varepsilon^\omega \omega^\halpha,\nonumber\\
 \sigma_\varepsilon^\tau =& \frac{1}{\Omega^2 \Gamma^2} 
 \left[\braket{:\widehat{T}_{\hatt\hatt}:_\Omega} + 
 \braket{:\widehat{T}_{\hvarphi\hvarphi}:_\Omega} + 
 \frac{1 + \rho^2 \Omega^2}{\rho\Omega} 
 \braket{:\widehat{T}_{\hatt\hvarphi}:_\Omega} \right], \nonumber\\
 \sigma_\varepsilon^\omega =& -\frac{1}{\Omega\Gamma}\left[
 \braket{:\widehat{T}_{\hatt\hatz}:_\Omega} + 
 \rho\Omega \braket{:\widehat{T}_{\hvarphi\hatz}:_\Omega}\right],
 \label{eq:SET_w_def}
\end{align}
where $\sigma_\varepsilon^\tau$ and $\sigma_\varepsilon^\omega$ are the circular and vortical heat conductivities.
In the case of $\Pi^{\halpha\hsigma}$, the orthogonality to $u^\halpha$ and
the tracelessness condition, together with the consideration that 
$\braket{:\widehat{T}_{\halpha\hrho}:_\Omega} = 0$ for all $\alpha \neq \rho$, 
allow $\Pi^{\halpha\hsigma}$ to be written as:
\begin{equation}
 \Pi^{\halpha\hsigma} = A a^\halpha a^\hsigma + 
 B \tau^\halpha \tau^\hsigma + C \omega^\halpha \omega^\hsigma
 + D(\tau^\halpha \omega^\hsigma + \tau^\hsigma \omega^\halpha).
\end{equation}
Noting that $\braket{:\widehat{T}_{\hrho\hrho}:_\Omega} = 
P + \Pi_{\hrho\hrho}$ and $\braket{:\widehat{T}_{\hatz\hatz}:_\Omega} = 
P + \Pi_{\hatz\hatz}$ are equal by virtue of Eq.~\eqref{eq:SET_equal}, 
it can be seen that $\Pi_{\hrho\hrho} = \Pi_{\hatz\hatz}$, such that
\begin{equation}
 C = \rho^2 \Omega^2 A.
\end{equation}
The tracelessness condition gives 
\begin{equation}
 A = -\frac{1}{2} B \Omega^2 \Gamma^4 = \frac{\omega^2}{2} B \Rightarrow 
 C =\frac{a^2}{2} B.
\end{equation}
Thus, only two degrees of freedom are required to 
describe $\Pi^{\halpha\hsigma}$, which are introduced 
below:
\begin{equation}
 \Pi^{\halpha\hsigma} = \Pi_1\left(\tau^\halpha \tau^\hsigma + \frac{\omega^2}{2} a^\halpha a^\hsigma +\frac{a^2}{2} \omega^\halpha \omega^\hsigma\right) +  \Pi_2(\tau^\halpha \omega^\hsigma + 
 \tau^\hsigma \omega^\halpha),
\end{equation}
where $\Pi_1$ and $\Pi_2$ can be obtained from 
the components of the SET, e.g., through:
\begin{equation}
 \Pi_1 = \frac{2(P - \braket{:\widehat{T}_{\hatz\hatz}:_\Omega})}
 {\Omega^4 \Gamma^6(\Gamma^2 - 1)}, \qquad 
 \Pi_2 = -\frac{\braket{:\widehat{T}_{\hvarphi\hatz}:_\Omega} + 
 \rho\Omega \braket{:\widehat{T}_{\hatt\hatz}:_\Omega}}{\rho \Omega^4 \Gamma^5}.
\end{equation}

\subsection{Small mass limit}\label{sec:SET:M0}

Following the procedure which is by now familiar, 
the t.e.v.s of the components of the SET are computed 
in the small mass regime. Starting from their expressions 
with respect to the rotating vacuum, given in Eq.~\eqref{eq:SET_aux},
the transition to the Minkowski vacuum can be performed.
After expanding the Fermi-Dirac factors via Eq.~\eqref{eq:M0_exp},
the summation over $m$ and integration over $k$ can be 
performed using Eqs.~\eqref{eq:M0_summ} and \eqref{eq:M0_intk}:
\begin{multline}
 \begin{pmatrix}
  \braket{:\widehat{T}_{\hatt\hatt}:_M} & 
  \braket{:\widehat{T}_{\hrho\hrho}:_M} \\
  \braket{:\widehat{T}_{\hvarphi\hvarphi}:_M}  &
  -\braket{:\widehat{T}_{\hatt\hvarphi}:_M}/\rho \Omega
 \end{pmatrix} = \frac{1}{2\pi^2} \sum_{\lambda = \pm \frac{1}{2}}
 \sum_{j = 0}^\infty (\rho \Omega)^{2j}
 \sum_{n = 0}^\infty \frac{\Omega^{2n} s_{n+j,j}^+}{(2n+2j+1)!} \\
 \times \int_M^\infty dE\, p^{2j+1} \frac{d^{2n+2j}}{dE^{2n+2j}} 
 \left[\frac{1}{e^{\beta_0(E - \mu_{\lambda,0})} + 1} + 
 \frac{1}{e^{\beta_0(E + \mu_{\lambda,0})} + 1}\right] \\\times 
 \begin{pmatrix}
  \displaystyle \frac{2n+2j+1}{2j+1} E^2 & 
  \displaystyle \frac{2n+2j+1}{(2j+1)(2j+3)} p^2 \smallskip\\
  \displaystyle \frac{2n+2j+1}{2j+3} p^2 &
  \displaystyle \frac{p^2}{2j+3} + E^2 + \frac{j}{4(\rho p)^2} \left[p^2 + E^2(2j+1)\right]
 \end{pmatrix},\label{eq:SET_M0_aux}
\end{multline}
while $\braket{:\widehat{T}_{\hatz\hatz}:_M} = \braket{:\widehat{T}_{\hrho\hrho}:_M}$, 
as shown in Eq.~\eqref{eq:SET_equal}. The $M$ subscript indicates that the above 
expressions are computed with respect to the Minkowski vacuum, as discussed in 
Sec.~\ref{sec:therm:tevs}. 
In obtaining the expression for $\braket{:\widehat{T}_{\hatt\hvarphi}:_M}$,
integration by parts, Eq.~\eqref{eq:M0_summ_rec} and the relation $s_{n,0}^+ = 2^{-2n}$ were used. After changing the integration variable to $p$, Eq.~\eqref{eq:M0_exp2} can be used to analyse the small mass regime. 
After employing integration by parts $2j$ times in the integral with 
respect to $p$, it is not difficult to see that the summation over $n$
terminates at a finite value of $n$. This is because
\begin{equation}
 \int_0^\infty dp\, p^3 \frac{d^{2n}}{dp^{2n}} 
 \left[\frac{1}{e^{\beta_0(p - \mu_{\lambda,0})} + 1} + 
 \frac{1}{e^{\beta_0(p + \mu_{\lambda,0})} + 1}\right] = 
 \begin{cases}
  \displaystyle \frac{2}{\beta_0^4} I_1^+(\beta_0 \mu_{\lambda,0}), & n = 0, \smallskip\\
  \displaystyle \frac{12}{\beta_0^2} I_0^+(\beta_0 \mu_{\lambda,0}), & n = 1, \smallskip\\
  6, & n = 2, \\
  0, & n > 2.
 \end{cases}
\end{equation}
The last equality follows from noting that 
$(1 + e^{-a})^{-1} + (1 + e^{a})^{-1} = 1$.
Taking into account that the $n = 2$ term makes a purely vacuum 
contribution, the following exact results are obtained for the 
t.e.v.s with respect to the rotating vacuum:
\begin{align}
 \braket{:\widehat{T}_{\hatt\hatt}:_\Omega} =& 
 \Gamma^2 E_{\rm RKT} + P_{\rm RKT} (\Gamma^2 -1)
 + \frac{\Omega^2 \Gamma^2}{8} \left[
 T^2 + \frac{3(\mu_V^2 + \mu_H^2)}{\pi^2}\right]
 \left(\frac{8}{3} \Gamma^4 - \frac{16}{9} \Gamma^2 + \frac{1}{9}\right)\nonumber\\
 \braket{:\widehat{T}_{\hatt\hvarphi}:_\Omega} =& -\rho \Omega \Gamma^2 \Bigg\{
 E_{\rm RKT} + P_{\rm RKT} + \frac{2 \Omega^2 \Gamma^2}{9} \left[
 T^2 + \frac{3(\mu_{V}^2 + \mu_H^2)}{\pi^2}\right]
 \left(\frac{3}{2} \Gamma^2 - \frac{1}{2}\right)\Bigg\},\nonumber\\
 \braket{:\widehat{T}_{\hrho\hrho}:_\Omega} =& 
 P_{\rm RKT} + \frac{\Omega^2 \Gamma^2}{24} \left[
 T^2 + \frac{3(\mu_V^2 + \mu_H^2)}{\pi^2}\right]
 \left(\frac{4}{3} \Gamma^2 - \frac{1}{3}\right),\nonumber\\
 \braket{:\widehat{T}_{\hvarphi\hvarphi}:_\Omega} =& 
 \Gamma^2 P_{\rm RKT} + E_{\rm RKT} (\Gamma^2 - 1) 
 + \frac{\Omega^2 \Gamma^2}{24} \left[T^2 + 
 \frac{3(\mu_V^2 + \mu_H^2)}{\pi^2}\right]
 \left(8\Gamma^4 - 8 \Gamma^2 + 1\right),\label{eq:SETr}
\end{align}
while $\braket{:\widehat{T}_{\hatz\hatz}:_\Omega} = 
\braket{:\widehat{T}_{\hrho\hrho}:_\Omega}$. The $O(M^2)$ 
corrections computed from Eq.~\eqref{eq:SET_M0_aux} using 
the technique described in Eq.~\eqref{eq:M0_exp2} are absorbed 
in the RKT predictions $E_{\rm RKT}$ and $P_{\rm RKT}$ for the 
energy density and pressure, given in Eq.~\eqref{eq:RKT_M_corr}.
The following expressions can be obtained:
\begin{align}
 E =& E_{\rm RKT} + \Delta E, &
 P =& P_{\rm RKT} + \Delta P, \nonumber\\
 \Delta E =& \frac{\bm{a}^2+3\bm{\omega}^2}{24}\left[T^2 + \frac{3(\mu_V^2 + \mu_H^2)}{\pi^2}\right] + 
 O(M^4), & 
 \Delta P =& \frac{1}{3} \Delta E + O(M^4),\nonumber\\
 \sigma_\varepsilon^\tau =& -\frac{1}{18}\left[T^2 + \frac{3(\mu_V^2 + \mu_H^2)}{\pi^2}\right] + O(M^4), &
 \Pi_1 =& O(M^4).
 \label{eq:SET_EPkt_M}
\end{align}
Comparing Eq.~\eqref{eq:SET_EPkt_M} and Eq.~\eqref{eq:AC_so_M}, it is interesting to 
note that $\sigma_\varepsilon^\tau = -\frac{1}{3} \sigma^\omega_A$.

There is a set of components which is non-vanishing only when 
$\mu_{H,0} \neq 0$. These can be computed using:
\begin{multline}
 \begin{pmatrix}
  \braket{:\widehat{T}_{\hatt\hatz}:_\Omega} \\
  \braket{:\widehat{T}_{\hvarphi\hatz}:_\Omega}
 \end{pmatrix} = \frac{1}{8\pi^2 \rho} 
 \sum_{\lambda = \pm \frac{1}{2}} 2\lambda
 \sum_{j = 0}^\infty (\rho \Omega)^{2j}
 \sum_{n = 0}^\infty \frac{\Omega^{2n} s_{n+j,j}^+}{(2n+2j+1)!}
 \int_M^\infty dE\, p^{2j} \\
 \times \frac{d^{2n+2j}}{dE^{2n+2j}} 
 \left[\frac{1}{e^{\beta_0(E - \mu_{\lambda,0})} + 1} + 
 \frac{1}{e^{\beta_0(E + \mu_{\lambda,0})} + 1}\right] \\\times 
 \begin{pmatrix}
  \displaystyle -\rho \Omega\frac{2j+4}{2j+3} [2(j + 1)E^2 + p^2] \smallskip\\
  \displaystyle 2j \frac{2n+2j+1}{2j+1} p^2
 \end{pmatrix}.\label{eq:SET_M0_aux1}
\end{multline}
As in the previous subsections, the sum over $n$ does not terminate 
at finite $n$. Instead, the first two terms in this sum give:
\begin{align}
 \braket{:\widehat{T}_{\hatt\hatz}:_\Omega} =& 
 -\frac{\Omega \Gamma^3}{\pi^2} \sum_{\lambda} 2\lambda 
 \left\{T^3
 I_{1/2}^+(\beta \mu_\lambda) + \frac{\Omega^2 T}{12} \Gamma^2
 (6\Gamma^2 - 5) I_{-1/2}^+(\beta \mu_\lambda) \right.\nonumber\\
 & \left. - \frac{M^2 T}{8(\Gamma^2 - 1)}\left[2\Gamma^2 - 1 - 
 \frac{{\rm arcsinh}(\rho \Omega \Gamma)}{\rho \Omega \Gamma^2}\right] 
 I_{-1/2}^+(\beta \mu_\lambda) + O(\Omega^4,M^4, \Omega^2 M^2)\right\},\nonumber\\
 \braket{:\widehat{T}_{\hvarphi\hatz}:_\Omega} =&
 \frac{\rho\Omega^2 \Gamma^3}{\pi^2} \sum_{\lambda} 2\lambda 
 \left\{T^3 I_{1/2}^+(\beta \mu_\lambda) + 
 \frac{\Omega^2 T}{12} \Gamma^2 (6\Gamma^2 - 1) 
 I_{-1/2}^+(\beta \mu_\lambda)\right.\nonumber\\
 &\left. - \frac{M^2 T}{8(\Gamma^2 - 1)} \left[2\Gamma^2 -1 
 -\frac{{\rm arcsinh}(\rho\Omega\Gamma)}{\rho \Omega \Gamma^2}\right] 
 I_{-1/2}^+(\beta \mu_\lambda) + O(\Omega^4, M^4, \Omega^2 M^2)\right\}.
\end{align}
From the above, the coefficients $\sigma_\varepsilon^\omega$ and $\Pi_2$ can be obtained:
\begin{align}
 \sigma_\varepsilon^\omega =& \frac{T^3}{\pi^2} 
 \sum_{\lambda = \pm \frac{1}{2}} 2\lambda I_{1/2}^+(\beta \mu_\lambda) +
 \frac{T}{12\pi^2}(\bm{\omega}^2 + \bm{a}^2) \sum_{\lambda = \pm \frac{1}{2}} 2\lambda I_{-1/2}^+(\beta \mu_\lambda) \nonumber\\
 & - \frac{M^2 T}{8\pi^2(\Gamma^2 - 1)} \left[2\Gamma^2 -1 
 -\frac{{\rm arcsinh}(\rho\Omega\Gamma)}{\rho \Omega \Gamma^2}\right] 
 \sum_{\lambda = \pm \frac{1}{2}} 2\lambda I_{-1/2}^+(\beta \mu_\lambda)+ O(\Omega^4, M^4, \Omega^2 M^2), \nonumber\\
  \Pi_2 =& -\frac{T}{3\pi^2} \sum_{\lambda = \pm\frac{1}{2}} 2\lambda I_{-1/2}^+(\beta\mu_\lambda) + O(\Omega^2, M^2).
  \label{eq:SET_Piko_aux}
\end{align}
The exact expressions for $I_{\pm 1/2}^+$ given in Eq.~\eqref{eq:FD_Half}
can be used to obtain:
\begin{align}
 \sigma_\varepsilon^\omega =&
 -\frac{T^3}{\pi^2} \left[{\rm Li}_3(-e^{-\beta \mu_+})-
 {\rm Li}_3(-e^{-\beta\mu_-}) +
 {\rm Li}_3(-e^{\beta \mu_+}) -
 {\rm Li}_3(-e^{\beta \mu_-})\right] \nonumber\\
 &+ \frac{T}{12\pi^2} \left\{\bm{\omega}^2 + \bm{a}^2 - 
 \frac{3M^2}{2(\Gamma^2 - 1)} \left[2\Gamma^2 -1 
 -\frac{{\rm arcsinh}(\rho\Omega\Gamma)}{\rho \Omega \Gamma^2}\right] \right\}
 \ln \left[\frac{\cosh\frac{\beta}{2}(\mu_V + \mu_H)}
 {\cosh\frac{\beta}{2}(\mu_V - \mu_H)}\right]\nonumber\\
 & + O(\Omega^4, M^4, \Omega^2 M^2),\nonumber\\
 \Pi_2 =& -\frac{T}{3\pi^2} \ln \left[\frac{\cosh\frac{\beta}{2}(\mu_V + \mu_H)}
 {\cosh\frac{\beta}{2}(\mu_V - \mu_H)}\right] + O(\Omega^2,M^2).
 \label{eq:SET_Piko}
\end{align}
The large temperature limit of the above expressions is:
\begin{align}
 \sigma_\varepsilon^\omega =& \frac{4\mu_V \mu_H T}{\pi^2} \ln 2 + 
 \frac{\mu_V \mu_H}{6\pi^2 T}\left(\mu_V^2 + \mu_H^2 + \frac{\bm{\omega}^2 + \bm{a}^2}{4}\right) \nonumber\\
 &- \frac{\mu_V \mu_H M^2}{16\pi^2 T(\Gamma^2 - 1)} \left[2\Gamma^2 -1 
 -\frac{{\rm arcsinh}(\rho\Omega\Gamma)}{\rho \Omega \Gamma^2}\right]
 + O(T^{-3}),\nonumber\\
 \Pi_2 =& -\frac{\mu_V \mu_H}{6\pi^2 T} + O(T^{-3}).
 \label{eq:SET_Piko_T}
\end{align}
It is interesting to note that $\sigma_\varepsilon^\omega$ differs from $Q_A$, given
in Eq.~\eqref{eq:AC_Q_st_T}, only through the mass correction and the 
terms proportional to $\bm{\omega}^2$ and $\bm{a}^2$.

The results derived in Eq.~\eqref{eq:SET_Piko} are valid 
only at small $\Omega$. Expressions for $\sigma_\varepsilon^\omega$ and $\Pi_2$ 
which are exact on the rotation axis can be derived by noting that
\begin{equation}
 \sigma_\varepsilon^\omega\rfloor_{\rho = 0} = -\frac{1}{\Omega\Gamma} \left.
 \braket{:\widehat{T}_{\hatt\hatz}:_\Omega}\right\rfloor_{\rho = 0}, \qquad 
 \Pi_2\rfloor_{\rho = 0} = -\left(\frac{\braket{:\widehat{T}_{\hvarphi\hatz}:_\Omega}}
 {\rho \Omega^4\Gamma^5} + 
 \frac{\braket{:\widehat{T}_{\hatt\hatz}:_\Omega}}{\Omega^3 \Gamma^5}\right)_{\rho = 0}.
\end{equation}
Starting from Eq.~\eqref{eq:SET_aux}:
\begin{multline}
 \begin{pmatrix}
 \braket{:\widehat{T}_{\hatt\hatz}:_\Omega} \\
 \braket{:\widehat{T}_{\hvarphi\hatz}:_\Omega} / \rho \Omega 
 \end{pmatrix} = \frac{1}{8\pi^2}
 \sum_{\lambda = \pm \frac{1}{2}} 2\lambda 
 \sum_{m = -\infty}^\infty \int_M^\infty dE\, p\,
 \left[\frac{1}
 {e^{\beta_0(\widetilde{E} - \mu_{\lambda;0})} + 1} + 
 \frac{1}{e^{\beta_0(\widetilde{E} + \mu_{\lambda;0})} + 1}\right]\\\times
 \int_{0}^p dk 
 \begin{pmatrix}
  -E\left(1 + \frac{k^2}{p^2}\right) J_m^-(q\rho) \\
  \frac{1}{\rho^2\Omega} \left[m J_m^-(q\rho) - \frac{1}{2} J_m^+(q\rho)\right]
 \end{pmatrix}
 \label{eq:SET_M0_H}
\end{multline}
and noting that $J_{\pm 1/2}^-(z) = \pm 1$ and
\begin{equation}
 \left[m J_m^-(q\rho) - \frac{1}{2} J_m^+(q\rho)\right]_{m = \pm \frac{1}{2}} = -\left[m J_m^-(q\rho) - \frac{1}{2} J_m^+(q\rho)\right]_{m = \pm \frac{3}{2}} = -\left(\frac{q\rho}{2}\right)^2 + \dots,
\end{equation}
the following expressions can be obtained:
\begin{align}
 \braket{:\widehat{T}_{\hatt\hatz}:_\Omega}_{\rho = 0} =& -\frac{1}{3\pi^2} \sum_{\lambda = \pm \frac{1}{2}} 2\lambda 
 \int_M^\infty dE\, E p^2 \left[f^-_{\beta_0}\left(\mu_{\lambda;0} + \frac{\Omega}{2}\right) - 
 f^-_{\beta_0}\left(\mu_{\lambda;0} - \frac{\Omega}{2}\right)\right],\nonumber\\
 \left.\frac{\braket{:\widehat{T}_{\hvarphi\hatz}:_\Omega}}{\rho \Omega} \right\rfloor_{\rho = 0} =& 
 -\frac{1}{24\pi^2 \Omega} \sum_{\lambda = \pm \frac{1}{2}} 2\lambda 
 \sum_{\sigma_\Omega =\pm 1} \int_M^\infty dE\, p^4 \nonumber\\
 & \times \left[
 f^+_{\beta_0}\left(\mu_{\lambda;0} + \sigma_\Omega \frac{\Omega}{2}\right) - 
 f^+_{\beta_0}\left(\mu_{\lambda;0} + \sigma_\Omega \frac{3\Omega}{2}\right)\right],
\end{align}
where the notation $f^\pm_{\beta_0}(a)$ was introduced in Eq.~\eqref{eq:fpm_def}.
Using Eq.~\eqref{eq:RKT_exp_aux3} to expand $f^\pm_{\beta_0}(a)$ and 
Eq.~\eqref{eq:RKT_exp_aux4} to perform the sum over helicities, the 
following result can be obtained:
\begin{align}
 \braket{:\widehat{T}_{\hatt\hatz}:_\Omega}_{\rho =0} =& \frac{1}{\pi^2 \beta_0^4}
 \sum_{\sigma_\Omega = \pm 1} \sigma_\Omega 
 \sum_{\sigma_V = \pm 1} \sigma_V
 \sum_{\sigma_H = \pm 1} \sigma_H \Bigg[
 {\rm Li}_4(-e^{\zeta_{1/2}}) \nonumber\\
 & + \beta_0 M {\rm Li}_3(-e^{\zeta_{1/2}}) + 
 \frac{(\beta_0 M)^2}{3} {\rm Li}_2(-e^{\zeta_{1/2}})\Bigg], 
 \nonumber\\
 \left.\frac{\braket{:\widehat{T}_{\hvarphi\hatz}:_\Omega}}{\rho \Omega}\right\rfloor_{\rho = 0} =& 
 -\frac{1}{2\pi^2 \Omega \beta_0^5} \sum_{\sigma_\Omega = \pm 1} \sum_{\sigma_V = \pm 1} \sigma_V
 \sum_{\sigma_H = \pm 1} \sigma_H \Bigg\{
 {\rm Li}_5(-e^{\zeta_{3/2}}) - {\rm Li}_5(-e^{\zeta_{1/2}}) \nonumber\\
 &+ \beta_0 M [{\rm Li}_4(-e^{\zeta_{3/2}}) - {\rm Li}_4(-e^{\zeta_{1/2}})] +
 \frac{(\beta_0 M)^2}{3} [{\rm Li}_3(-e^{\zeta_{3/2}}) - {\rm Li}_3(-e^{\zeta_{1/2}})]\Bigg\},
 \label{eq:SET_Piko_axis}
\end{align}
where $\zeta_m = \beta_0 (\sigma_V \mu_V + \sigma_H \mu_H + m \sigma_\Omega \Omega - M)$.
At large temperatures, the following limits can be obtained:
\begin{align}
 \sigma_\varepsilon^\omega\rfloor_{\rho = 0} =& \frac{4 \mu_V \mu_H T}{\pi^2} \ln 2 + 
 \frac{\mu_V \mu_H}{6 \pi^2 T} \left(\mu_V^2 + \mu_H^2 +  \frac{\Omega^2}{4} - M^2\right) + O(T^{-3}),\nonumber\\
 \Pi_2\rfloor_{\rho = 0} =& -\frac{\mu_V \mu_H}{6 \pi^2 T} + 
 \frac{\mu_V \mu_H}{72\pi^2 T^3} \left(\mu_V^2 + \mu_H^2 + 
 \frac{11\Omega^2}{20} - M^2\right) + O(T^{-4}),
 \label{eq:SET_Piko_axis_T}
\end{align}
in agreement with Eq.~\eqref{eq:SET_Piko_T}.

In summary, the SET for the Dirac field can be written as:
\begin{align}
 \braket{:\widehat{T}_{\halpha\hsigma}:_\Omega} =& 
 \left(E_{\rm RKT} + \Delta E\right) u_\halpha u_\hsigma - 
 (P_{\rm RKT} + \Delta P) \Delta_{\halpha\hsigma} + \Pi_{\halpha\hsigma} + 
 u_{\halpha} W_{\hsigma} + u_\hsigma W_\halpha,\nonumber\\
 E_{\rm RKT} =& 
 \left[\frac{7\pi^2 T^4}{60}
 + \frac{T^2}{2}(\mu_V^2 + \mu_H^2)
 + \frac{\mu_V^4 + 6\mu_V^2 \mu_H^2 + \mu_H^4}{4\pi^2}\right]- \frac{M^2}{12} \left[T^2 + 
 \frac{3(\mu_V^2 + \mu_H^2)}{\pi^2} \right],\nonumber\\
 \Delta E =& \frac{\bm{a}^2+3\bm{\omega}^2}{24}\left[T^2 + \frac{3(\mu_V^2 + \mu_H^2)}{\pi^2}\right],\qquad 
 \Delta P = \frac{1}{3} \Delta E, \nonumber\\
 \Pi_{\halpha\hsigma} =& -\frac{\mu_V \mu_H}{6\pi^2 T} (\tau_{\halpha} \omega_{\hsigma} + \omega_\halpha \tau_\hsigma),\nonumber\\
 W_\halpha =& -\frac{1}{18}\left[T^2 + \frac{3(\mu_V^2 + \mu_H^2)}{\pi^2} \right] \tau_\halpha + 
 \frac{4\mu_V \mu_H T}{\pi^2} (\ln 2) \omega_\halpha.
 \label{eq:SET_summary}
\end{align}
Only the leading order terms with respect to $T$ of 
$\Pi_{\halpha\hsigma}$ and $\sigma_\varepsilon^\omega$ were included above. 
The corrections due to the mass are of order $O(M^4)$. 
It is worth noting that, in the limit when 
$\mu_H = M = 0$, the results in Eq.~\eqref{eq:SET_summary} reproduce 
the vanishing axial potential limit of Eq.~(4.5) and Tables~2 and 3 
from Ref.~\cite{buzzegoli18}.

\subsection{Numerical analysis}\label{sec:SET:num}

\begin{figure}
    \centering
\begin{tabular}{cc}
    \includegraphics[width=0.45\linewidth]{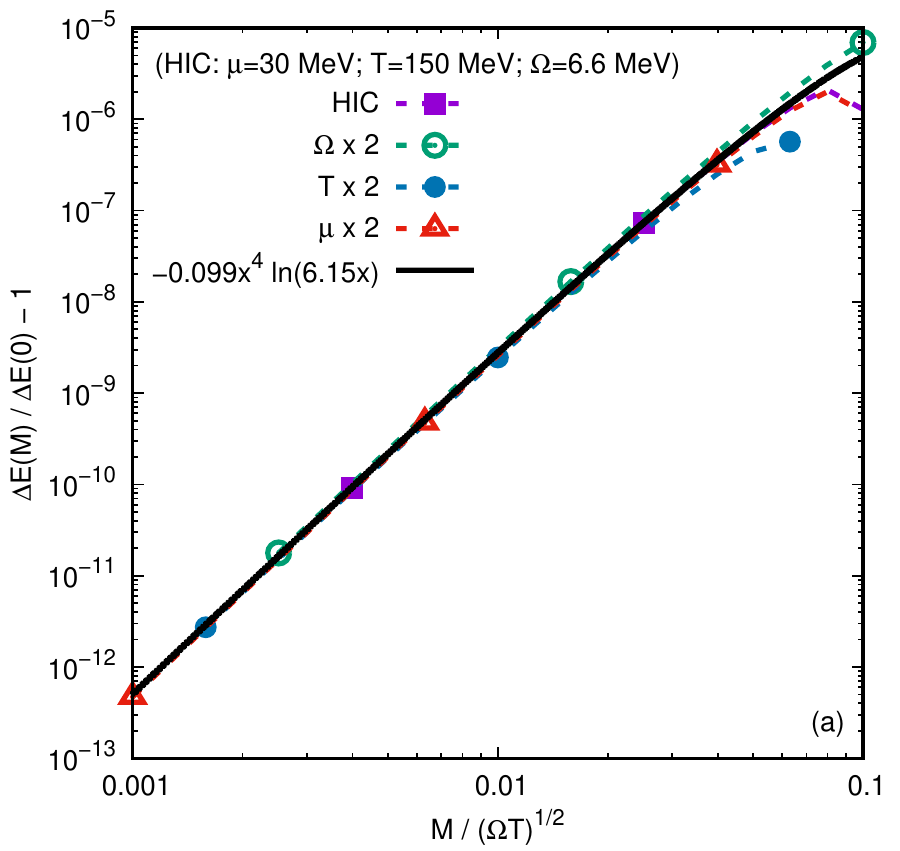} & 
    \includegraphics[width=0.45\linewidth]{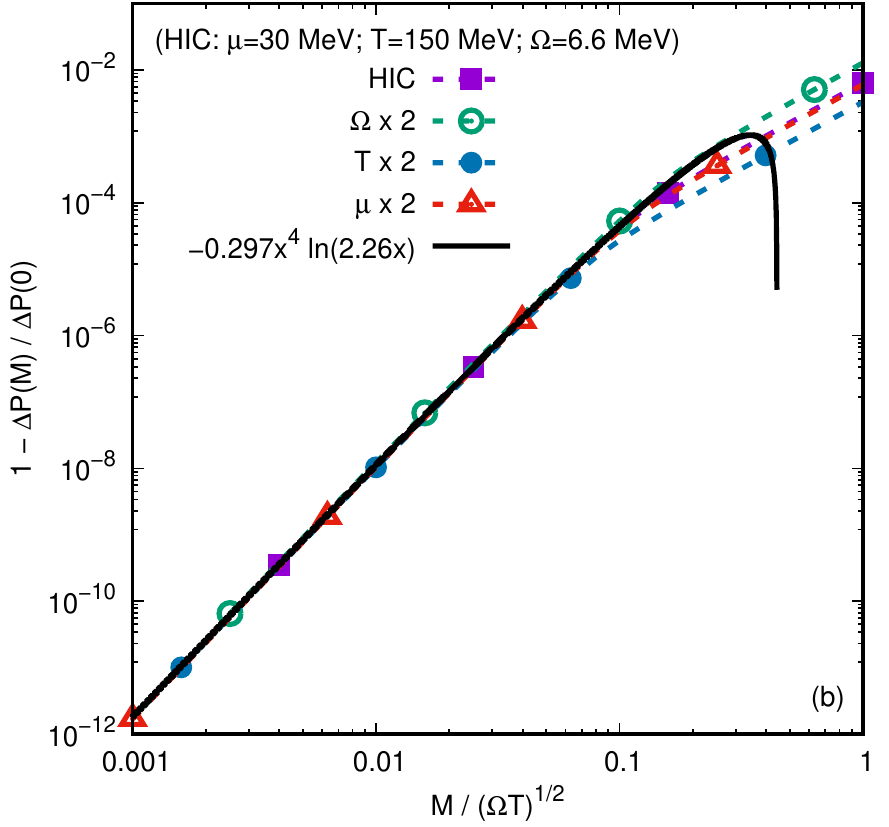} \\
    \includegraphics[width=0.45\linewidth]{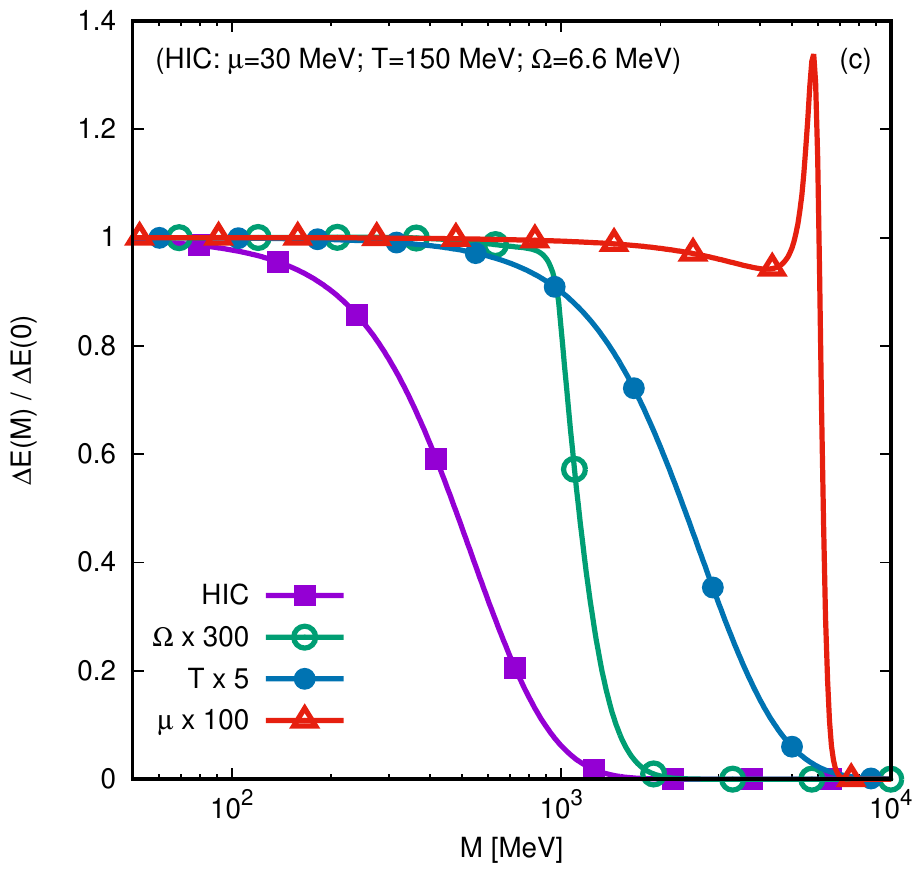} & 
    \includegraphics[width=0.45\linewidth]{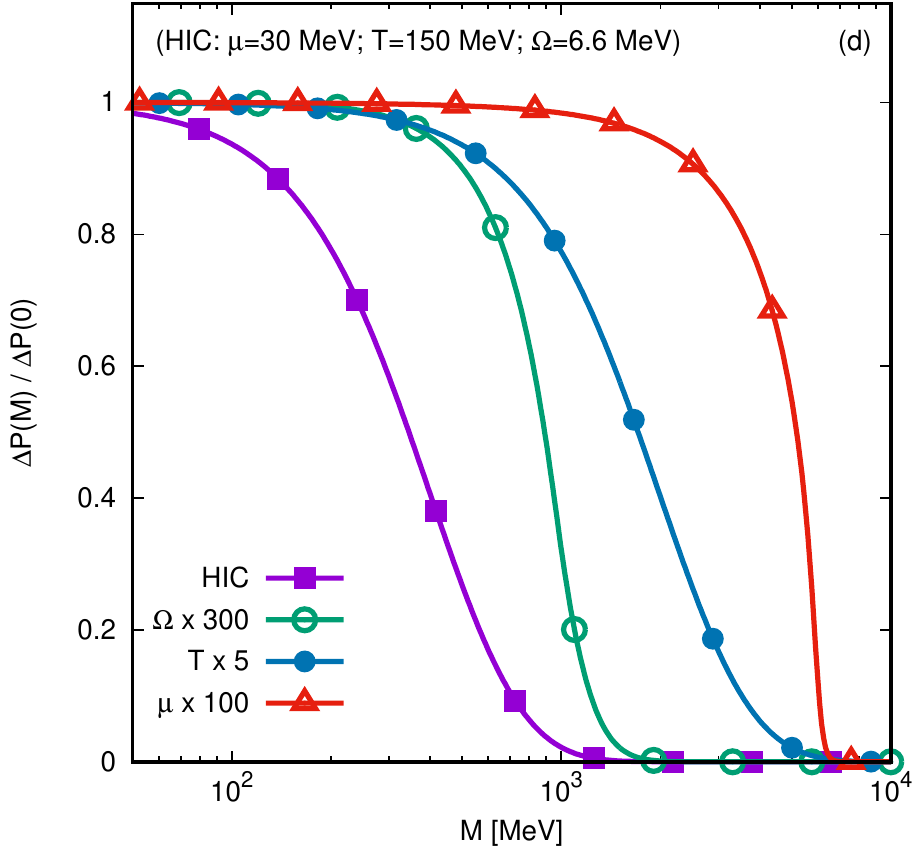} 
\end{tabular}
\caption{(Top line) Relative mass corrections (a) 
$\Delta E(M) / \Delta E(0) - 1$ and (b) $1 - \Delta P(M) / \Delta P(0)$ 
represented with respect to $x = M / \sqrt{T \Omega}$. The solid black 
lines represent the best fit curves of the function $\alpha x^4 \ln (\beta x)$. 
(Bottom line) Ratios (c) $\Delta E(M) /\Delta E(0)$ and 
(d) $\Delta P(M) / \Delta P(0)$ between the quantum corrections 
$\Delta E = E - E_{\rm RKT}$ and $\Delta P = P - P_{\rm RKT}$ at mass $M$ 
and at vanishing mass, represented with respect to $M$ for various parameter 
choices.}
\label{fig:SET_DE_DP}
\end{figure}

In the beginning of this section, an analysis of the analytic
expressions for the quantum corrections $\Delta E$ and $\Delta P$ is 
presented. The massless limit results are summarised in 
Eq.~\eqref{eq:SET_summary} and the analysis of the preceeding section 
suggests that they are valid up to $O(M^4)$. This is confirmed in 
Figs.~\ref{fig:SET_DE_DP}(a) and \ref{fig:SET_DE_DP}(b), where the 
relative mass corrections $\Delta E(M) / \Delta E(0) - 1$ and 
$1 - \Delta P(M) / \Delta P(0)$ are represented with respect to 
the mass parameter. The numerical experiments 
indicate that the mass corrections behave to leading order as 
$\alpha x^4 \ln(\beta x)$, where $\alpha$ and $\beta$ are numerical coefficients, 
while $x = M / \sqrt{\Omega T}$. At large masses, the relations
in Eq.~\eqref{eq:SET_summary} can be expected to break 
down. Figs.~\ref{fig:SET_DE_DP}(c) and \ref{fig:SET_DE_DP}(d) show 
the influence of the parameters $\Omega$, $T$ and $\mu$ on the value 
of the mass where the breakdown occurs. Starting from the set of 
values pertaining to heavy ion collisions (labelled HIC), when the 
massless results stay valid up to about $100\ {\rm MeV}$, it can 
be seen that increasing $\Omega$, $\mu$ and $T$ enhances their domain 
of validity. In the case of the energy correction $\Delta E$, a spike 
can be seen in the case when $\mu = 3\ {\rm GeV} \gg \Omega = 6.6\ {\rm MeV}$. 
This spike is akin to the one observed for the charge correction, 
$\Delta Q_\pm$, shown in Fig.~\ref{fig:CC_Qst}(c).

\begin{figure}
\centering
\begin{tabular}{cc}
    \includegraphics[width=0.45\linewidth]{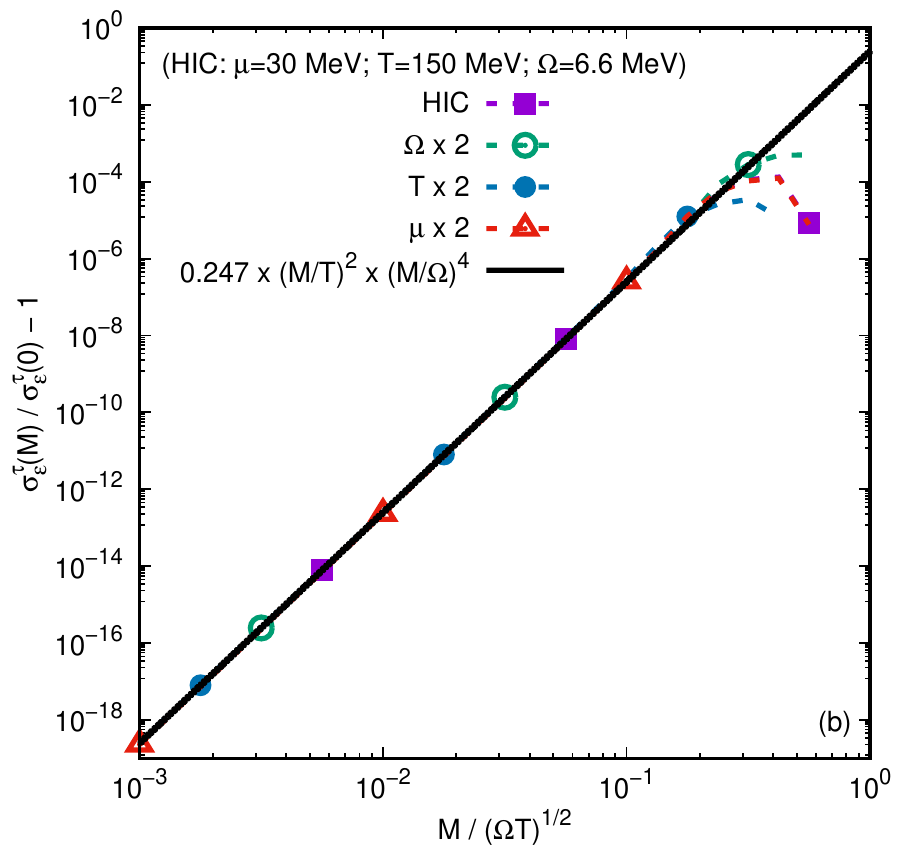} & 
    \includegraphics[width=0.45\linewidth]{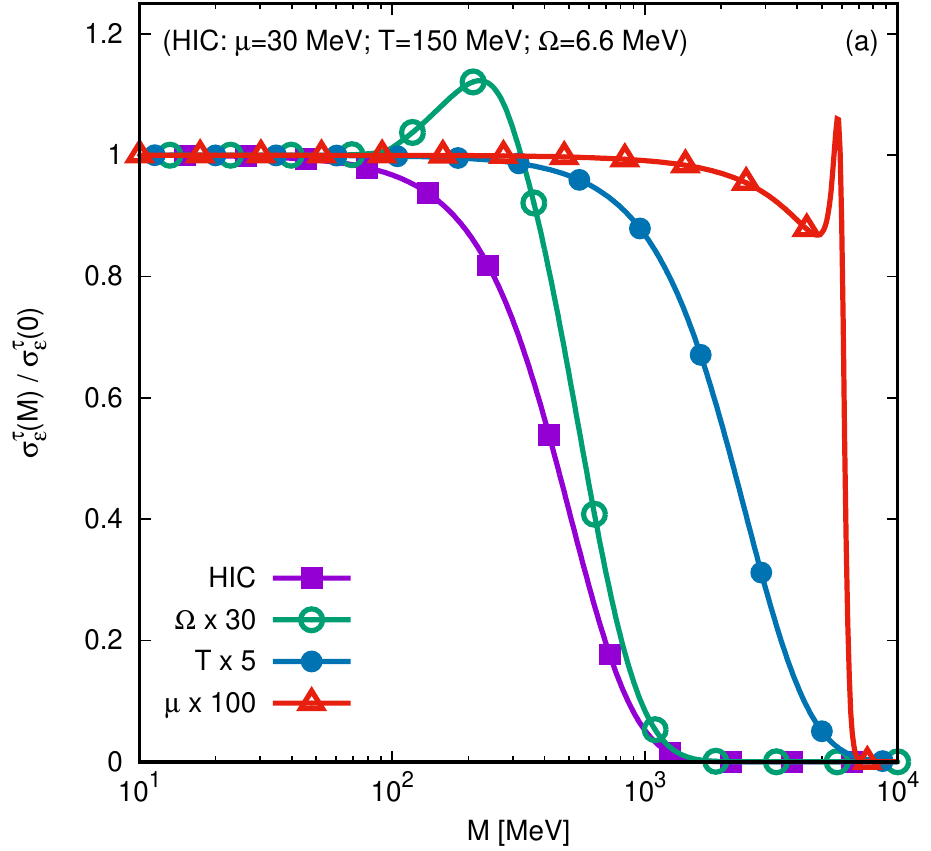} 
\end{tabular}
\caption{(a) Relative mass correction $\sigma_\varepsilon^\tau(M) / \sigma_\varepsilon^\tau(0) - 1$, 
represented with respect to $x = M / \sqrt{T \Omega}$. The solid black line 
represents the best fit curve of the function $\alpha x^4$. (b) Ratio 
$\sigma_\varepsilon^\tau(M) / \sigma_\varepsilon^\tau(0)$, represented with respect to $M$ for 
various parameter choices. 
}
\label{fig:SET_kt}
\end{figure}

Turning to the circular heat conductivity $\sigma_\varepsilon^\tau$,
whose massless limit is given in Eq.~\eqref{eq:SET_summary},
Fig.~\ref{fig:SET_kt}(a) confirms that the leading order mass correction, 
computed as $\sigma_\varepsilon^\tau(M) / \sigma_\varepsilon^\tau(0) - 1$, is of order $O(M^4)$. 
Fig.~\ref{fig:SET_kt}(b) confirms the validity of the constitutive relation 
for massess up to $\sim 100\ {\rm MeV}$ for the HIC values of 
$\mu$, $T$ and $\Omega$. The spike seen when $M \rightarrow \mu$ at large 
$\mu$ is akin to the one observed for $\Delta Q_{\pm}$, $\sigma^\tau_\pm$ and 
$\Delta E$ in Figs.~\ref{fig:CC_Qst}(c), \ref{fig:CC_Qst}(d) and 
\ref{fig:SET_DE_DP}(c). The peak observed in the curve corresponding to 
$\Omega \simeq 2\ {\rm GeV}$ is akin to the one observed for the axial vortical 
conductivity $\sigma^\omega_A$ in Fig.~\ref{fig:AC_so}(b).

\begin{figure}
\centering
\begin{tabular}{cc}
    \includegraphics[width=0.45\linewidth]{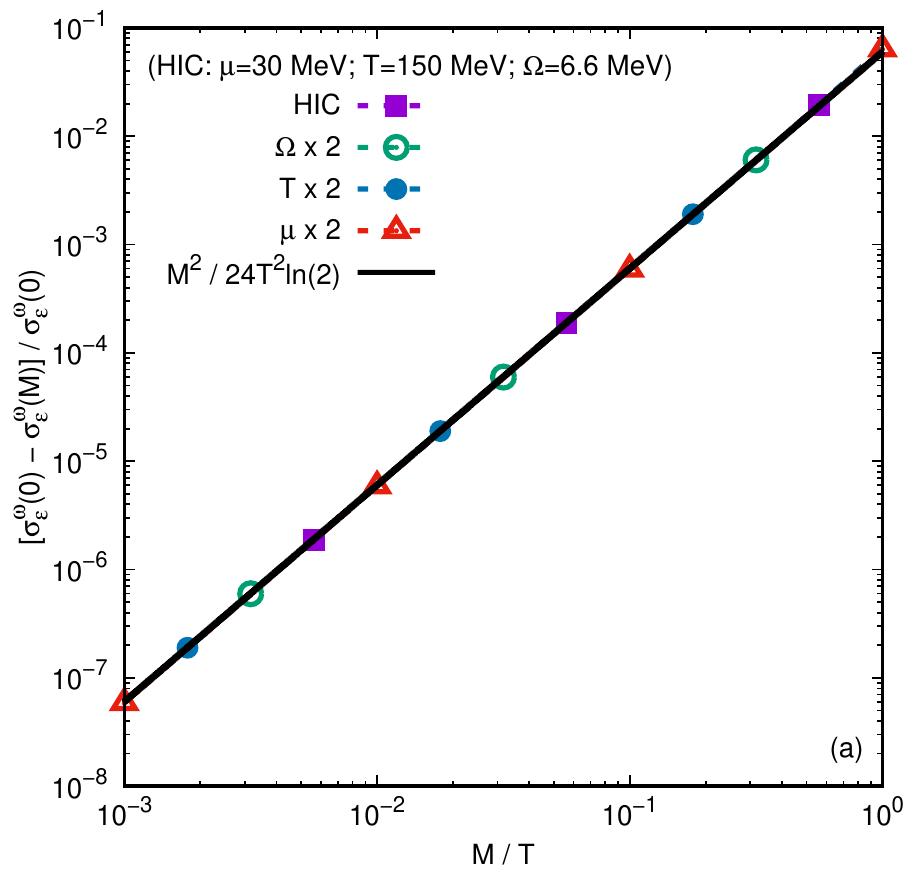} & 
    \includegraphics[width=0.45\linewidth]{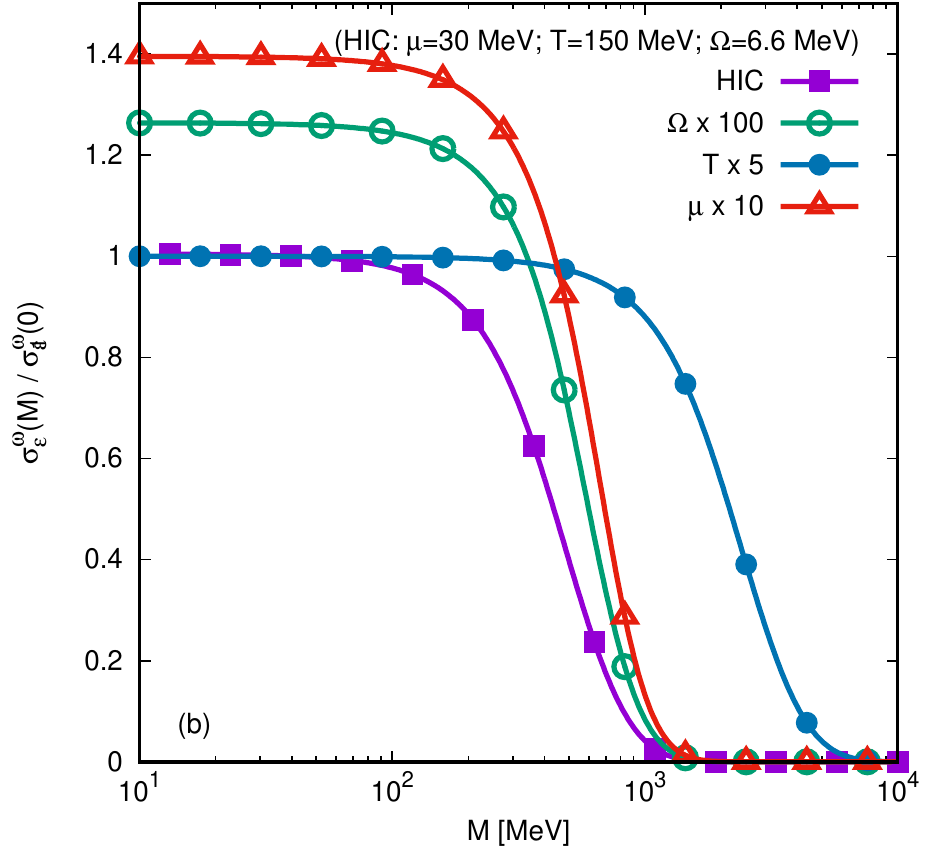} \\
    \includegraphics[width=0.45\linewidth]{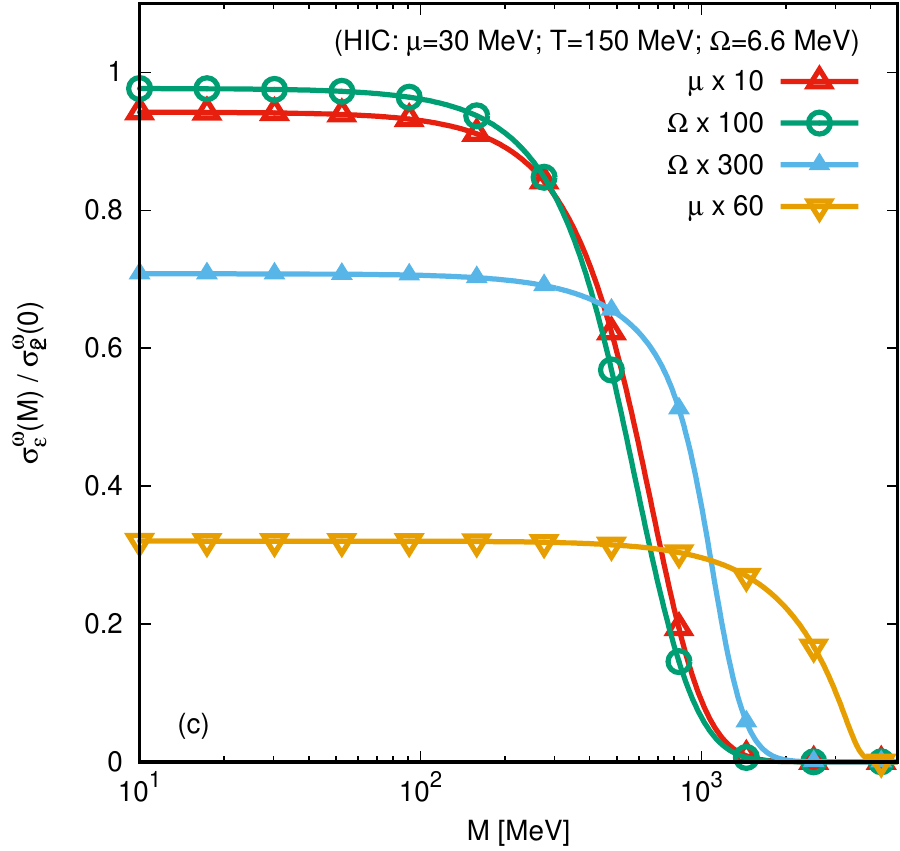} & 
    \includegraphics[width=0.45\linewidth]{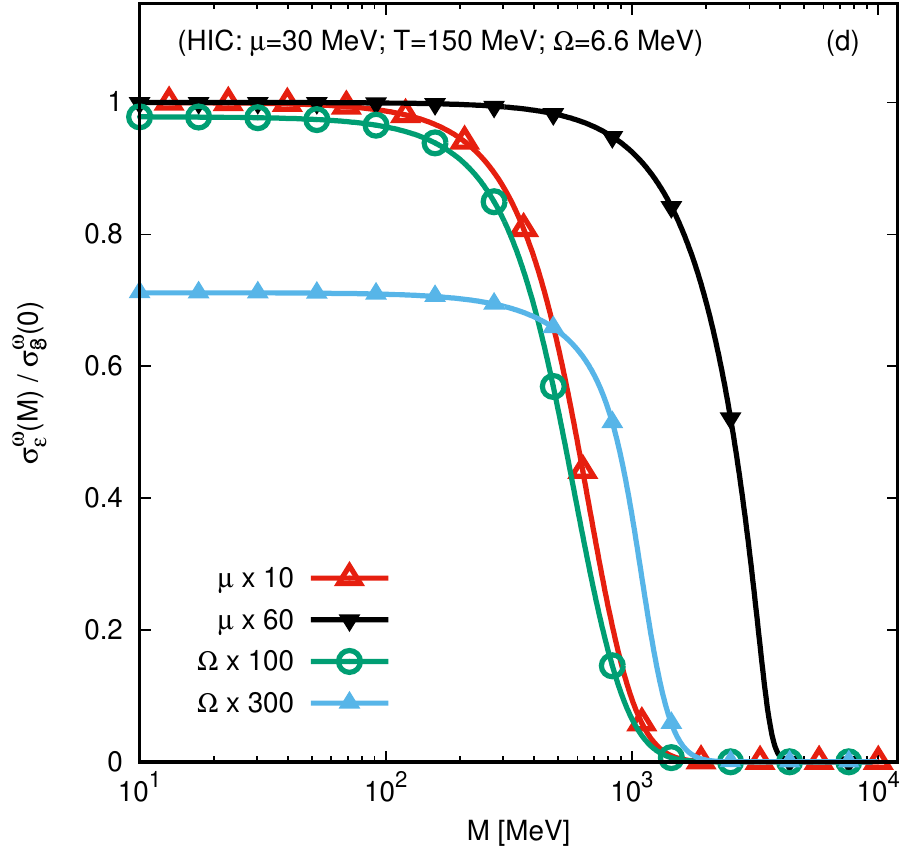}
\end{tabular}
\caption{(a) Relative mass correction $1 - \sigma_\varepsilon^\omega(M) / \sigma_\varepsilon^\omega(0)$, 
represented with respect to $M / T$. The solid black line represents the 
prediction given in Eq.~\eqref{eq:SET_Piko_axis_T}. (b)-(d) Ratio 
$\sigma_\varepsilon^\omega(M) / \sigma_\varepsilon^{\omega;n}(0)$ between $\sigma_\varepsilon^\omega(M)$ 
computed using Eq.~\eqref{eq:SET_Piko_axis} and the approximations 
$\sigma_\varepsilon^{\omega;n}(0)$ given in Eqs.~\eqref{eq:SET_ko_O1}, 
\eqref{eq:SET_ko_O2} and \eqref{eq:SET_ko_O3} for (b) $n = 1$, 
(c) $n = 2$ and (d) $n = 3$. 
}
\label{fig:SET_ko}
\end{figure}

The heat conductivity $\sigma_\varepsilon^\omega$ along the vector $\omega^\halpha$
is considered in Fig.~\ref{fig:SET_ko}. This quantity can be obtained in 
closed form 
on the rotation axis, for any mass, and is given in terms of the 
polylogarithm functions in Eq.~\eqref{eq:SET_Piko_axis}. First, the 
leading order contribution coming from the mass is considered. 
Figure~\ref{fig:SET_ko}(a) presents the relative mass correction, 
$1 - \sigma_\varepsilon^\omega(M) / \sigma_\varepsilon^\omega(0)$, with respect to the ratio 
$M / T$. The second order contribution predicted in 
Eqs.~\eqref{eq:SET_Piko_T} and \eqref{eq:SET_Piko_axis_T} is confirmed.

The expression for $\sigma_\varepsilon^\omega$ derived in Eq.~\eqref{eq:SET_Piko_axis} 
is valid only on the rotation axis and is thus inapplicable for studying 
the properties of $\sigma_\varepsilon^\omega$, e.g., in the vicinity of the speed of 
light surface, as well as the interplay between the vorticity 
$\omega^\hatz = \Omega \Gamma^2$ and acceleration 
$a^\hrho = -\rho \Omega \Gamma^2$, since the latter vanishes on the 
rotation axis. Another drawback concerns the complexity of the polylogarithm 
functions, which make the physical properties of $\sigma_\varepsilon^\omega$ difficult 
to assess. To this end, three levels of approximations are employed, 
following the approach taken for $\sigma^\omega_\pm(M)$ and $Q_A(M)$, 
which are represented in Figs.~\ref{fig:CC_so}(b-d) and \ref{fig:AC_Q}(b-d), 
respectively. The approximations are denoted using $\sigma_\varepsilon^{\omega;n}$, with 
$1 \le n \le 3$, and are based on Eq.~\eqref{eq:SET_Piko}. The cases 
$n = 1$ and $n = 2$ are obtained from the high temperature expansion given 
in Eq.~\eqref{eq:SET_Piko_T}. For the $n=  1$ term, only the leading order 
contribution is taken into account:
\begin{equation}
 \sigma_\varepsilon^{\omega;1} = \frac{4\mu_H \mu_V T}{\pi^2} \ln 2.
 \label{eq:SET_ko_O1}
\end{equation}
As can be seen in Fig.~\ref{fig:SET_ko}(b), this approximation works well 
for the HIC parameters, breaking down when $M \gtrsim T$. This is confirmed 
by looking at the curve corresponding to $T \simeq 750\ {\rm MeV}$ (blue line 
and solid circles). However, the results for higher vorticity 
($\Omega = 660\ {\rm MeV}$ for the green line and empty circles) and higher 
chemical potential ($\mu =300\ {\rm MeV}$ for the red line and empty triangles) 
show a discrepancy.

The $n = 2$ approximation takes into account the $T^{-1}$ correction 
appearing in Eq.~\eqref{eq:SET_Piko_T}, but disregards the mass term:
\begin{equation}
 \sigma_\varepsilon^{\omega; 2} = \frac{4\mu_H \mu_V T}{\pi^2} \ln 2 + 
 \frac{\mu_H \mu_V}{6\pi^2 T}\left(\mu_H^2 + \mu_V^2 + \frac{\bm{\omega}^2 + \bm{a}^2}{4}\right).
 \label{eq:SET_ko_O2}
\end{equation}
It should be noted that on the rotation axis, Eq.~\eqref{eq:SET_ko_O2} 
coincides with the massless limit of Eq.~\eqref{eq:SET_Piko_axis_T}, which 
represents the high temperature expansion of the exact result in 
Eq.~\eqref{eq:SET_Piko_axis}. Figure~\ref{fig:SET_ko}(c) shows that this 
second approximation works well for the two problematic cases discussed 
in Fig.~\ref{fig:SET_ko}(b), i.e., $\Omega = 660\ {\rm MeV}$ and 
$\mu = 300\ {\rm MeV}$. Further increasing the vorticity 
($\Omega \simeq 2\ {\rm GeV}$ for the blue line with filled upper triangles) 
or the chemical potential ($\mu = 0.9\ {\rm GeV}$ for the orange line and lower, 
filled triangles) again gives rise to discrepancies. The final approximation 
considered here is obtained by setting $M = 0$ in Eq.~\eqref{eq:SET_Piko}:
\begin{multline}
 \sigma_\varepsilon^{\omega;3} = -\frac{T^3}{\pi^2} \left[{\rm Li}_3(-e^{-\mu_+/T})-
 {\rm Li}_3(-e^{-\mu_-/T}) - 
 {\rm Li}_3(-e^{\mu_+/T}) +
 {\rm Li}_3(-e^{\mu_-/T})\right] \\
 + \frac{T}{12\pi^2}(\bm{\omega}^2 + \bm{a}^2)
 \ln \left[\frac{\cosh\frac{1}{2T}(\mu_V + \mu_H)}
 {\cosh\frac{1}{2T}(\mu_V - \mu_H)}\right].
 \label{eq:SET_ko_O3}
\end{multline}
Fig.~\ref{fig:SET_ko}(d) shows that this approximation works  
well at high chemical potential, however, the discrepancy
at $\Omega = 2\ {\rm GeV}$ remains similar to that seen in 
Fig.~\ref{fig:SET_ko}(c). This discrepancy is due to the fact that 
Eq.~\eqref{eq:SET_Piko} represents a small $\Omega$ expansion which 
is valid only up to second order in $\Omega$. 

\begin{figure}
\centering
\begin{tabular}{c}
 \includegraphics[width=0.45\linewidth]{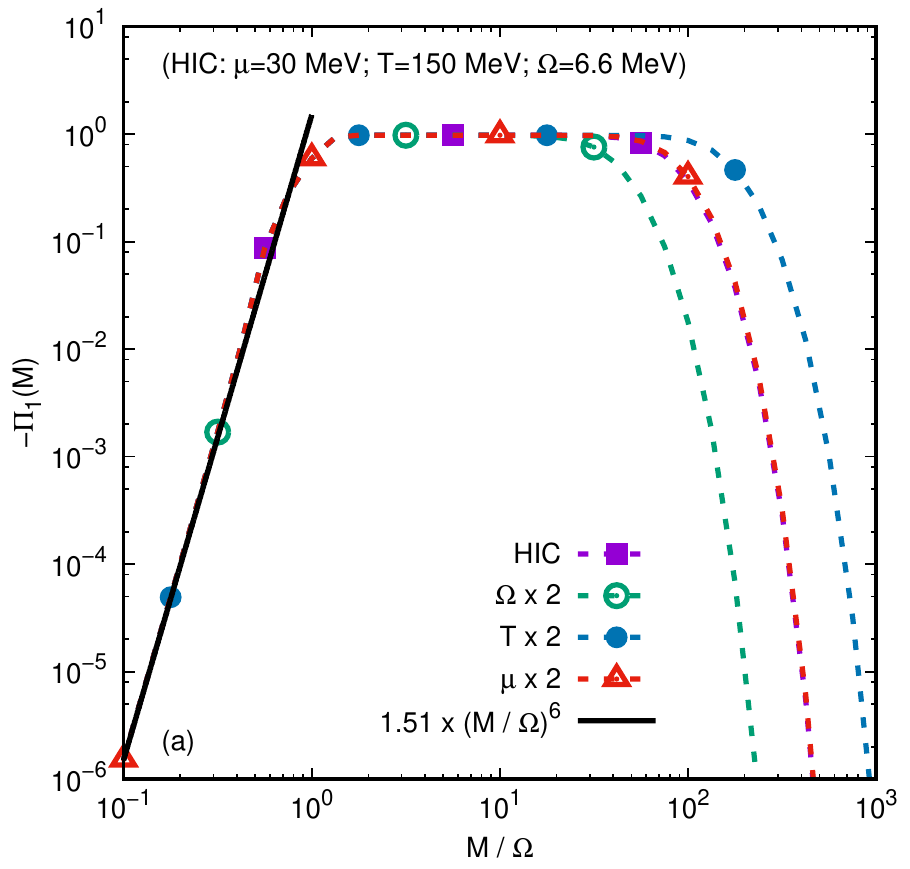} 
\end{tabular}
\caption{Dependence of $\Pi_1(M)$ on the ratio $M / \Omega$ for various 
values of the parameters. The solid black line shows the best fit of the 
expression $\alpha (M / \Omega)^6$ to the numerical data. 
}
\label{fig:SET_Pi1}
\end{figure}

In the case of the coefficient $\Pi_1$, the analysis in 
the previous subsection indicated that $\Pi_1$ vanishes up to 
$O(M^4)$. Fig.~\ref{fig:SET_Pi1} shows that the leading order 
contribution to $\Pi_1$ is of order $(M / \Omega)^6$.

\begin{figure}
\centering
\begin{tabular}{c}
    \includegraphics[width=0.45\linewidth]{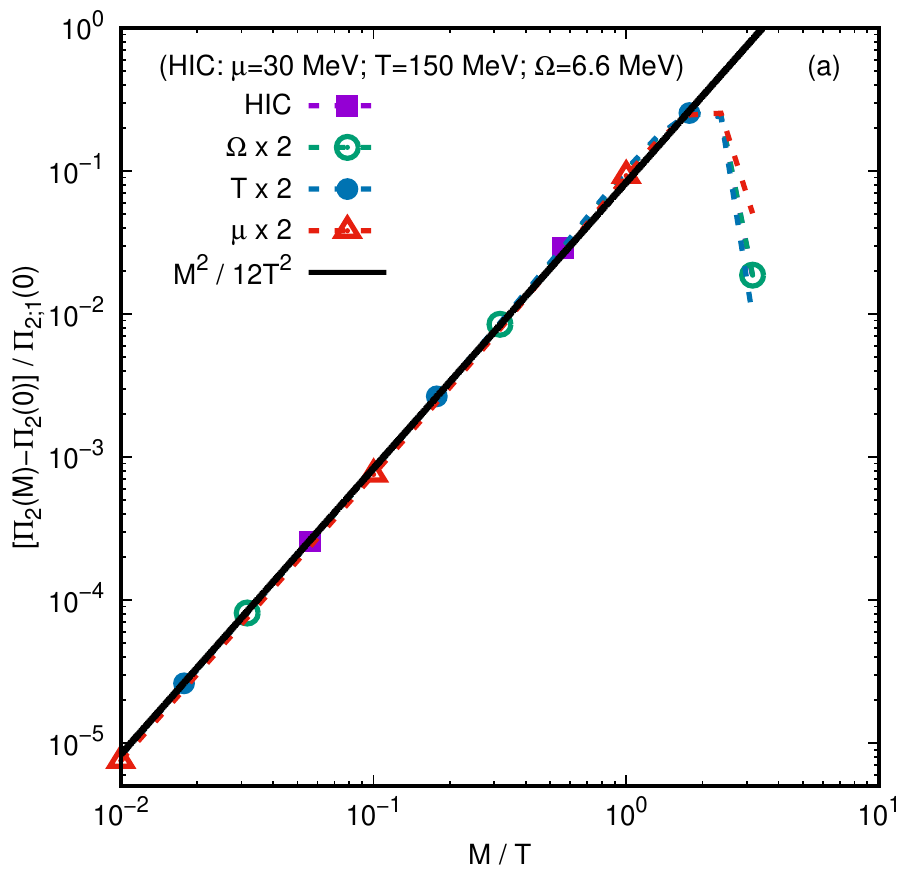}
\end{tabular}
\begin{tabular}{cc}
    \includegraphics[width=0.45\linewidth]{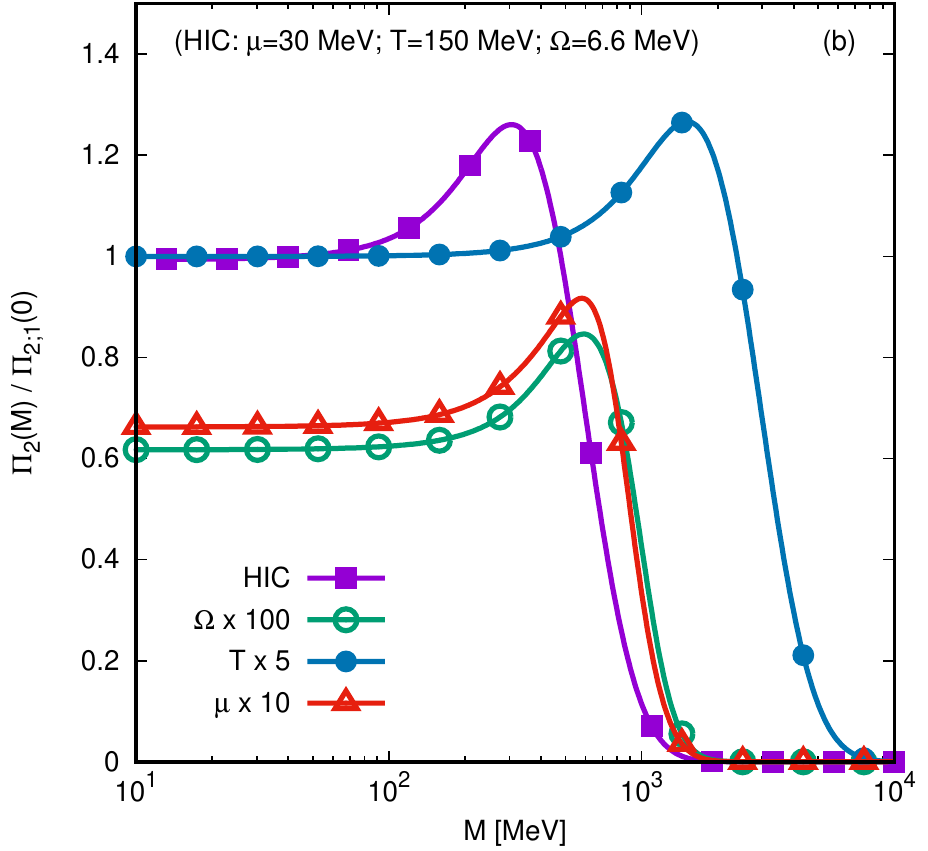} & 
    \includegraphics[width=0.45\linewidth]{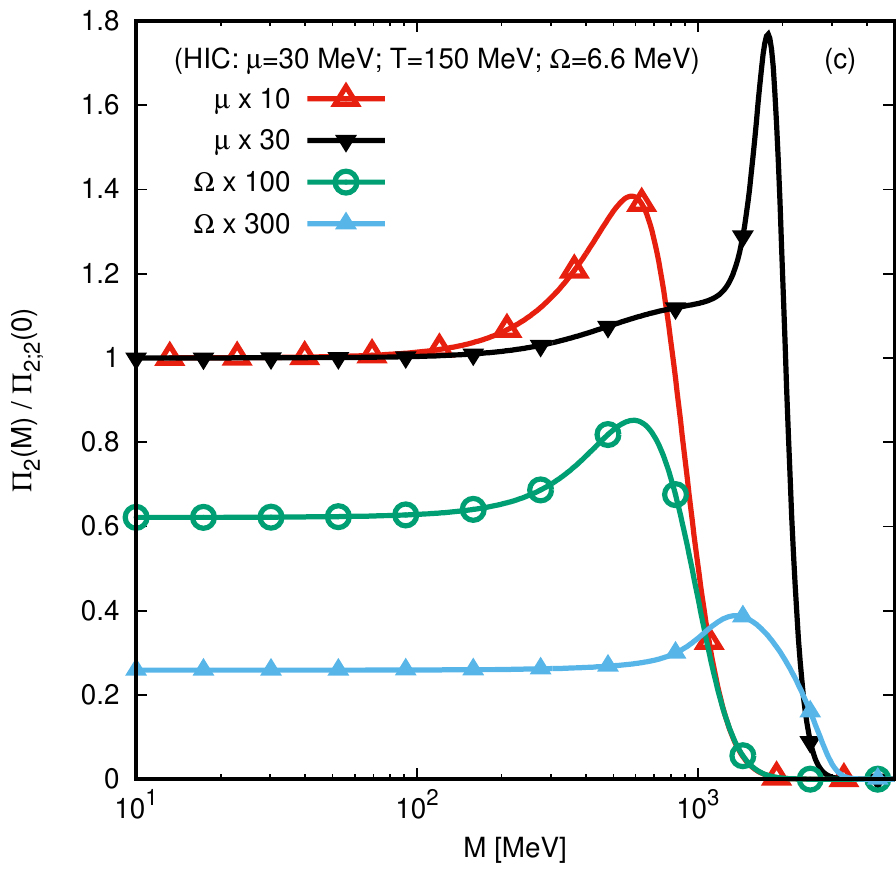}
\end{tabular}
\caption{(a) Relative mass correction $[\Pi_2(M) - \Pi_2(0)] / \Pi_{2;1}$, 
represented with respect to $M / T$. The solid black line represents the 
prediction given in Eq.~\eqref{eq:SET_Piko_axis_T}. (b)-(c) Ratios 
$\Pi_2(M) / \Pi_{2;n}(0)$ between $\Pi_2(M)$ computed using 
Eq.~\eqref{eq:SET_Piko_axis} and the approximations $\Pi_{2;n}(0)$ 
given in Eqs.~\eqref{eq:SET_Pi2_O1} and \eqref{eq:SET_Pi2_O2} for 
(b) $n = 1$ and (c) $n = 2$, respectively. 
}
\label{fig:SET_Pi2}
\end{figure}

In the case of $\Pi_2$, the same situation as for $\sigma_\varepsilon^\omega$
is encountered: Eq.~\eqref{eq:SET_Piko_axis} gives the exact value 
of $\Pi_2$ on the rotation axis, for any mass, while 
Eq.~\eqref{eq:SET_Piko} gives its value up to $O(\Omega^2)$ for any 
value of $\rho$. The mass is predicted through Eqs.~\eqref{eq:SET_Piko_T} 
and \eqref{eq:SET_Piko_axis_T} to contribute a correction of order $O(M^2)$. 
This is confirmed in Fig.~\ref{fig:SET_Pi2}(a). Next, the high temperature 
approximation given in Eq.~\eqref{eq:SET_Piko_T} is considered, which is 
labelled according to the convention in this section as $\Pi_{2;1}$:
\begin{equation}
 \Pi_{2;1} = -\frac{\mu_H \mu_V}{6\pi^2 T}.
 \label{eq:SET_Pi2_O1}
\end{equation}
As demonstrated in Fig.~\ref{fig:SET_Pi2}(b), this approximation is 
valid for the HIC conditions, but loses applicability when either 
$\Omega$ or $\mu$ are increased. The second approximation is the one 
derived in Eq.~\eqref{eq:SET_Piko}, which is valid up to $O(\Omega^2)$:
\begin{equation}
 \Pi_{2;2} = -\frac{T}{3\pi^2} \ln \left[\frac{\cosh\frac{\beta}{2}(\mu_V + \mu_H)}
 {\cosh\frac{\beta}{2}(\mu_V - \mu_H)}\right] + O(\Omega^2,M^2).
 \label{eq:SET_Pi2_O2}
\end{equation}
Fig.~\ref{fig:SET_Pi2}(c) shows that this approximation is valid at 
high values of the chemical potential ($\mu = 0.9\ {\rm GeV}$ is shown 
with black lines and lower, filled triangles), but remains inaccurate at 
high vorticities. This is because Eq.~\eqref{eq:SET_Piko} is derived as 
a small $\Omega$ expansion, which can be expected to be inaccurate at 
high values of $\Omega$.

\section{Conclusions}\label{sec:conc}

\begin{table}
\begin{center}
\renewcommand*\arraystretch{2}
\begin{tabular}{r|c|l|l}
 & & Leading order & Reference \\\hline\hline
 $J^\mu_{V/H}$ & $\sigma^\tau_{V/H}$ & $\dfrac{\mu_{V/H}}{6\pi^2} + O(M^4)$ & 
 \eqref{eq:CC_Qst} \\
 & $\sigma^\omega_{V/H}$ & $\dfrac{2\mu_{H/V} T}{\pi^2} \ln 2 + O(T^{-1})$ & 
 \eqref{eq:CC_so_Omega}, \eqref{eq:CC_so_axis} \\[2pt]\hline 
 $J^\mu_A$ & $\sigma^\tau_A$ & $\dfrac{\mu_V \mu_H}{12\pi^2 T} \left(1 - \dfrac{4M^2}{\Omega^2}\right) + O(T^{-2})$ & \eqref{eq:AC_Q_st}, \eqref{eq:AC_Q_st_axis} \\
 & $\sigma^\omega_A$ & $\dfrac{T^2}{6} + \dfrac{\mu_V^2 + \mu_H^2}{2\pi^2} + O(M^4)$ & \eqref{eq:AC_so_M} \\[2pt]\hline
 $T^{\mu\nu}$ & $\sigma_\varepsilon^\tau$ & $-\dfrac{T^2}{18} - \dfrac{\mu_V^2 + \mu_H^2}{6\pi^2} + O(M^4)$ & \eqref{eq:SET_EPkt_M}\\
 & $\sigma_\varepsilon^\omega$ & $\dfrac{4 \mu_V \mu_H T}{\pi^2} \ln 2 + O(T{^-1})$ & \eqref{eq:SET_Piko},\eqref{eq:SET_Piko_axis}\\
 & $\Pi_1$ & $O(M^4)$ & \eqref{eq:SET_EPkt_M}\\
 & $\Pi_2$ & $\dfrac{4 \mu_V \mu_H T}{\pi^2} \ln 2 + O(T^{-1})$ & \eqref{eq:SET_Piko},\eqref{eq:SET_Piko_axis}\\[2pt]\hline\hline
\end{tabular}
\caption{Summary of the anomalous transport coefficients for the 
vector/helical charge currents $J^\mu_{V/H}$, axial current $J^\mu_A$ and 
stress-energy tensor $T^{\mu\nu}$. Only the leading order terms are shown in this table.
Higher order terms can be found in the equations given as references. When a second equation
is provided, it gives the exact value of the corresponding quantity on the rotation axis.
\label{tab:summary}}
\end{center}
\end{table}

In this paper, the properties of massive Dirac fermions under rotation 
at finite temperature and finite chemical potential were considered. 
Aside from the chemical potential associated to the vector charge ($\mu_V$), a 
helicity chemical potential ($\mu_H$) accounting for the polarisation imbalance 
was taken into account. A simple relativistic kinetic theory (RKT) model incorporating 
the helicity imbalance was proposed to serve as a background classical theory. 
Within this kinetic framework, the vector and helicity charge currents (VCC and HCC) 
and the stress-energy tensor (SET) have the form corresponding to the perfect fluid, 
being characterised by the charge densities $Q^{\rm RKT}_V$ and $Q^{\rm RKT}_H$, 
energy density $E_{\rm RKT}$ and pressure $P_{\rm RKT}$.

The quantum thermal expectation values (t.e.v.s) were computed 
by performing an exact trace over Fock space using the 
mode basis introduced in Refs.~\cite{ambrus14plb,ambrus16prd,ambrus19lnp} 
with respect to which the thermal weight function $\hvrho$ is diagonal.
For all quantities involved, a small mass expansion was derived analytically,
valid at any distance between the rotation axis and the speed of light surface (SLS). 
Compared to the RKT results, the quantum t.e.v.s exhibit corrections at the 
level of $Q_V$, $Q_H$, $E$ and $P$, which are proportional to the square of the 
vorticity ($\bm{\omega}^2 = \Omega^2 \Gamma^4$) or of the acceleration 
[$\bm{a}^2 = \Omega^2 \Gamma^2(\Gamma^2 - 1)$].
Furthermore, the anomalous contributions to the charge currents and 
SET were highlighted. The corresponding transport coefficients are
summarised to leading order for convenience in Table~\ref{tab:summary}. 
The results derived in this paper agree in the massless limit with 
the vanishing axial chemical potential ($\mu_A$) limit 
of the results derived in Refs. \cite{helican} and \cite{buzzegoli18} 
(the $\mu_H \rightarrow 0$ limit is necessary to compare with the latter 
reference). 

It was shown that, under the conditions which are prevalent in the 
quark-gluon plasma formed in relativistic heavy ion collisions 
experiments (the HIC conditions), the constitutive relations are in 
general robust, in the sense that they remain valid for masses up to 
the thermal energy ($M \lesssim 150\ {\rm MeV}$). The only exception 
was seen in the case of the circular axial charge conductivity $\sigma_A^\tau$, 
which exhibits a peculiar mass dependence of the form $M^2 / \Omega^2$.
The mass correction $\delta \sigma_A^\tau = \sigma_A^\tau(M) - \sigma_A^\tau(0)$ 
becomes dominant over the massless value when $M \simeq \Omega$.

As pointed out in Sec.~\ref{sec:hel:anomal}, the transport laws involving the 
helicity chemical potential $\mu_H$ strongly suggest a new type of anomalies 
in quantum electrodynamics (QED), which may be investigated by considering 
the $AHH$ and $HTV$ triangle diagrams, involving vector ($V$), axial ($A$),
helical ($H$) and graviton ($T$) vertices.

\acknowledgments
Discussions with Prof. M. Chernodub (Universit\'e de Tours, France) and
Prof. E. Winstanley (University of Sheffield, UK) are gratefully acknowledged.
This work was supported by a Grant from the Romanian 
National Authority for Scientific Research and Innovation, CNCS-UEFISCDI,
project number PN-III-P1-1.1-PD-2016-1423. 

\appendix
\section{Fermi-Dirac integration formulae}
\label{app:FD}

In this section, the procedure for computing the integrals 
involving the Fermi-Dirac distribution function for 
massless particles appearing in the relativistic kinetic 
theory (RKT) formulation discussed in Sec.~\ref{sec:RKT}
and in the quantum field theory (QFT) formulation discussed 
in Sections~\ref{sec:CC}--\ref{sec:SET} is presented.

The first set of integrals that will be discussed 
can be introduced as follows:
\begin{align}
 I^-_n(a) =& \frac{1}{2} 
 \int_0^\infty dx\, x^{2n}\left(\frac{1}{e^{x-a}+1} - 
 \frac{1}{e^{x + a} + 1}\right),\nonumber\\
 I^+_n(a) =& \frac{1}{2} 
 \int_0^\infty dx\, x^{2n+1}\left(\frac{1}{e^{x-a}+1} + 
 \frac{1}{e^{x + a} + 1}\right). \label{eq:FD_aux}
\end{align}
An expansion of the Fermi-Dirac factors in powers of $a$ can 
be considered, as follows:
\begin{equation}
 \frac{1}{e^{x \pm a} + 1} = 
 \sum_{j = 0}^\infty \frac{(\pm a)^j}{j!} 
 \frac{d^j}{dx^j} \left(\frac{1}{e^x + 1}\right).
\end{equation}
Inserting this expansion in Eq.~\eqref{eq:FD_aux} gives:
\begin{align}
 I^-_n(a) =& - \int_0^\infty dx\, x^{2n}
 \sum_{j = 0}^\infty \frac{a^{2j+1}}{(2j+1)!} 
 \frac{d^{2j+1}}{dx^{2j+1}} 
 \left(\frac{1}{e^x + 1}\right),\nonumber\\
 I^+_n(a) =& \int_0^\infty dx\, x^{2n+1}
 \sum_{j = 0}^\infty \frac{a^{2j}}{(2j)!} 
 \frac{d^{2j}}{dx^{2j}} 
 \left(\frac{1}{e^x + 1}\right).
 \label{eq:FD_aux2}
\end{align}
It can now be assumed that the summation and integration signs 
can be interchanged. This assumption will prove to be 
justified by the structure of the result. 
Integration by parts can be performed in the above 
expressions, yielding:
\begin{align}
 I^-_n =& \sum_{j = 0}^{n-1} \binom{2n}{2j+1}
 a^{2j+1} \int_0^\infty dx \frac{x^{2n-2j-1}}{e^x+1} -
 \sum_{j= 0}^\infty \frac{a^{2n+2j+1}(2n)!}{(2n+2j+1)!} 
 \int_0^\infty dx\,\frac{d^{2j+1}}{dx^{2j+1}} 
 \left(\frac{1}{e^x+1}\right),\nonumber\\
 I^+_n =& \sum_{j = 0}^n \binom{2n+1}{2j} a^{2j} 
 \int_0^\infty dx\frac{x^{2n+1-2j}}{e^x+1} -
 \sum_{j= 0}^\infty \frac{a^{2(n+j+1)} (2n+1)!}{(2n+2j+2)!}
 \int_0^\infty dx\, \frac{d^{2j+1}}{dx^{2j+1}} 
 \left(\frac{1}{e^x+1}\right),
 \label{eq:FD_aux3}
\end{align}
where the explicit dependence of $I^\pm_n$ on $a$ was dropped for 
brevity. The second terms appearing above can be integrated, yielding:
\begin{equation}
 \int_0^\infty dx\, \frac{d^{2j+1}}{dx^{2j+1}} 
 \left(\frac{1}{e^x+1}\right) = -\frac{d^{2j}}{dx^{2j}}
 \left.\left(\frac{1}{e^x+1}\right)\right\rfloor_{x = 0}.
 \label{eq:FD_intaux}
\end{equation}
Noting that
\begin{equation}
 \frac{1}{e^x+1} + \frac{1}{e^{-x} + 1} = \frac{1}{2},
\end{equation}
it can be seen that $(e^x + 1)^{-1}$ admits the 
following series representation about $x = 0$:
\begin{equation}
 \frac{1}{e^x + 1} = \frac{1}{2} + \sum_{\ell = 0}^\infty 
 \frac{a_\ell}{(2\ell + 1)!} x^{2\ell + 1},
 \label{eq:FD_series}
\end{equation}
where the exact expression for $a_\ell$ is not 
important.\footnote{It is known that 
$a_{\ell} = E_{2\ell + 1}(0)$ is related 
to the Euler function $E_n(z)$.\cite{olver10}}
The only even term in the series 
\eqref{eq:FD_series} is the leading 
$1/2$, while all other terms are odd with respect to 
$x$. Thus, Eq.~\eqref{eq:FD_intaux} is non-vanishing
only when $j = 0$, such that:
\begin{equation}
 \int_0^\infty dx\, \frac{d^{2j+1}}{dx^{2j+1}} 
 \left(\frac{1}{e^x+1}\right) =
 \begin{cases}
  -\frac{1}{2}, & j = 0,\\
  0, & \text{ otherwise.}
 \end{cases}
\end{equation}
Thus, Eq.~\eqref{eq:FD_aux3} reduces to:
\begin{align}
 I^-_n(a) =& \sum_{j = 0}^{n-1} \binom{2n}{2j+1}
 a^{2j+1} \int_0^\infty dx \frac{x^{2n-2j-1}}{e^x+1} +
 \frac{a^{2n+1}}{2(2n+1)},\nonumber\\
 I^+_n(a) =& \sum_{j = 0}^n \binom{2n+1}{2j} a^{2j} 
 \int_0^\infty dx\frac{x^{2n+1-2j}}{e^x+1} +
 \frac{a^{2(n+1)}}{4(n+1)}.
 \label{eq:FD_aux4}
\end{align}

The terms in the sums appearing in Eq.~\eqref{eq:FD_aux4} 
are standard Fermi-Dirac integrals. The first few 
of these integrals can be computed exactly:
\begin{equation}
 \int_0^\infty \frac{x\, dx}{e^x + 1} = \frac{\pi^2}{12}, \qquad 
 \int_0^\infty \frac{x^3 dx}{e^x + 1} = \frac{7\pi^4}{120}.
\end{equation}
Thus, the following results can be obtained:
\begin{gather}
 I^-_0(a) = \frac{a}{2}, \qquad 
 I^-_1(a) = \frac{a \pi^2}{6} + \frac{a^3}{6},\qquad 
 I^-_2(a) = \frac{7 a \pi^4}{30} + \frac{a^3 \pi^2}{3} + 
 \frac{a^5}{10},\nonumber\\
 I^+_0(a) = \frac{\pi^2}{12} + \frac{a^2}{4}, \qquad 
 I^+_1(a) = \frac{7\pi^4}{120} + \frac{\pi^2 a^2}{4} + 
 \frac{a^4}{8}.
 \label{eq:FD}
\end{gather}

Aside from the integrals $I_{n}^\pm(a)$ considered in Eq.~\eqref{eq:FD_aux},
the analysis in the main text requires the computation of integrals when 
the subscript of $I$ is no longer an integer. Retaining the convention
that $n = 0, 1, 2, \dots$ is a natural number, the integrals $I_{n+1/2}^\pm(a)$
can be defined by analogy to Eq.~\eqref{eq:FD_aux} as follows:
\begin{align}
 I^-_{n+1/2}(a) =& \frac{1}{2} 
 \int_0^\infty dx\, x^{2n+1}\left(\frac{1}{e^{x-a}+1} - 
 \frac{1}{e^{x + a} + 1}\right),\nonumber\\
 I^+_{n+1/2}(a) =& \frac{1}{2} 
 \int_0^\infty dx\, x^{2n+2}\left(\frac{1}{e^{x-a}+1} + 
 \frac{1}{e^{x + a} + 1}\right).
 \label{eq:FD_Half_aux}
\end{align}
The equivalent of Eq.~\eqref{eq:FD_aux3} becomes:
\begin{align}
 I^-_{n+1/2} =& \sum_{j = 0}^{n} \binom{2n+1}{2j+1}
 a^{2j+1} \int_0^\infty dx \frac{x^{2n-2j}}{e^x+1} -
 \sum_{j= 0}^\infty \frac{a^{2n+2j+3}(2n+1)!}{(2n+2j+3)!} 
 \frac{d^{2j+1}}{dx^{2j+1}} 
 \left(\frac{1}{e^x+1}\right)_{x= 0},\nonumber\\
 I^+_{n-1/2} =& \sum_{j = 0}^n \binom{2n}{2j} a^{2j} 
 \int_0^\infty dx\frac{x^{2n-2j}}{e^x+1} -
 \sum_{j= 0}^\infty \frac{a^{2(n+j+1)} (2n)!}{(2n+2j+2)!}
 \frac{d^{2j+1}}{dx^{2j+1}} 
 \left(\frac{1}{e^x+1}\right)_{x=0},
 \label{eq:FD_Half_aux3}
\end{align}
where the arguments $a$ were dropped for brevity.
The integration with respect to $x$ can be performed in terms of 
the zeta function:
\begin{equation}
 \int_0^\infty \frac{dx\, x^{2n-2j}}{e^x + 1} = 
 (1 - 4^{j-n}) (2n-2j)! \zeta(2n + 1 - 2j),
\end{equation}
where the case $n = j$ is obtained via the following limit:
\begin{equation}
 \lim_{\nu \rightarrow 0} (1 - 4^\nu) \Gamma(1-2\nu) \zeta(1 - 2\nu) = \ln 2.
\end{equation}
For the second sum over $j$ in Eq.~\eqref{eq:FD_Half_aux3}, 
it is useful to note that:
\begin{equation}
 -\sum_{j= 0}^\infty \frac{a^{2j+k + 1}}{(2j+k + 1)!}
 \frac{d^{2j+1}}{dx^{2j+1}} 
 \left(\frac{1}{e^x+1}\right)_{x=0} = 
 \frac{1}{2} \sum_{j= 0}^\infty \frac{a^{j+k}}{(j+k)!}
 \frac{d^j}{dx^j} \left(\tanh\frac{x}{2}\right)_{x=0},
\end{equation}
where $\tanh \frac{x}{2} = (e^{-x}+1)^{-1}-(e^x+1)^{-1}$,
while $k = 2n+2$ for $I_{n+1/2}^-$ and $k = 2n+1$ for 
$I_{n-1/2}^+$.
The summation can be interpreted as a Maclaurin series
by noting that
\begin{equation}
 \frac{a^{j+k}}{(j+k)!} = \int_0^a da_1 
 \int_0^{a_1} da_2 \dots \int_0^{a_{k-1}} da_k \frac{a_k^j}{j!}.
\end{equation}
Using the relation:
\begin{equation}
 \sum_{j= 0}^\infty \frac{a^{j}}{j!}
 \frac{d^j}{dx^j} \left(\tanh\frac{x}{2}\right)_{x=0} = 
 \tanh\frac{a}{2},
\end{equation}
the following results can be obtained:
\begin{align}
 I^-_{n+1/2}(a) =& \sum_{j = 0}^{n} \frac{(2n+1)!}{(2j+1)!}
 a^{2j+1}(1-4^{j-n}) \zeta(2n+1-2j) \nonumber\\
 &+2^{2n+1}(2n+1)! \int_0^{\frac{a}{2}} da_1 \dots \int_0^{a_{2n+1}} 
 da_{2n+2} \tanh a_{2n+2},\nonumber\\
 I^+_{n-1/2}(a) =& \sum_{j = 0}^{n} \frac{(2n)!}{(2j)!}
 a^{2j} (1-4^{j-n}) \zeta(2n+1-2j) \nonumber\\
 &+ 2^{2n} (2n)! \int_0^{\frac{a}{2}} da_1 \dots \int_0^{a_{2n}} 
 da_{2n+1} \tanh a_{2n+1}.
 \label{eq:FD_Half_aux4}
\end{align}

For small values of $n$, Eq.~\eqref{eq:FD_Half_aux4} reduces to:
\begin{gather}
 I^-_{1/2}(a) = \frac{1}{2}[{\rm Li}_2(-e^{-a}) - {\rm Li}_2(-e^{a})],\qquad 
 I^-_{3/2}(a) = 3[{\rm Li}_4(-e^{-a}) - {\rm Li}_4(-e^{a})],\nonumber\\
 I^+_{-1/2}(a) = \ln\left(2 \cosh\frac{a}{2}\right), \qquad
 I^+_{1/2}(a) = -{\rm Li}_3(-e^{-a}) -{\rm Li}_3(-e^{a}),\nonumber\\
 I^+_{3/2}(a) = -12[{\rm Li}_5(-e^{-a}) + {\rm Li}_5(-e^{a})],
 \label{eq:FD_Half}
\end{gather}
where ${\rm Li}_n(z) = \sum_{k = 1}^\infty z^k / n^k$ is the
polylogarithm. The following relations were employed:
\begin{gather}
 \int_0^a dx\, \tanh x = \ln \cosh a, \qquad 
 \int_0^a dx\, \ln \cosh x = \frac{a^2}{2} - a\ln 2 + 
 \frac{\pi^2}{24} + \frac{1}{2} {\rm Li}_2(-e^{-2a}), \nonumber\\ 
 \int_0^a dx\, {\rm Li}_n(-e^{-2x}) = 
 -\frac{1}{2}(1 - 2^{-n}) \zeta(n+1) - \frac{1}{2} {\rm Li}_{n+1}(-e^{-2a}).
 \label{eq:polylog_int}
\end{gather}
For future convenience, the following properties of the polylogarithm are 
listed below:
\begin{align}
 {\rm Li}_2(-e^{-a}) + {\rm Li}_2(-e^a) =& -\frac{\pi^2}{6} - \frac{a^2}{2},\nonumber\\
 {\rm Li}_3(-e^{-a}) - {\rm Li}_3(-e^a) =& \frac{\pi^2 a}{6} + \frac{a^3}{6},\nonumber\\
 {\rm Li}_4(-e^{-a}) + {\rm Li}_4(-e^a) =&- -\frac{7\pi^4}{360} -
 \frac{\pi^2 a^2}{12} - \frac{a^4}{24}, \nonumber\\
 {\rm Li}_5(-e^{-a}) - {\rm Li}_5(-e^a) =& \frac{7\pi^4 a}{360} +
 \frac{\pi^2 a^3}{36} + \frac{a^5}{120}.
 \label{eq:polylog_refl}
\end{align}

\bibliographystyle{JHEP}
\bibliography{rotchem}

\end{document}